\documentclass[twocolumn]{aastex63}
\usepackage{amsmath}
\usepackage{graphicx}
\usepackage{float}
\usepackage[caption=false]{subfig}
\usepackage{txfonts}
\received{January 15, 2020}
\accepted{March 20, 2020}
\submitjournal{ApJ}

\shorttitle{3D Rad-MHD simulations of starspots}
\shortauthors{Panja et al.}

\begin{document}

\title{3D Radiative MHD simulations of starspots}

\correspondingauthor{Mayukh Panja}
\email{panja@mps.mpg.de}

\author[0000-0003-0832-4117]{Mayukh Panja}
\affiliation{Max Planck Institute for Solar System Research \\
Justus-Von-Liebig-Weg 3, D-37077, Göttingen, Germany \\}

\author[0000-0001-9474-8447]{Robert Cameron}
\affiliation{Max Planck Institute for Solar System Research \\
 Justus-Von-Liebig-Weg 3, D-37077, Göttingen, Germany \\}

\author[0000-0002-3418-8449]{Sami K. Solanki}
\affiliation{Max Planck Institute for Solar System Research \\
Justus-Von-Liebig-Weg 3, D-37077, Göttingen, Germany \\}
 \affiliation{School of Space Research, Kyung Hee University \\
Yongin, Gyenoggu-Do, 446-701,Korea\\}
 







\begin{abstract}
 There are no direct spatially resolved observations of spots on stars other than the Sun and starspot properties are inferred indirectly through lightcurves and spectropolarimetric data. We present the first self-consistent 3D radiative MHD computations of starspots on G2V, K0V and M0V stars, which will help to better understand observations of activity, variability and magnetic fields in late-type main-sequence stars. We used the MURaM code, which has been extensively used to compute "realistic" sunspots, for our simulations. We aim to study how fundamental starspot properties such as intensity contrast, temperature and magnetic field strength vary with spectral type. We first simulated in 2D, multiple spots of each spectral type to find out appropriate initial conditions for our 3D runs. We find that with increasing stellar effective temperature, there is an increase in the temperature difference between the umbra of the spot and its surrounding photosphere, from 350K on the M0V star to 1400K on the G2V star. This trend in our simulated starspots is consistent with observations. The magnetic field strengths of all the starspot umbrae are in the 3-4.5 kG range. The G2V and K0V umbrae have comparable magnetic field strengths around 3.5 kG, while the M0V umbra has a relatively higher field strength around 4 kG. We discuss the physical reasons behind both these trends. All of the three starspots develop penumbral filament-like structures with Evershed flows. The average Evershed flow speed drops from 1.32 km s$^{-1}$ in the G2V penumbra to 0.6 km s$^{-1}$ in the M0V penumbra.

\end{abstract}



\section{Introduction} \label{sec:intro}

What do spots on stars other than the Sun look like? How dark are they and how strong are their magnetic fields? The lack of direct spatially resolved observations of other stars makes it difficult to answer such questions. Some properties of starspots can be inferred using lightcurves and spectropolarimetric data. The most prevalent methods used to infer information about starspot temperatures and filling factors are lightcurve inversions \citep{light_curve_spot}, molecular bands modelling \citep{Huene87,Neff95,ONeal96,ONeal2004}, line depth ratios, \citep{gray_1996,Catalano02} and Doppler imaging \citep{Goncha77,Vogt83}. Recently, planetary transit lightcurves have  also been used to gain information about starspot properties \citep{Mancini2013,Espinoza19}. All of these methods, with the exception of molecular lines, suffer from the drawback that they can only detect spots that are large enough to leave an imprint on disk integrated quantities. Even for sufficiently large spots, it is difficult to distinguish between temperature contributions from the umbra and the penumbra. Additionally, it is worthwhile to note that different techniques have been known to produce very different spot temperatures for the same star, a notable example being spots on the G1.5V sun-like star Ek Draconis. While \cite{Dorren94} found a spot temperature of 5400 K using lightcurve modelling, \cite{Strassmeier98} and \cite{ONeal2004} reported spot temperatures of 4800 K and 3800 K using Doppler Imaging and modelling molecular bands, respectively. A comprehensive discussion on the various methods of observations of starspots and their advantages and disadvantages can be found in the reviews by \cite{Berdyugina2005} and \cite{Strassmeier2009}.

The measurement of magnetic field strengths on other stars is even more difficult as the lack of spatial resolution means the net circular polarisation tends to be cancelled out by oppositely directed magnetic fields. However, in rapidly rotating stars, if magnetic features of opposite polarities are sufficiently separated in longitude, the Doppler effect disentangles them in the Stokes V component of magnetically sensitive lines and this is exploited by the Zeeman Doppler Imaging technique to map magnetic fields on the stellar surface \citep{Semel89,Donati1990}. For slowly rotating stars if the line broadening due to the Zeeman effect is larger than the rotational broadening, and the surface coverage by such very strong fields is large enough, the magnetic field can be inferred from the amount of broadening \citep{Gray84}. For starspots, there is the added issue that, being dark, they provide little contribution to line profiles integrated over the stellar surface. This makes measuring their fields particularly challenging. However, molecular lines that form primarily inside starspots and have little contribution from quiet-star regions are being increasingly used to better constrain starspot magnetic field strengths \citep{Afram_Berdy_15, Afram_Berdy_19}.  For a review of stellar magnetic field measurements, see \cite{Reiners2012}.  

 The review by \cite{Berdyugina2005} compiled a list of starspot observations obtained by using various methods and, despite the many limitations placed on observations, she found a clear trend when she plotted starspot temperature contrast against stellar surface temperature. The cooler the star, the lower is the difference between the spot and quiet star temperature and the physical reasons for this are unclear.
 
All of the above-mentioned reasons point to the need for performing MHD simulations to better understand the physics of starspots and how it may differ from that of sunspots. Such simulations of  thermal and magnetic structures of spots on other stars could also be useful to interpret observations and may even help in understanding the underlying stellar dynamo processes.

Although 3D radiative hydrodynamic simulations of near-surface layers for stars other than the Sun had been performed as early as 1990 by \cite{NordandDravis90}, the first stellar simulations including magnetic fields were performed by \cite{Beeck0} and \cite{Wedemeyer_Mstar}. Subsequently, \cite{Beeck3,Beeck4}, studied the effects of the magnetic field on surface properties such as intensity contrast and granulation structure in the spectral types F through M, using the MURaM code \citep{Vogler2005}. They further generated synthetic disk integrated spectral line profiles that can be compared with observations. Similar numerical studies using the CO$^{5}$BOLD code, investigating the effects of magnetic fields on surface processes in cool stars, have been carried out by \cite{steiner_salhab_cobold} and \cite{cobold}.

Spots on the Sun have been extensively simulated \citep{rempel09b,rempel09a} using the MURaM code. These simulations have been fairly successful in reproducing the fundamental observed features of sunspots -  a dark umbra dotted with bright umbral dots, surrounded by  brighter penumbrae composed of filaments with thin dark cores, and the Evershed flow directed away from the umbral region towards the quiet Sun. Such simulations have been used to investigate the physical origins of many hitherto ill-understood observed properties of sunspots \citep{Rempel_2011b,Rempel_2011c,Rempel_2012,Rempel_2015_decay,Siu_Tapia_2018}. 

In this paper, we have used the MURaM code to perform the first-ever ab-initio simulations of spots on cool main-sequence stars other than the Sun, to investigate their fundamental properties, specifically - brightness relative to the stellar surface and magnetic field strength, as functions of stellar spectral type. As host stars we have considered a G2V, a K0V and an M0V star.
Before carrying out 3D simulations, we first performed 2D computationally inexpensive simulations to navigate the parameter space with the intent of a) identifying suitable initial conditions for our final 3D runs and b) testing the sensitivity of our results to the variations in the initial conditions. First, we describe the setup of the simulations in Section \ref{sec:sim}. We have presented a summary of the 2D simulations in Section \ref{sec:2d_res}, while more details are given in the Appendix. We then present the results of our 3D simulations in Section \ref{sec:3d_res}. 
 Subsequently in Section \ref{sec:disc}, we provide physical explanations for the existence of such trends. In Section \ref{sec:summary} we summarise the conclusions of this paper.

\section{Simulations}\label{sec:sim}

We have used the MURaM (Max-Planck University of Chicago Radiative MHD) code \citep{Vogler2005}, which solves the MHD equations along with the radiative transfer equation and an equation of state that takes into account the effects of partial ionization. The version of the code used was the one employed by \cite{Beeck1,Beeck2,Beeck3,Beeck4}. Since we do not generate synthetic line profiles in this study, we used the grey approximation for the radiative energy transport.

Table \ref{table:1} lists the dimensions and initial physical properties of the simulation boxes used for our 3D and 2D runs. We have simulated the spectral types - G2V, K0V, and M0V. The atmospheres of the M0V and K0V stars were obtained by starting from a solar atmosphere and changing the gravity (assumed constant throughout the computational domain) and entropy density of the plasma at the lower boundary until our desired effective temperatures were achieved. All the simulated stars are assumed to have solar metallicities. 

\subsection{3D Simulation Setup}
 \begin{figure}
   \centering
	\hspace*{-1.2cm}\includegraphics[width=10.0cm]{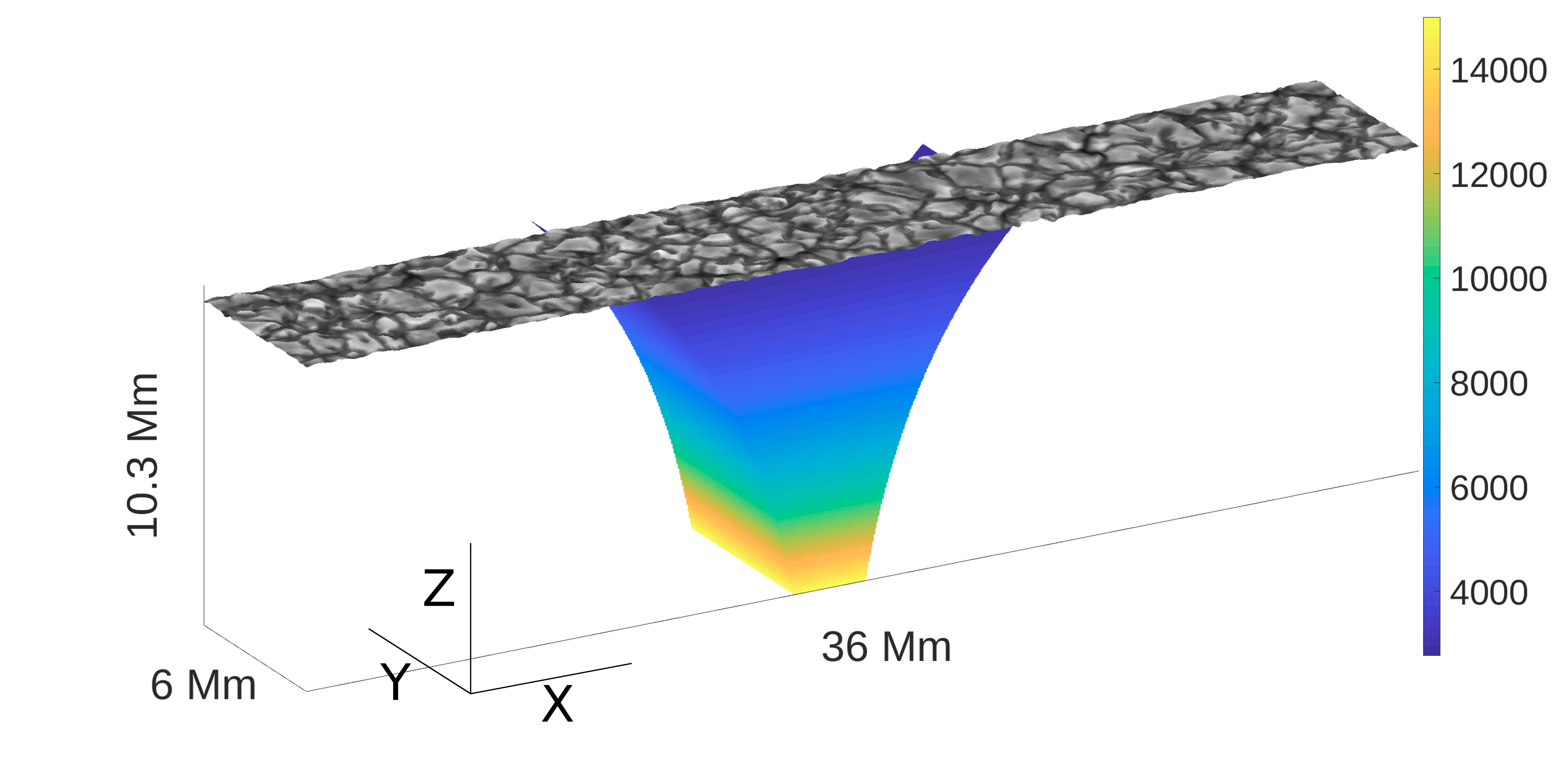}
           \caption{Initial magnetic structure for the G2V spot simulation. The depicted flux tube was introduced after the hydrodynamic run had achieved thermal equilibrium.  The colours on the flux tube show B$_{z}$ in Gauss (see right colorbar). The gray surface shows the emergent bolometric intensity at the time the flux tube was inserted.
               }
        \label{fig:fig1}
    \end{figure}

\begin{table}
\caption{Simulation Box Properties - 3D and 2D}             
\label{table:1}      
\flushleft                 
\begin{tabular}{c c |c c c c }        
\hline\hline                 
Spectral Type& & G2V & K0V & M0V\\ 
\hline\hline  
    &  &  &  &  \\
 log g\tablenotemark{a}& & 4.438 & 4.609  & 4.826 \\
   &  &  &  &  \\
  T$_{eff}$ [K] & 3D & 5824.5 & 4809.5  & 3871.6 \\
         & 2D & 5738.2 & 4894.3 & 3864.8 \\
     &  &  &  &  \\
\hline 
     &  &  &  &  \\

   Box Height [Mm] & 3D  & 10.3 & 6  & 1.3 \\
        &    2D        &  7.3    & "   & "     \\
        &  &  &  &  \\
   Photosphere [Mm]& 3D & 9.8 & 5.43  & 1.05 \\
            & 2D      & 6.4 & 5.44 & 1.04 \\
            &  &  &  &  \\
   Box Length\tablenotemark{b} [Mm]& 3D & 36 & 27  & 10 \\
    & 2D      & " & " & "    \\
    &  &  &  &  \\
  \# of $H_{p}$ \tablenotemark{c}(above,below)& 3D & 5, 11 & 8.2, 11.3  & 8.1, 8.2 \\
   & 2D & 6.3, 9.3 & 4, 11.1  & 8.6, 8.0 \\
   &  &  &  &  \\
   Resolution (hor,vert)[km]& 3D & 48, 17  & 36, 15  & 13.3, 4 \\ 
   & 2D & 48, 21.5  & "  & "\\
   &  &  &  &  \\
   

\hline                                   
\end{tabular}
 \\
\tablenotetext{a}{g is the surface gravity in cm$^{2}$ s$^{-1}$} 

\tablenotetext{b}{All boxes have a length:width ratio of 6:1}

\tablenotetext{c}{Number of pressure scale heights}
\end{table}

\begin{table}
\caption{Initial Magnetic Flux Tube Properties}             
\label{table:2}      
\centering                          
\begin{tabular}{c c|c c c c }        
\hline\hline                 
Spectral Type& & G2V & K0V & M0V\\ 
\hline\hline  

     & & & & \\
   B$_{opt}$[kG]& 3D & 3  & 3.2  & 7 \\
   & 2D (range) & 2.2 - 8.8 & 3.2 -12 & 3 - 12 \\
    & & & & \\
   B$_{bot}$[kG]& 3D & 15  & 14  & 15 \\
   & 2D (range) & 4 - 14 & 6 - 16 & 5 - 18 \\

\hline                                   
\end{tabular}
\end{table}

We started our simulations with hydrodynamical runs of the three spectral types of stars. Once the hydrodynamical runs were sufficiently relaxed, we put in wedge-shaped magnetic flux tubes as initial conditions for the magnetic field. The horizontal extents of the inserted flux tubes at the optical surface were chosen such that they covered a similar number of granules in all of the different spectral types.   Figure \ref{fig:fig1} shows the initial condition for the G2V starspot simulation.

 The initial vertical field strength ($B_{z}$) of the flux tube is dependent only on geometrical height and is prescribed by:

\begin{align}
\label{eqn:1}
 B_{z} & = B_{bot} \exp{\frac{-z}{\sigma}}, &  x\in [-w/2,w/2], \nonumber \\ 
      & = 0,                                & \mbox{otherwise}.
\end{align}

Here $z$ is the height from the lower boundary, $x$ is the longer horizontal dimension, and $w$ is the width of the flux tube at every z, determined such that the vertical flux through every height is constant.

At $z=0$ we set $B_{z}$ to $B_{bot}$ and at $z = h_{opt}$ (height of optical surface from lower boundary) , we set $B_{z}$ to $B_{opt}$, and this yields $\sigma$  to be

\begin{equation}
\sigma = h_{phot}/\log(\frac{B_{bot}}{B_{opt}}).     
\end{equation}

We choose B$_{y}$ (y being the shorter horizontal dimension) to be zero initially, everywhere. 
 Thus the $\nabla \cdot B = 0$ constraint demands that ${\frac {\partial B_{x}}{\partial x}}+{\frac {\partial B_{z}}{\partial z}} = 0 $ and we calculate $B_{x}$ as follows: 
 
\begin{equation}
\frac {\partial B_{z}}{\partial z} =  -\frac{B_{bot}}{\sigma}\exp{\frac{-z}{\sigma}} = -\frac{Bz}{\sigma}. 
\end{equation}
Therefore,
\begin{equation}
\frac {\partial B_{x}}{\partial x} =  \frac{Bz}{\sigma}. 
\end{equation}

With $B_{z}$ being independent of x, and $\sigma$ being a constant, the integration is straightforward and 
\begin{equation}
B_{x} = \frac{Bz}{\sigma} x,
\end{equation}
with $B_{x}$ =0 at $x =0$ (the centre of the  flux tube). 

Our choices of $B_\mathrm{bot}$ and $B_\mathrm{opt}$ have been tabulated in Table \ref{table:2}. Our 3D boxes extended to similar pressure scale depths and we chose similar values of  $B_\mathrm{bot}$ for all the three stars - 15,14, and 15 kG for the G2V, K0V, and M0V stars respectively. We picked values of $B_\mathrm{opt}$ that one would naively guess from just the surface pressures  – 3 and 3.2 kG for the G2V and K0V stars as they have comparable surface pressures. For the M0V star which has a surface pressure 5 times that of the G2V quiet star, a higher $B_\mathrm{opt}$ of 7 kG, which is roughly 3*sqrt(5), was chosen.

The magnetic field at the upper boundary was matched to a potential field configuration and the upper boundary was kept open to plasma flows. At the lower boundary, the flow velocity was artificially set to zero inside the flux tube, effectively "tying down" the flux tube to the lower boundary. This also mimics the physical effects of the flux tube extending below the lower boundary as heat flow by means of convection is prohibited.

 \begin{figure}
   \centering
	\hspace*{-0.4cm}\includegraphics[width=10.0cm]{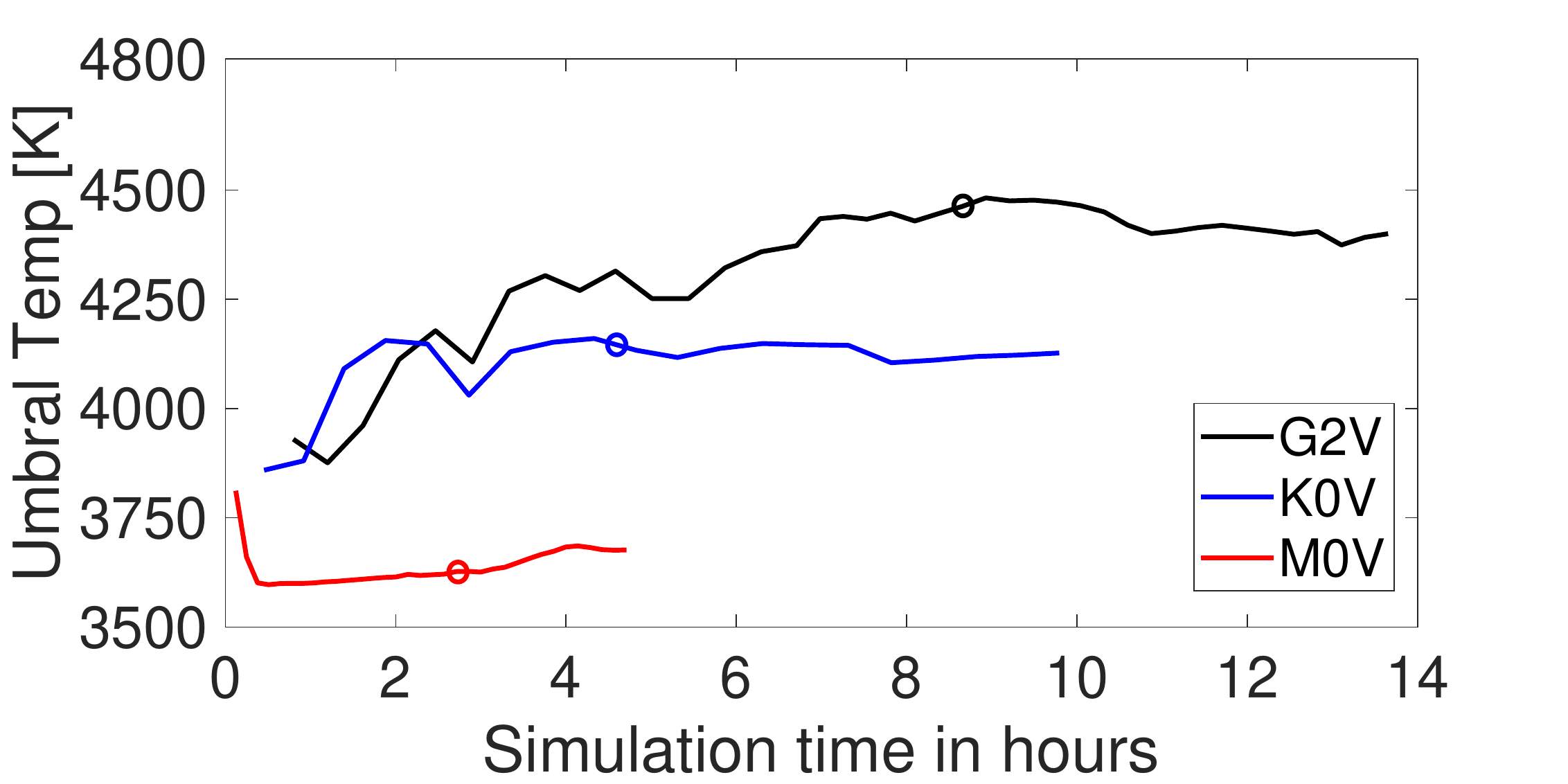}
	\hspace*{-0.4cm}\includegraphics[width=10.0cm]{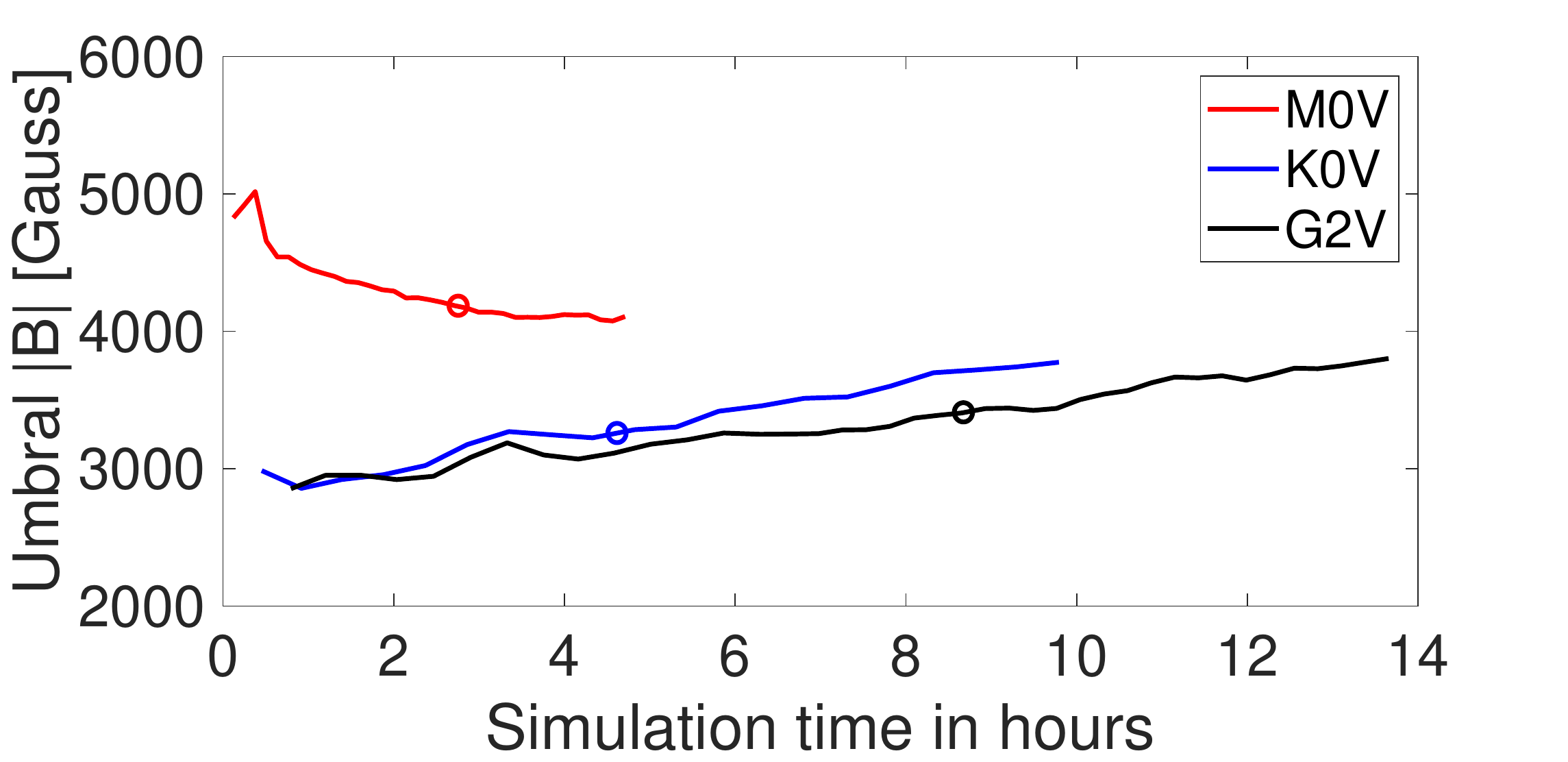}
           \caption{Evolution of average umbral temperatures (top) and average umbral magnetic field strengths (bottom) with time for the three simulated starspots. The circles mark the points in time when our snapshots were taken, which were after a similar number of granule lifetimes. }
        \label{fig:fig2}
    \end{figure}
    
     \begin{figure}
   \centering
	\hspace*{0.2cm}\includegraphics[width=10.0cm]{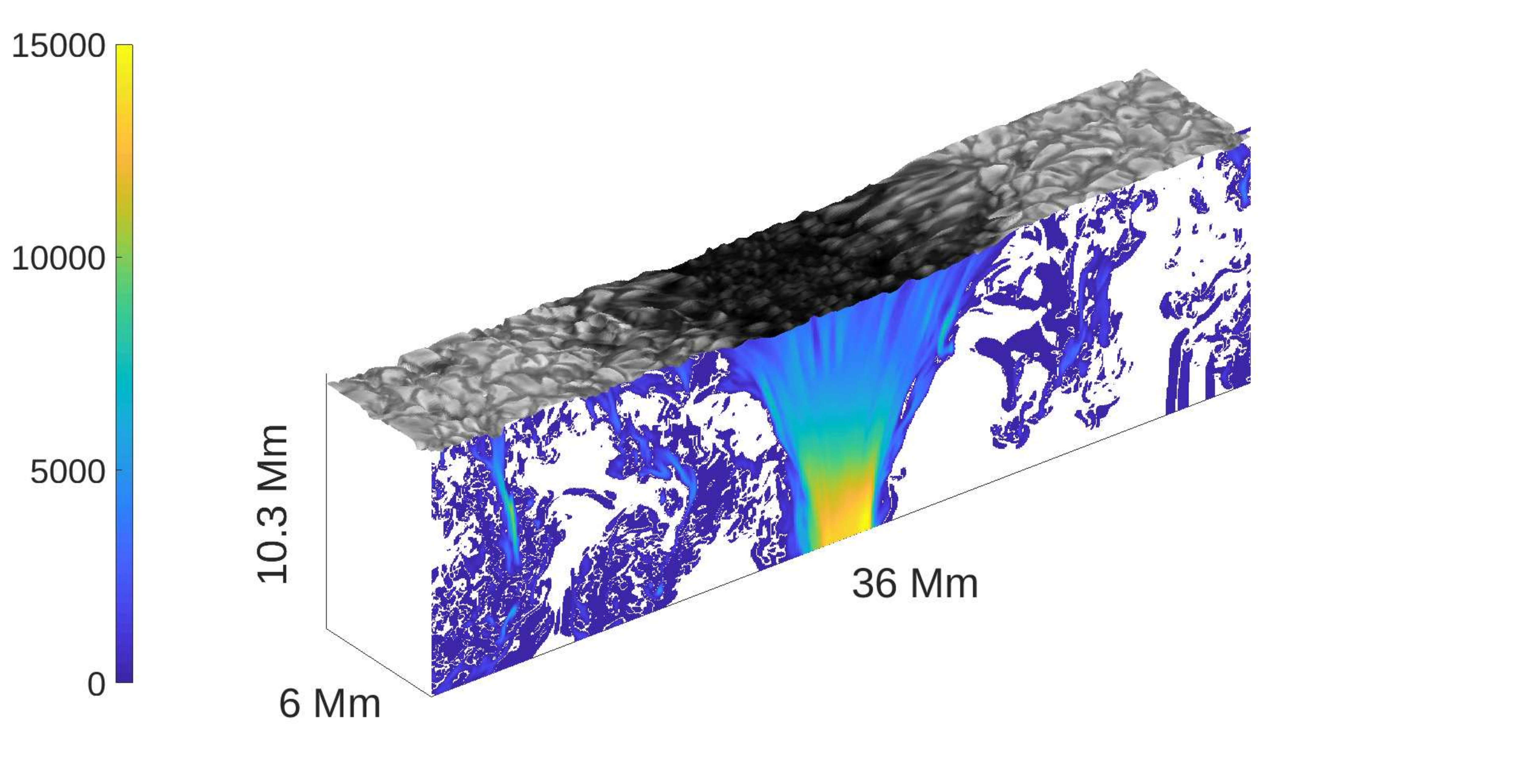}
           \caption{A 3D rendition of the simulated  G2V star spot at the timestep of our analysis. The top grayscale surface shows the bolometric intensity  and the vertical slice shows B$_{z}$. The corrugations of the top surface represent the variations in the geometric height of the optical surface ($\tau$ =1). The colorbar shows B$_{z}$ in Gauss.
               }
        \label{fig:fig3}
    \end{figure}

\subsubsection{\textit{Choosing a snapshot}}    
All of our simulated starspots are dynamical in nature, and they underwent several stages of evolution in the course of the simulation. After an initial highly dynamic phase, the magnetic flux tubes reach magnetostatic equilibrium with the surrounding fluid typically within the first hour of stellar run time. Subsequently, all of the three spots formed penumbral filament-like structures and the G2V and the K0V spots developed umbral dots. After a few additional hours, all of the flux tubes begin to get distorted underneath the surface by flute-like intrusions from the surrounding fluid. If allowed to develop for a sufficiently long time these intrusions manifest themselves at the optical surface as lightbridges. This process sets in at different times for the three spectral types, earliest for the M star and latest for the G star. Figure \ref{fig:fig2} shows the evolution of average umbral temperatures and average umbral magnetic field strengths of the three simulated spots with time.

We have chosen the time of our snapshots such that the umbral temperatures were well past their initial fluctuating phases and the spots had developed penumbral filament-like structures, but also sufficiently ahead of the umbrae of the spots getting too distorted by intruding lightbridges, for a meaningful analysis. A 3D rendition of the G2V starspot at the instance our snapshot was taken is depicted in Figure \ref{fig:fig3}. 
       
The G2V, K0V, and M0V snapshots were taken after 8.7, 4.6 and 2.7 hours of stellar runtime respectively. It is important to note that the timescales of granule evolution are different for different spectral types. A typical granule lifetime on the Sun is on average about 6 minutes whereas on the M0 star the granules last around 2.5 minutes. \citep{Beeck2}. Consequently, the snapshots were taken after a similar number of granule  lifetimes - 80, 60 and 65 respectively for the G2V, K0V, and M0V stars.

\subsection{2D Simulation Setup}
The initial conditions used for the 2D simulations were the 2D analogs of the conditions used for the 3D runs, with the flux tubes being inserted in a 2D hydro-dynamical run. The computational ease afforded by having one less dimension allowed us to vary the two free parameters- $B_\mathrm{bot}$ and $B_\mathrm{opt}$, of the inserted flux tubes to simulate multiple spots for each spectral type. This allowed us to test the sensitivity of our results to changes in our chosen initial parameters. We simulated 24 spots in 2D, 8 for each spectral type. First, we kept $B_\mathrm{opt}$ constant (2.2, 3.2 and 3 kG for the G2V, K0V and M0V cases respectively) and explored a range of values of $B_\mathrm{bot}$. We started with $B_\mathrm{bot}$s of 4 (G2V), 6 (K0V) and 5 (M0V) kG and increased them roughly by a factor of 3, in steps of 2-3 kG. This constituted 6 of the 8 simulated spots for each spectral type. Then, keeping $B_\mathrm{bot}$ constant we increased $B_\mathrm{opt}$ by factors of 2 and 4. Table \ref{table:2} lists our choices for $B_\mathrm{bot}$ and $B_\mathrm{opt}$. A detailed description of the 2D runs is given in the Appendix. 

\begin{figure}
  
   \centering
   \hspace*{-0.2cm}\includegraphics[scale=0.35]{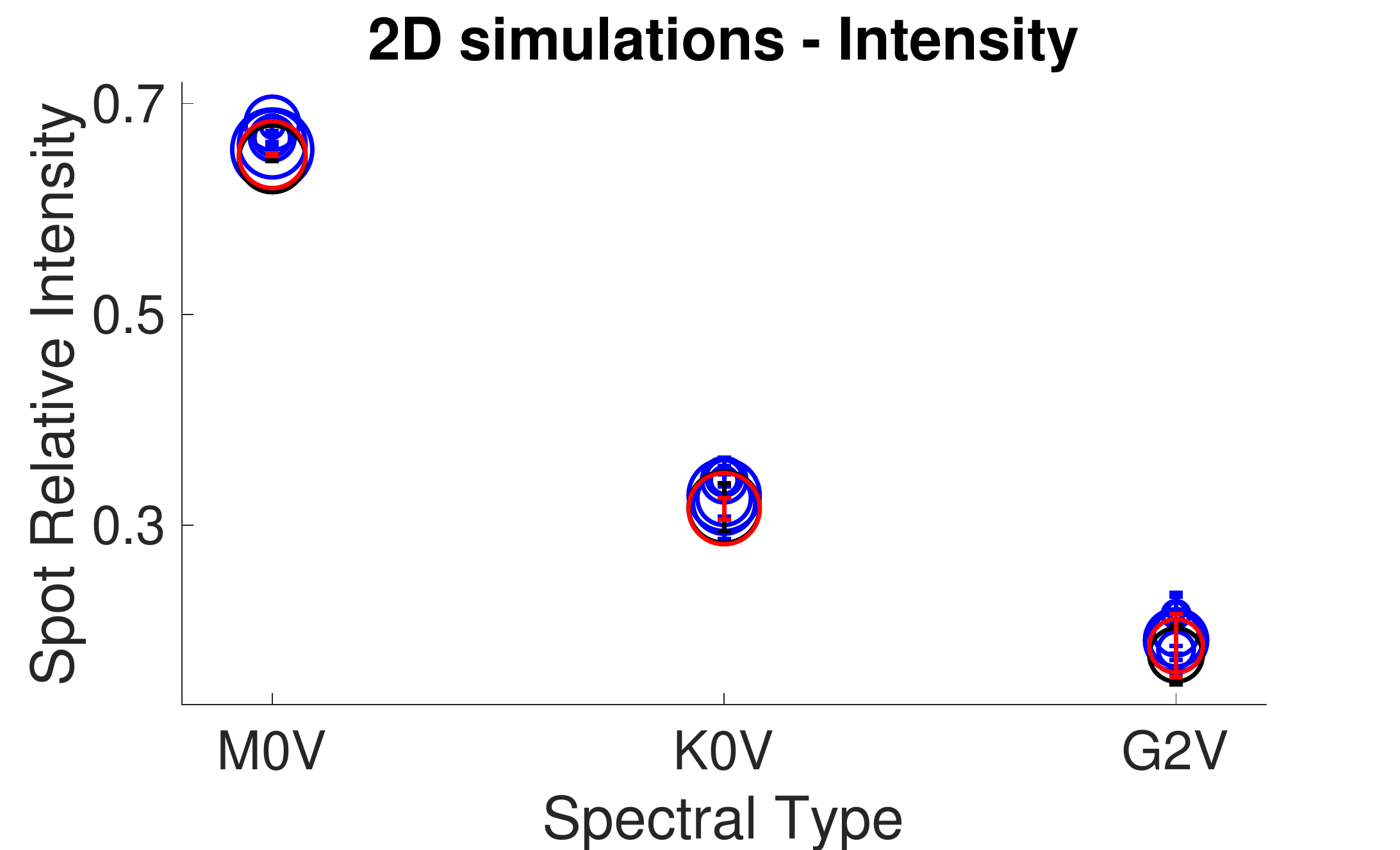}
   \hspace*{-0.2cm}\includegraphics[scale=0.35]{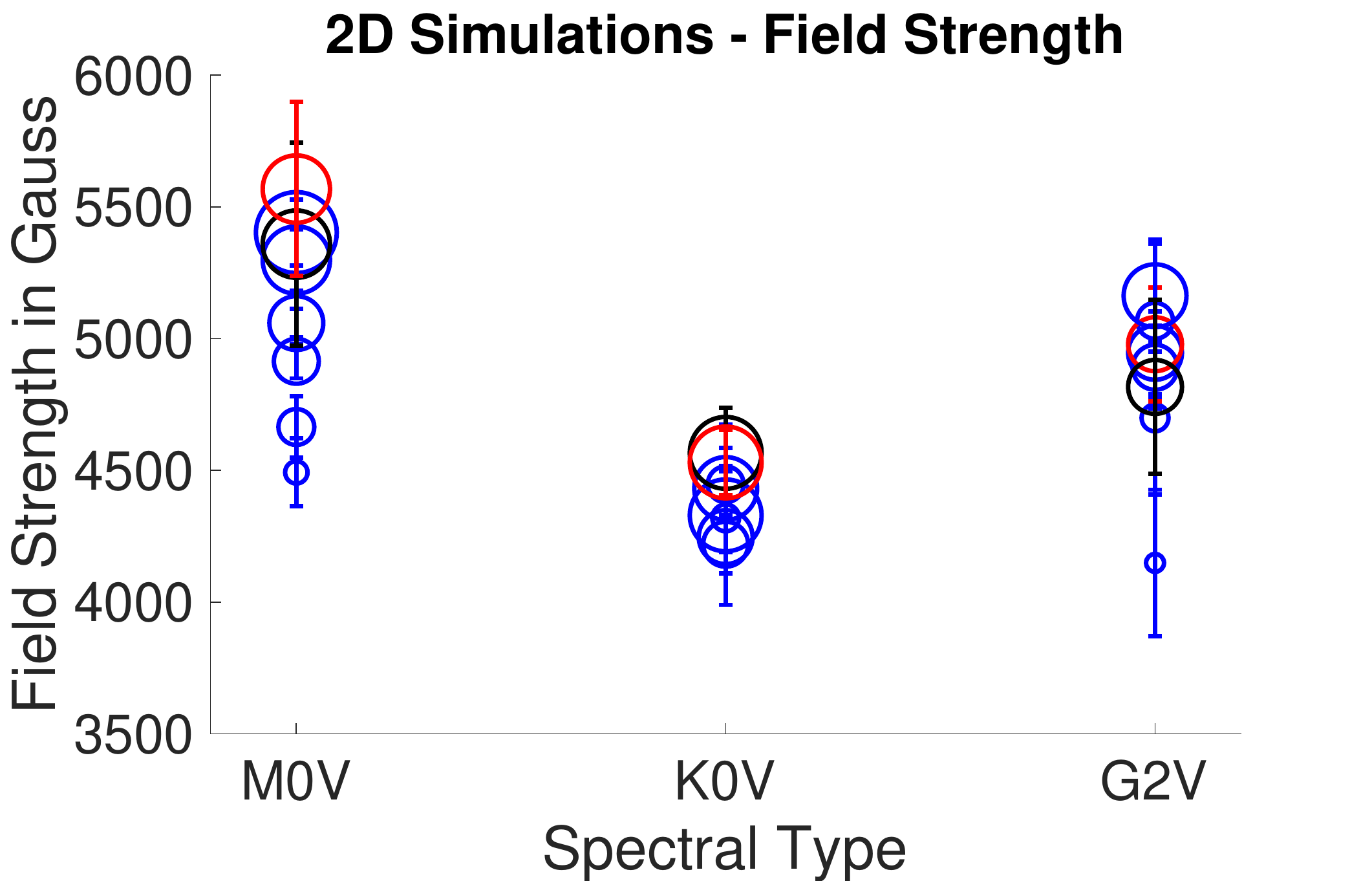} 
   
\caption{Upper Panel: Average umbral intensities, normalized by their quiet star intensities, of the 2D simulations, plotted against spectral type. Lower Panel: Average umbral field strengths at the optical surface. The size of the circles, in both the upper and lower panels, indicate the strength of $B_\mathrm{bot}$ while the color indicates the strength of $B_\mathrm{opt}$. Blue circles indicate runs with original $B_\mathrm{opt}$, while red and black correspond to runs where $B_\mathrm{opt}$ was increased by factors of 2 and 4 respectively. See Section 2.2 and the Appendix for more details.}
             \label{fig:fig4}
    \end{figure}

\begin{figure*}
  
   \centering
   \hspace*{-2cm}\includegraphics[scale=0.4]{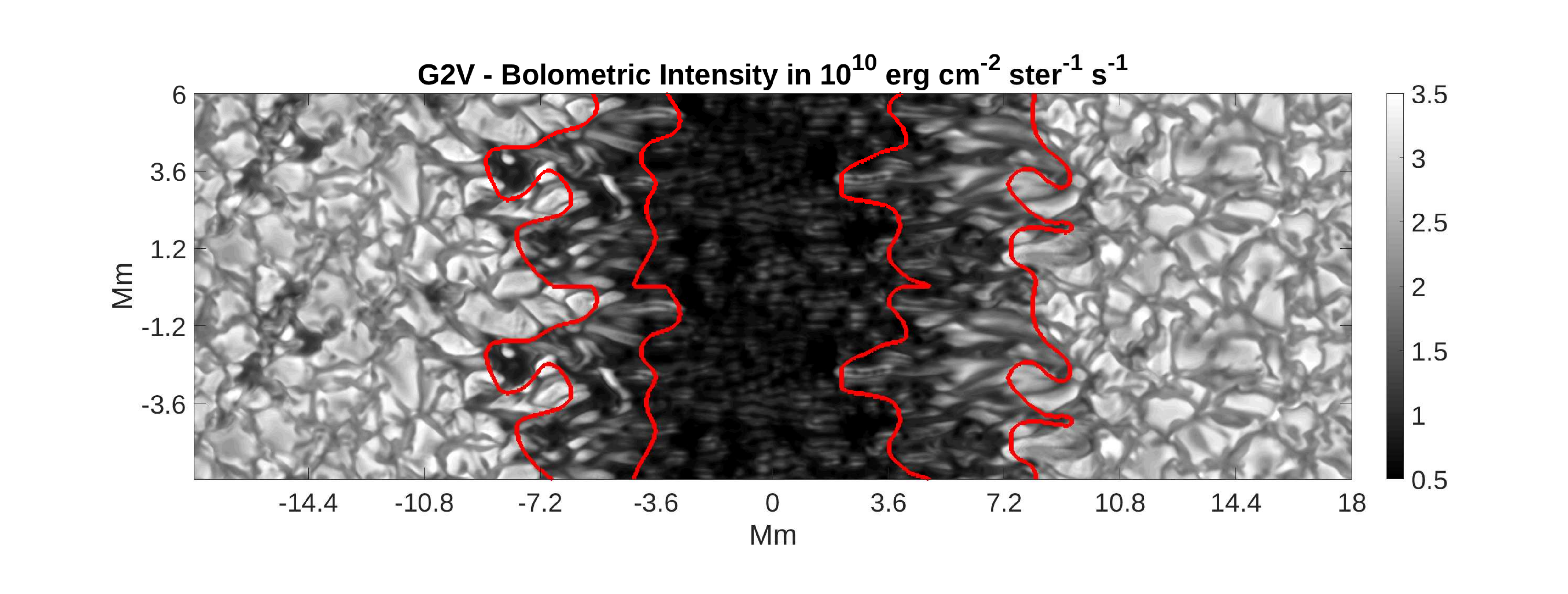} 
   \hspace*{-2cm}\includegraphics[scale=0.4]{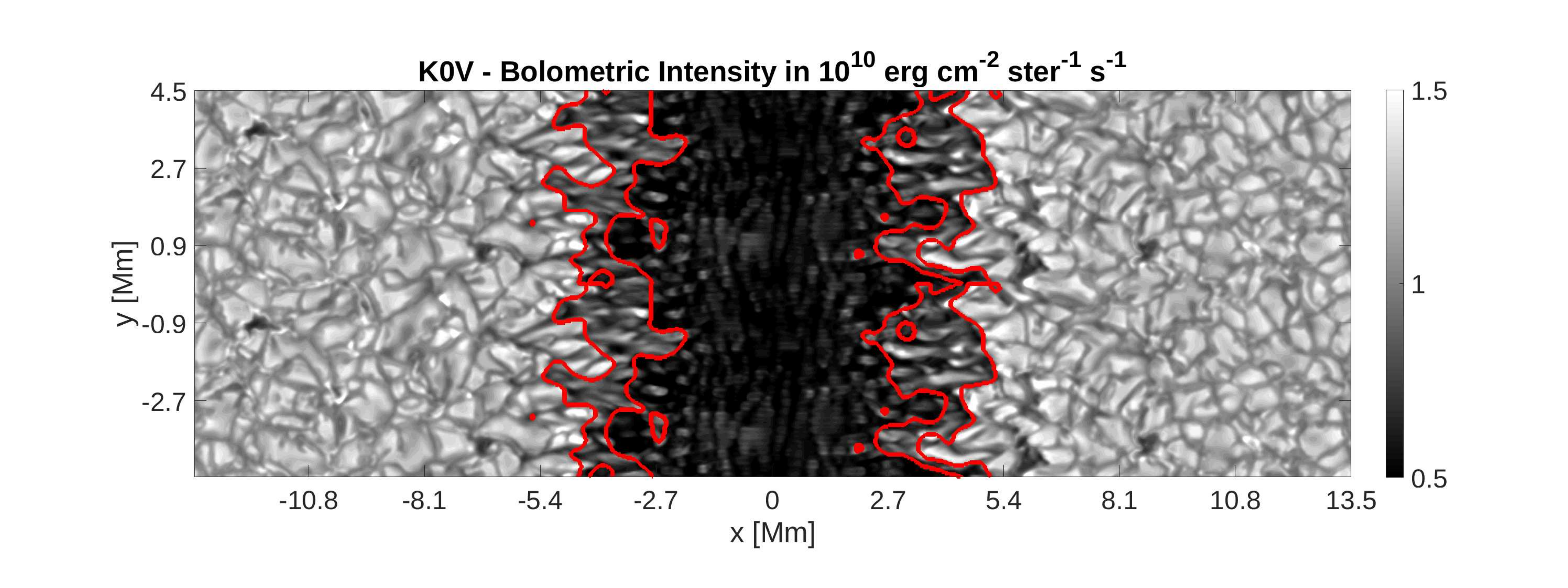}
   \hspace*{-2cm}\includegraphics[scale=0.4]{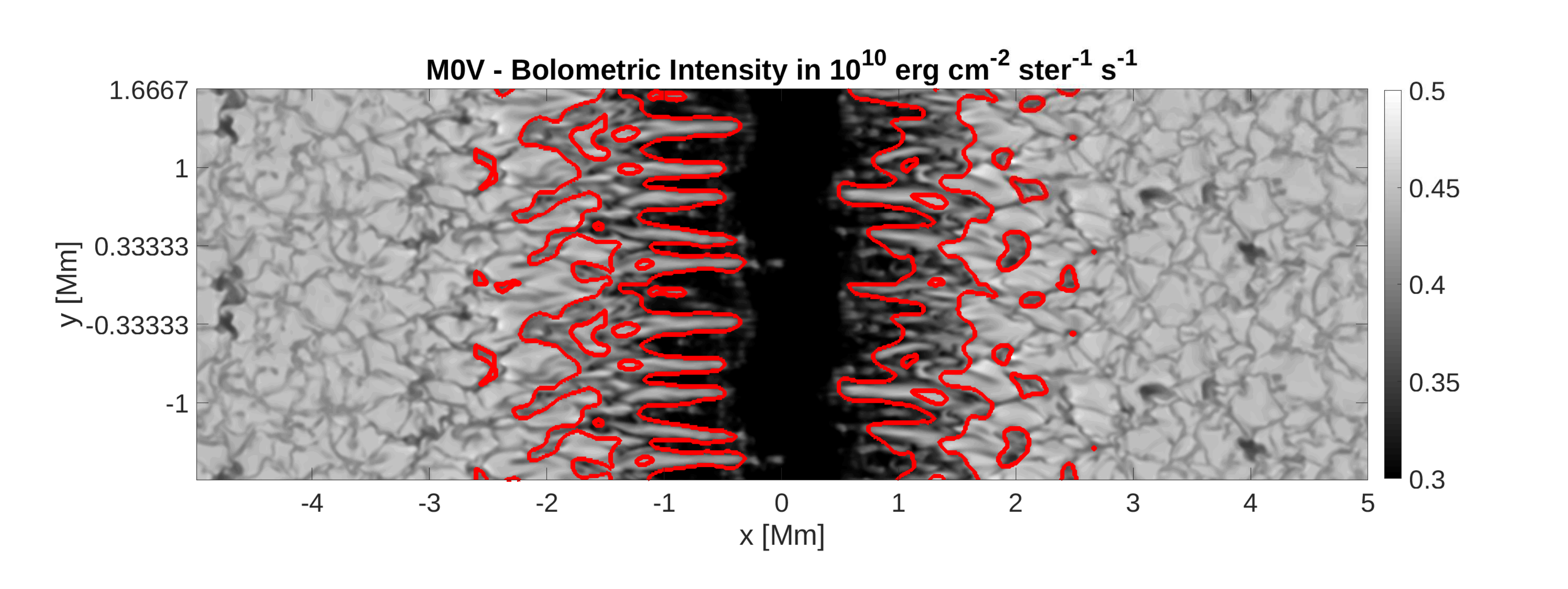}
 
\caption{Bolometric intensity images of the simulated G2V, K0V and M0V starspots, repeated in the y-directon. The colorbar shows the intensity in units of 10$^{10}$ erg cm$^{-2}$ ster$^{-1}$ s$^{-1}$. The red contours mark the boundaries of the penumbra.}
              \label{fig:fig5}
    \end{figure*}

 \begin{figure}
   \centering
   
	\hspace*{-1.5cm}\includegraphics[width=10.5cm]{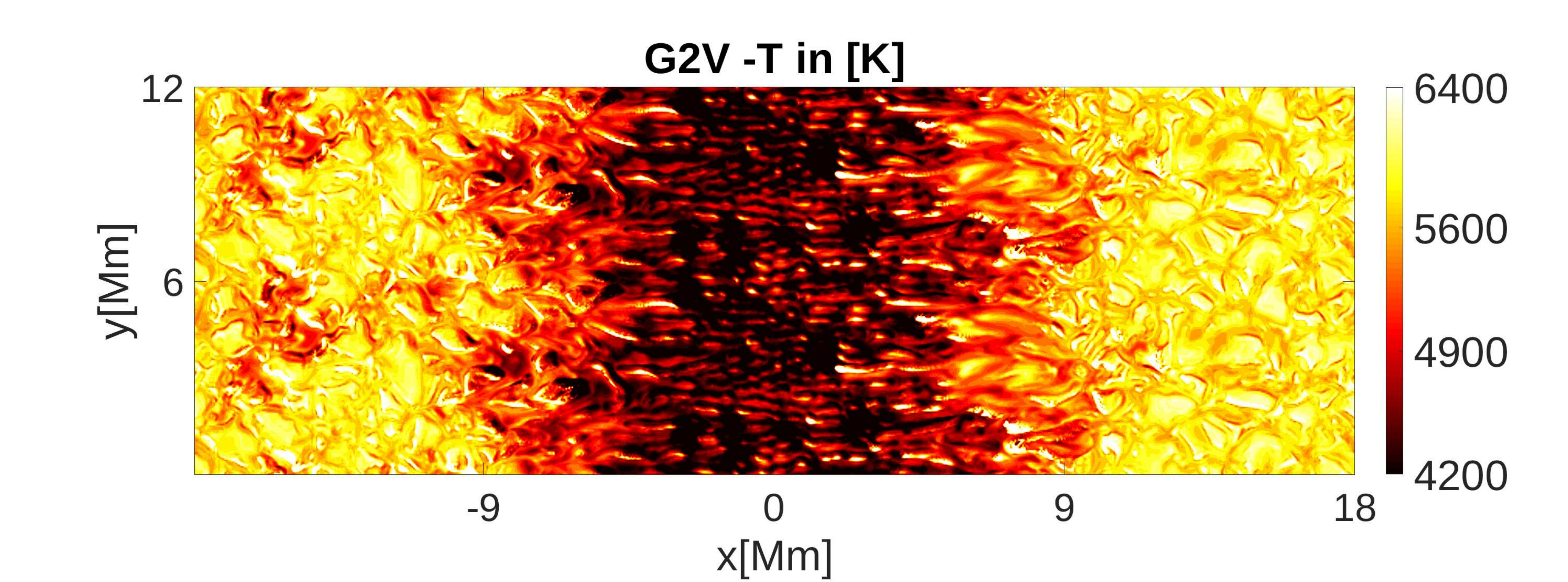}
    \hspace*{-1.5cm}\includegraphics[width=10.5cm]{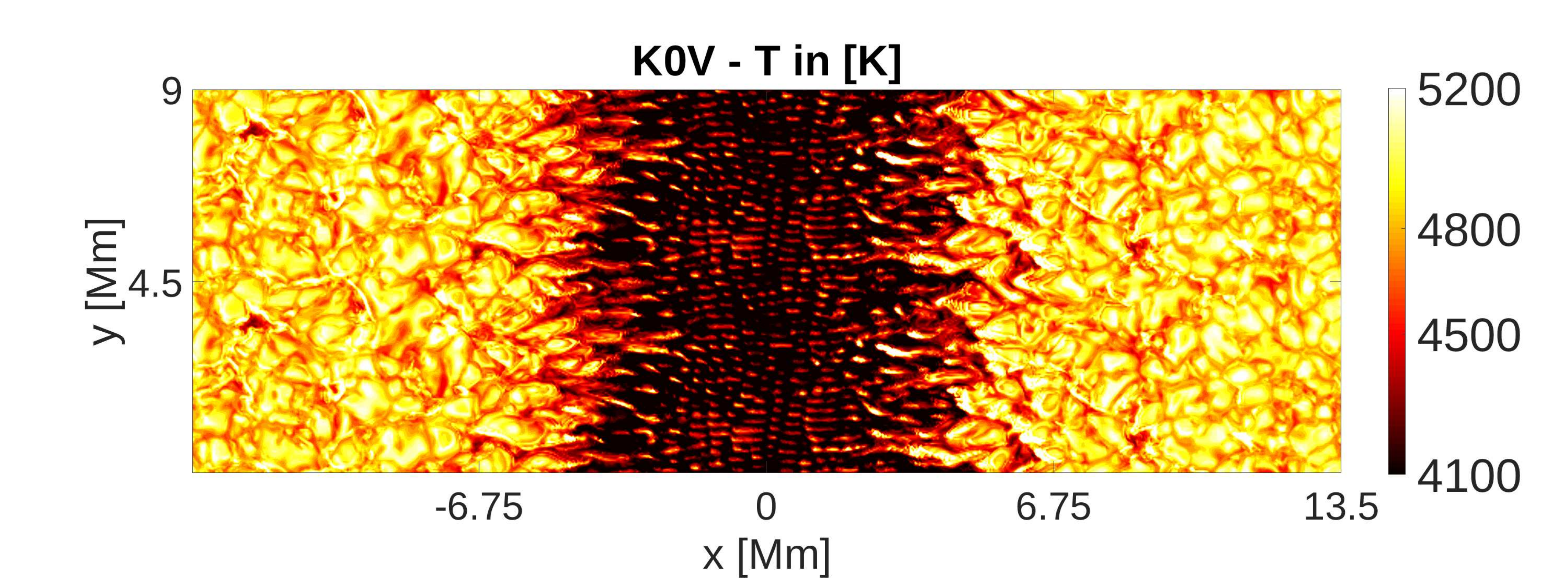}
    \hspace*{-1.5cm}\includegraphics[width=10.5cm]{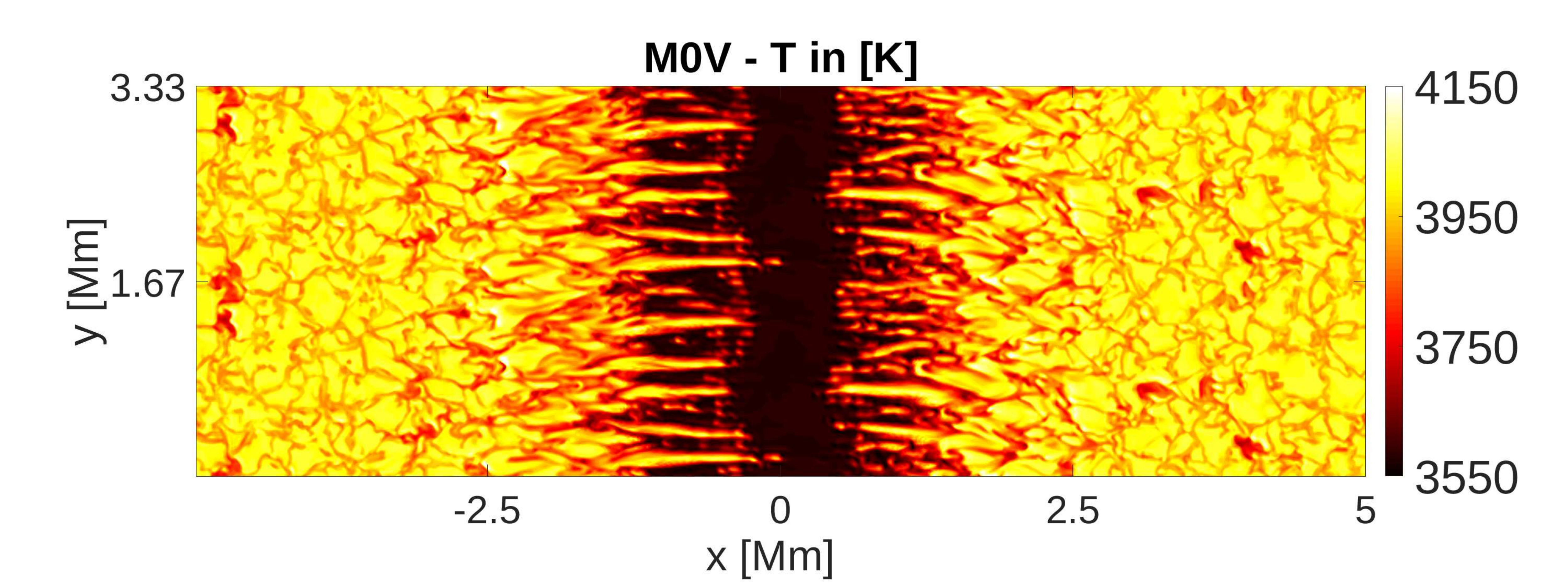}
    \caption{Temperature maps, in Kelvin (see colour bars at right of the individual panels), at the $\tau$ =1 surfaces of the three simulated stars - G2V, K0V and M0V.
               }
        \label{fig:fig6}
    \end{figure}
    
\begin{figure}
  
   \centering
   \hspace*{-0.2cm}\includegraphics[scale=0.35]{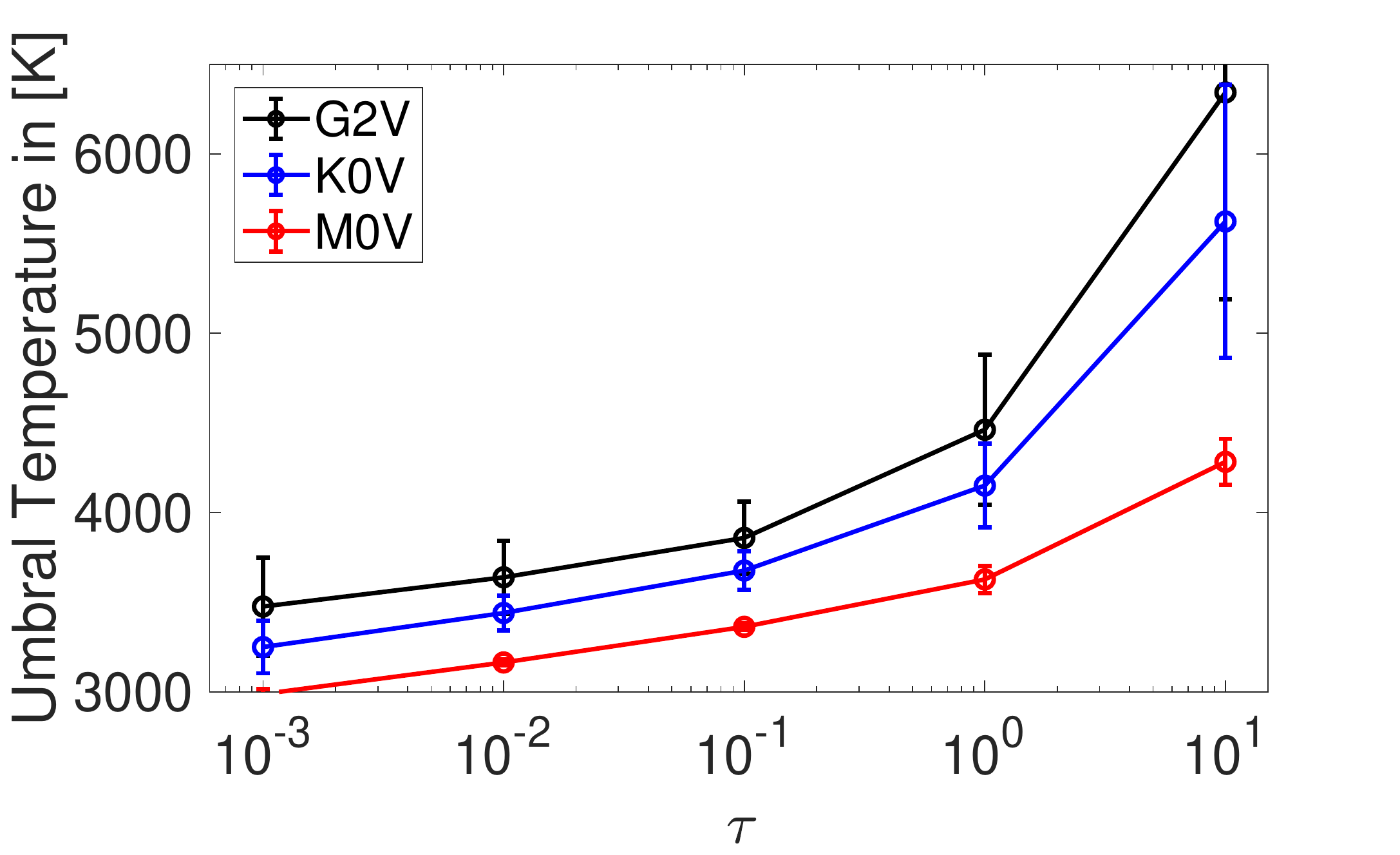}
\hspace*{-0.2cm}\includegraphics[scale=0.35]{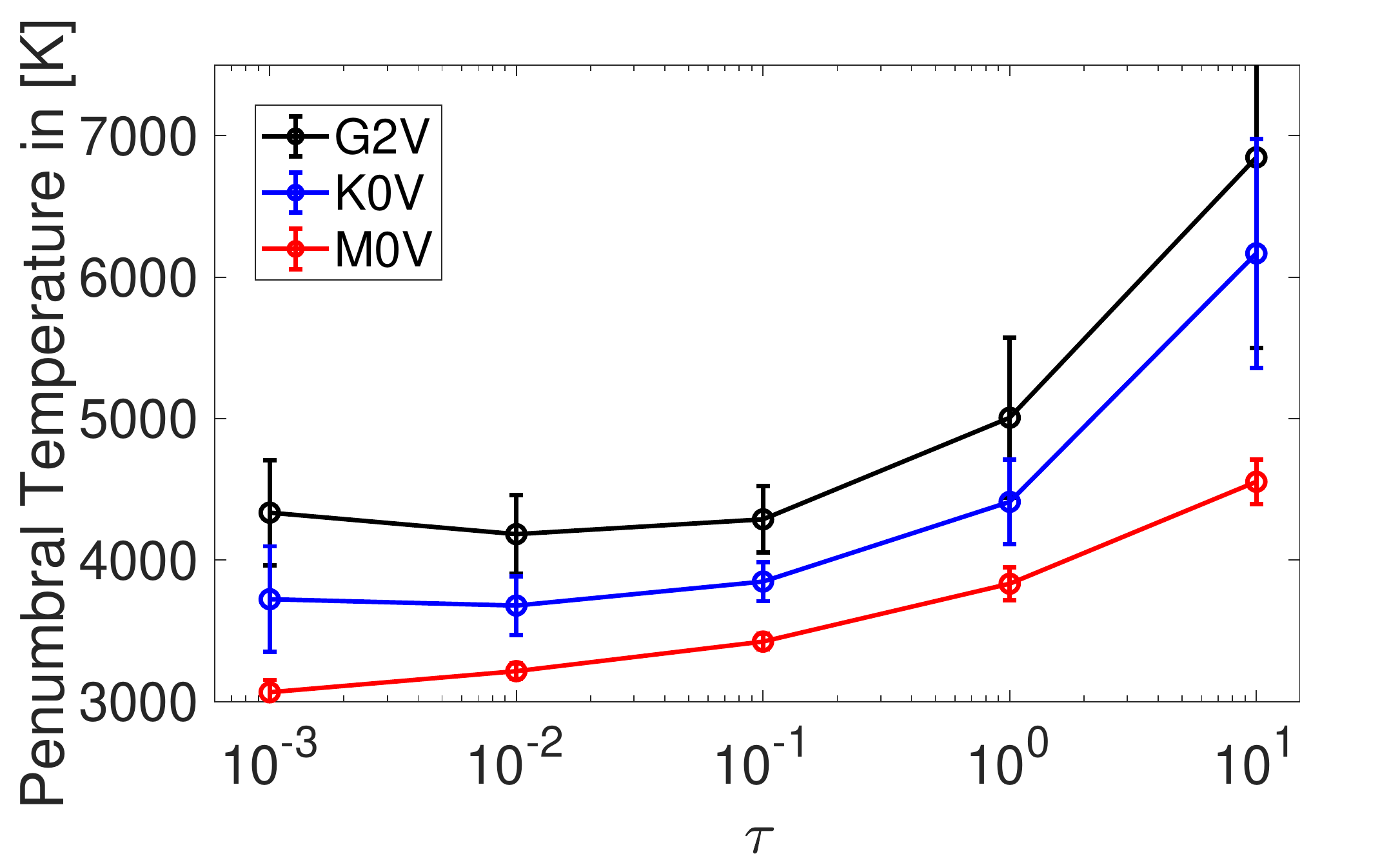}
 \hspace*{-0.2cm}\includegraphics[scale=0.35]{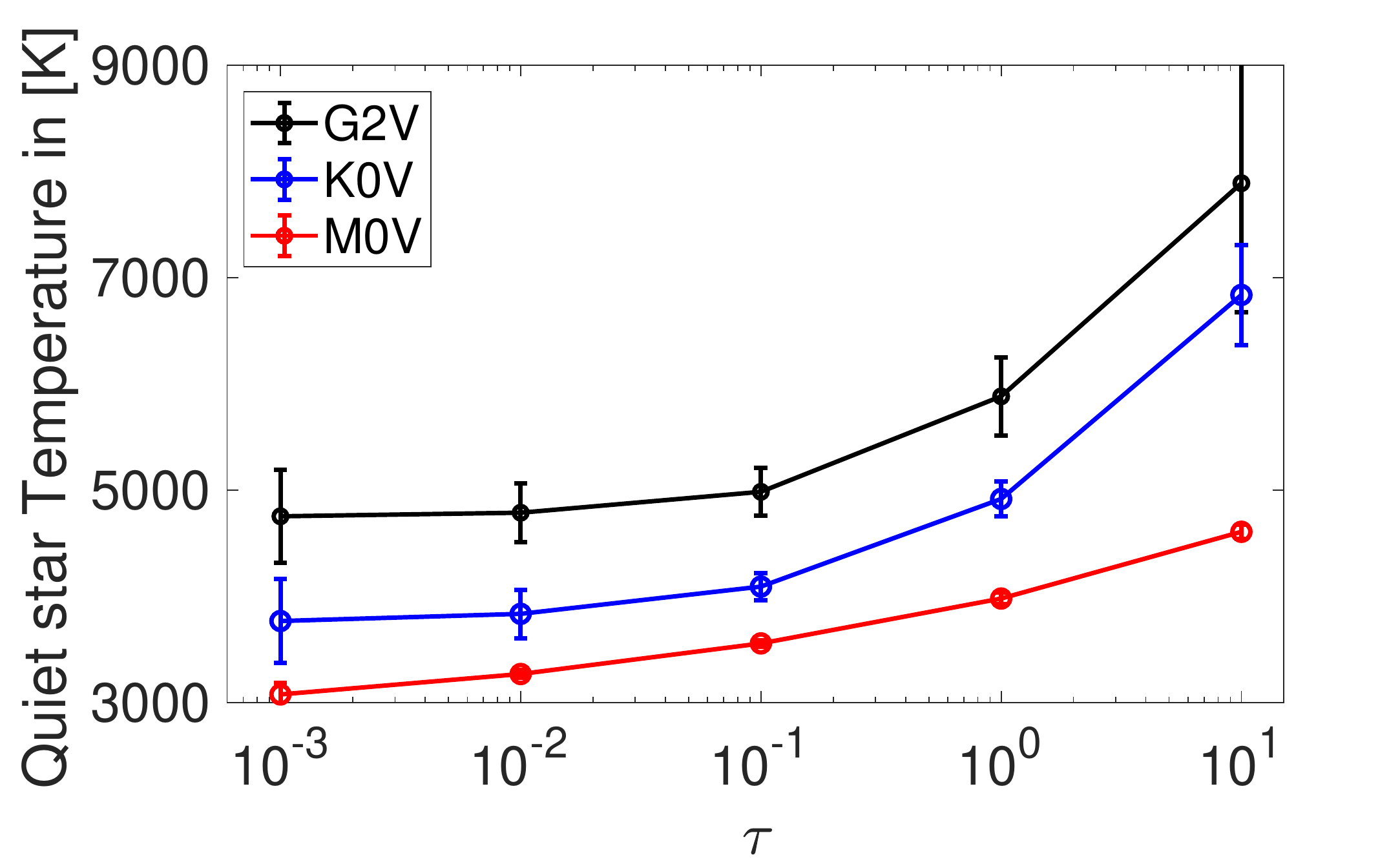}
   
\caption{Top to bottom: Umbral, penumbral and quiet star temperatures, in Kelvin, averaged over different iso-$\tau$ surfaces. The error bars show the standard deviations of the computed averages.}.
              \label{fig:fig7}
    \end{figure}
    
\begin{figure*}
  
   \centering
   \hspace*{-0.2cm}\includegraphics[scale=0.43]{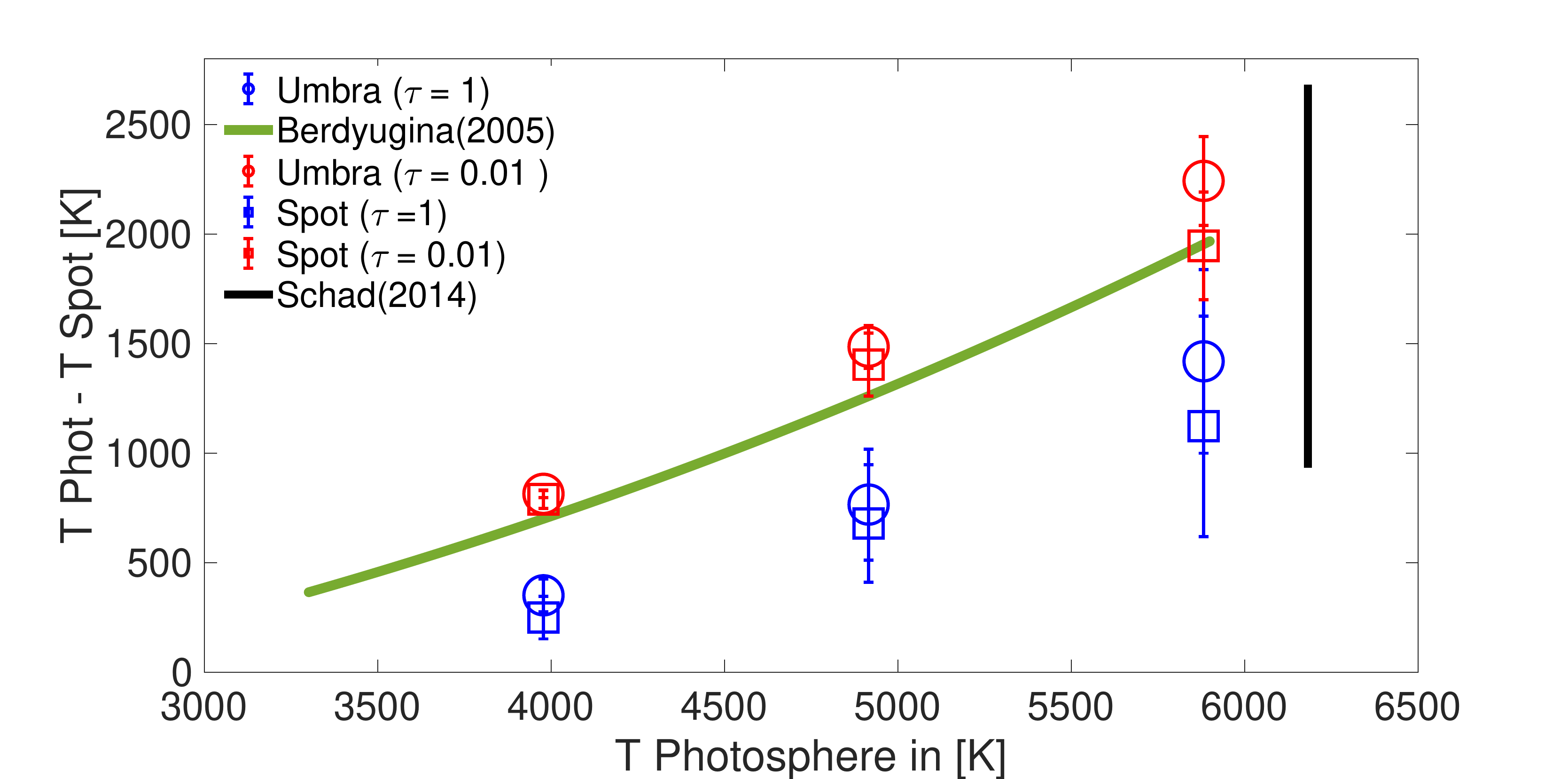}
   \hspace*{-0.2cm}\includegraphics[scale=0.43]{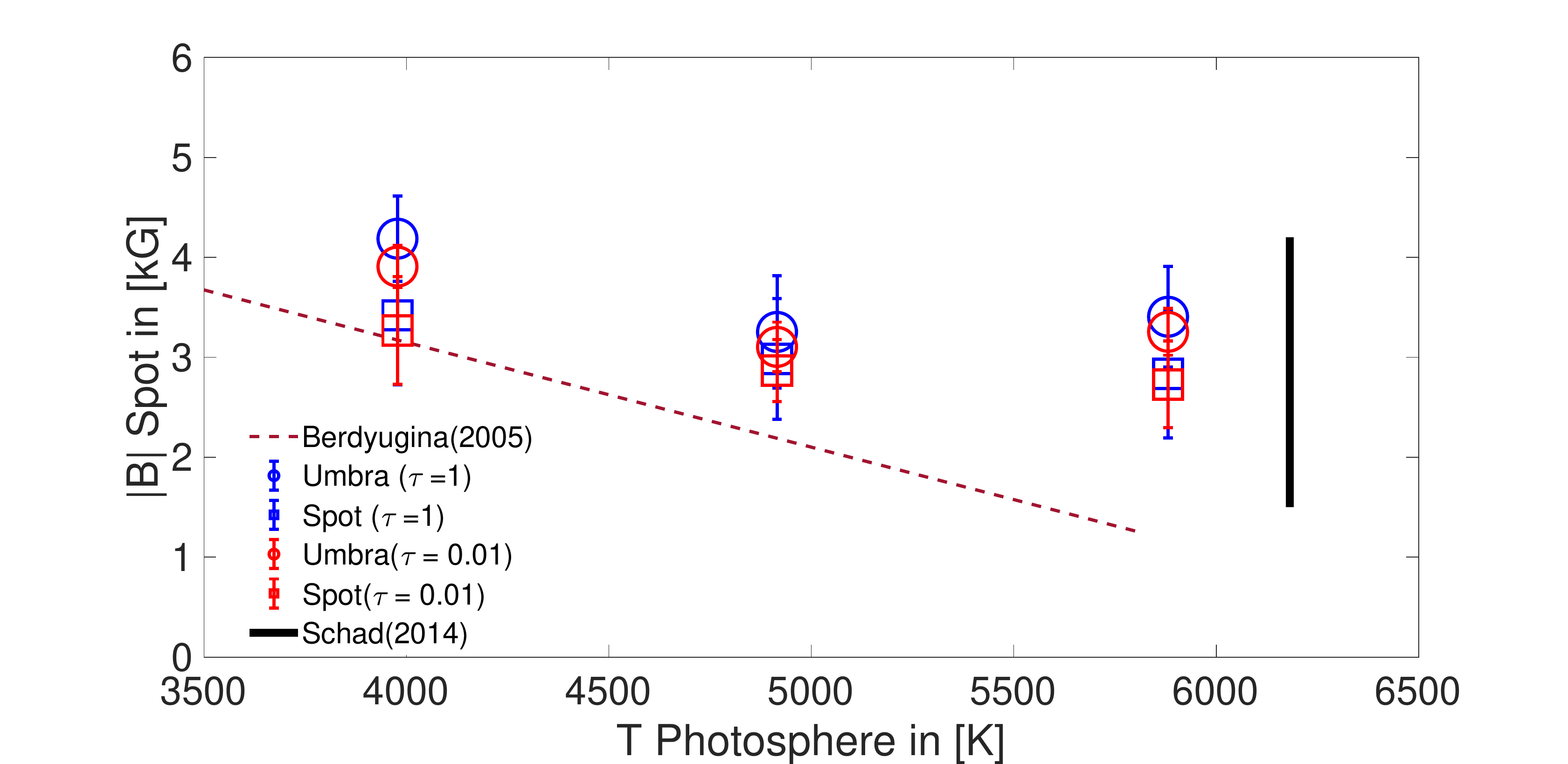}
   
\caption{Top Panel: Spot (both with and without penumbra) temperature contrasts ($T_{quiet(\tau=1)} - T_{spot(\tau=1,0.01)}$) plotted against photospheric temperature ($T_{quiet(\tau=1)}$). Lower Panel: Spot (both with and without penumbra) magnetic field strengths plotted against photospheric temperature at different $\tau$ heights. The error bars show the standard deviations of the computed averages. The green line in the top panel has been reproduced from Figure 7 of \cite{Berdyugina2005} and is a fit to observed starspot temperatures. The red dashed line in the lower panel has been reproduced from Figure 8 of \cite{Berdyugina2005} and is a fit to magnetic field strengths measured on stellar surfaces. The black line in both the panels represent the range of umbral values (also includes pores) measured on the Sun taken from \cite{Schad2014} (lower panel, Figure 2).} 
              \label{fig:fig8}
    \end{figure*}

     \begin{figure}
   \centering
	\hspace*{-0.8cm}\includegraphics[width=10.5cm]{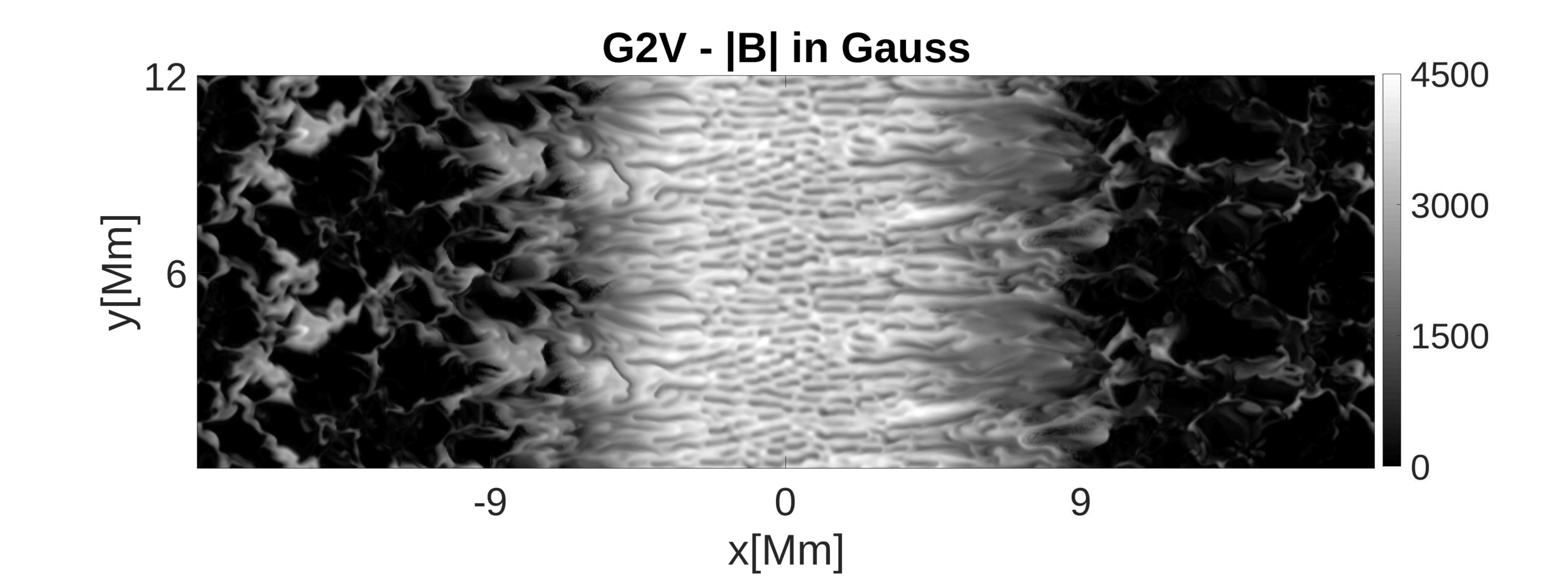}
    \hspace*{-0.8cm}\includegraphics[width=10.5cm]{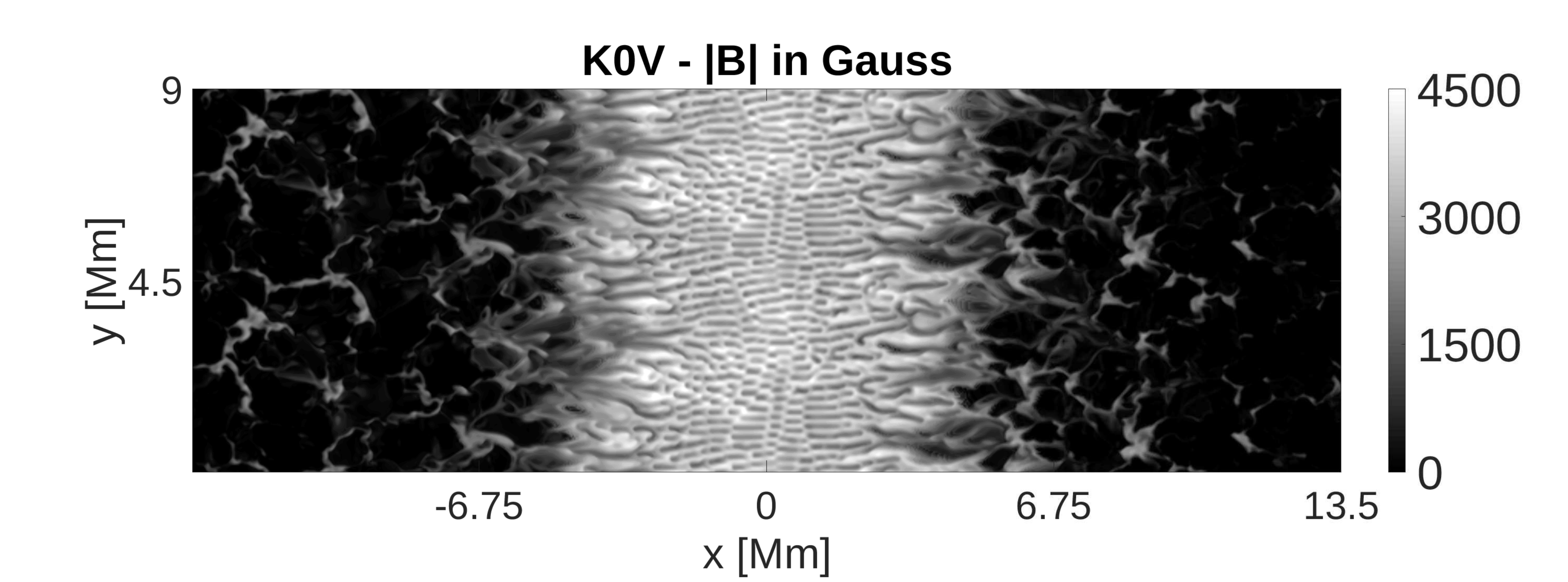}
    \hspace*{-0.8cm}\includegraphics[width=10.5cm]{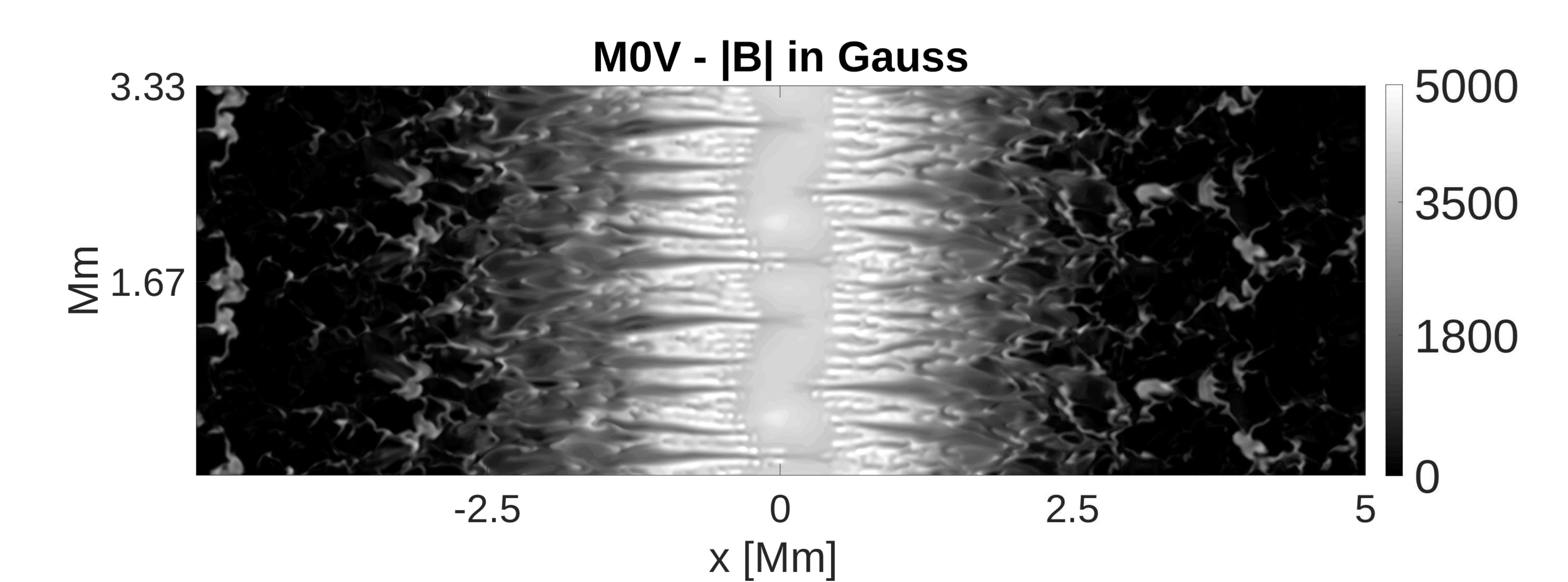}
      \caption{Magnetic field strengths, in Gauss, at the $\tau$=1  surfaces of the three simulated stars - G2V, K0V and M0V.
              }
        \label{fig:fig9}
    \end{figure}

   \begin{figure}
  
   \centering
   
\hspace*{-0.2cm}\includegraphics[scale=0.35]{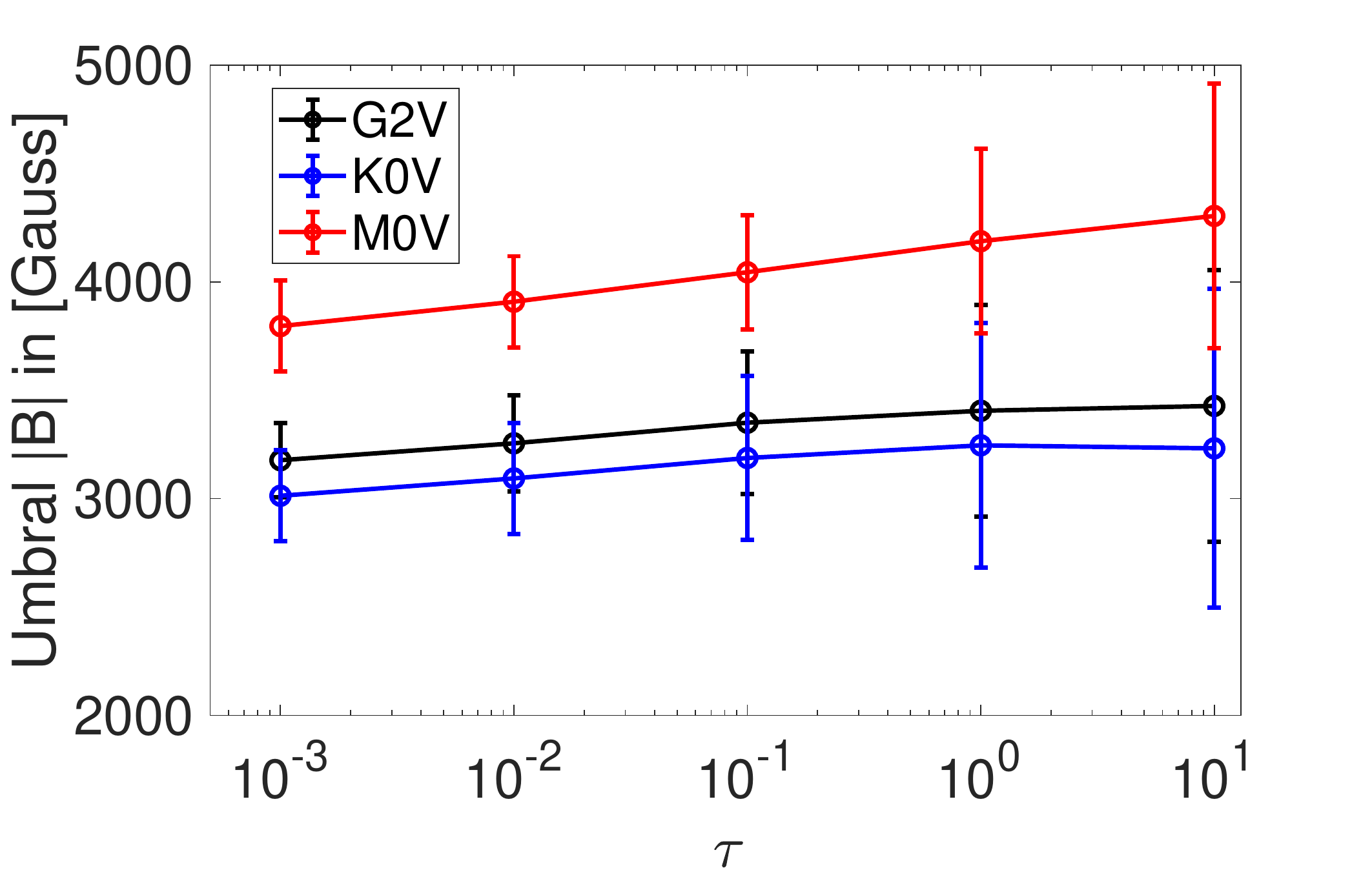}
   
\caption{Umbral magnetic field strengths, averaged over different iso-$\tau$ surfaces. Black: G2V, Blue: K0V, Red: M0V. The error bars show the standard deviations of the computed averages.}.
               \label{fig:fig10}%
    \end{figure} 
    
        \begin{figure}
   \centering
	\hspace*{-0.0cm}\includegraphics[width=10.5cm]{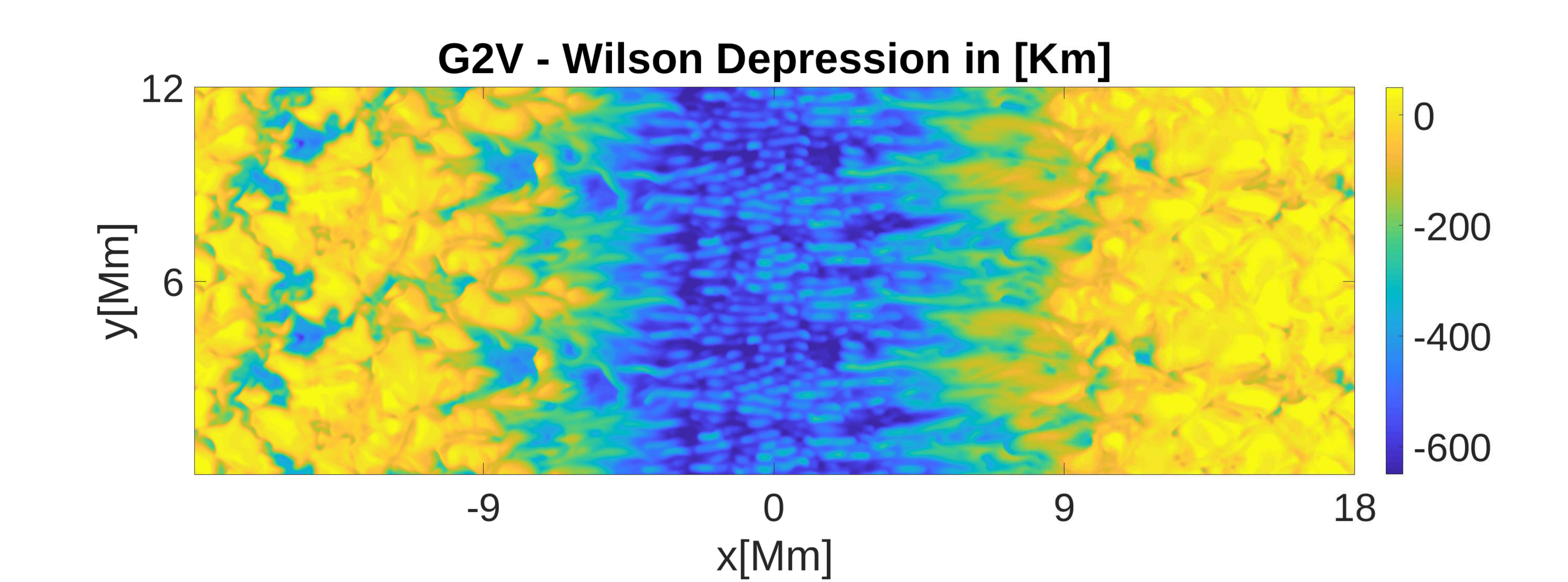}
    \hspace*{-0.0cm}\includegraphics[width=10.5cm]{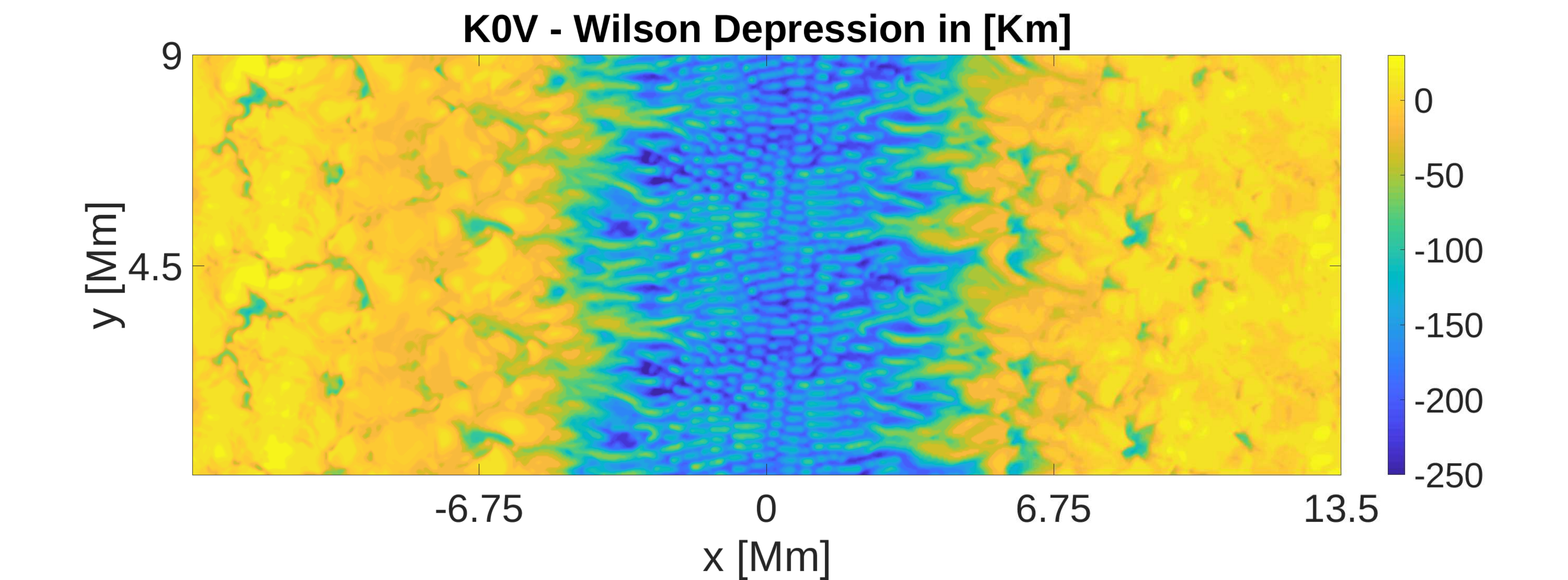}
    \hspace*{-0.0cm}\includegraphics[width=10.5cm]{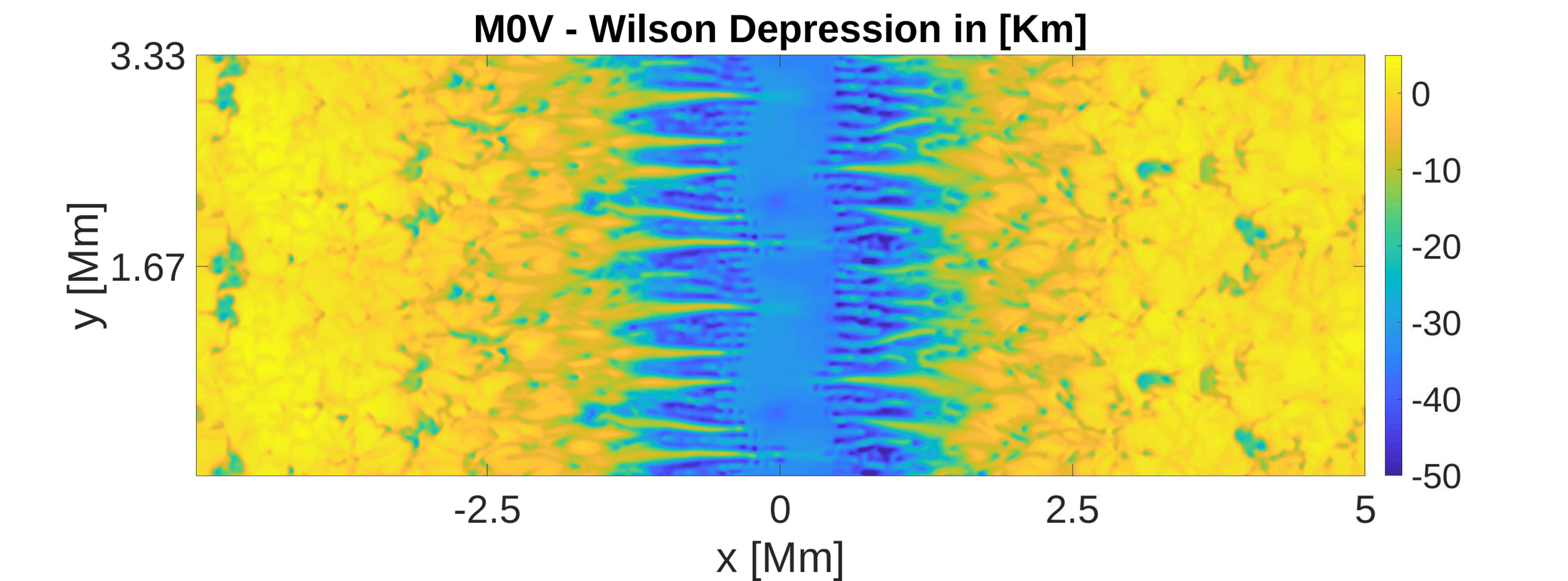}
      \caption{Geometric height maps of the $\tau$=1 surface, in km, of the the three simulated stars - G2V, K0V and M0V. Here, zero corresponds to the average height of the quiet star $\tau$=1 surface.
              }
         \label{fig:fig11}
    \end{figure}

    \begin{figure*}
   \centering
	\hspace*{-3.0cm}\includegraphics[width=24.0cm]{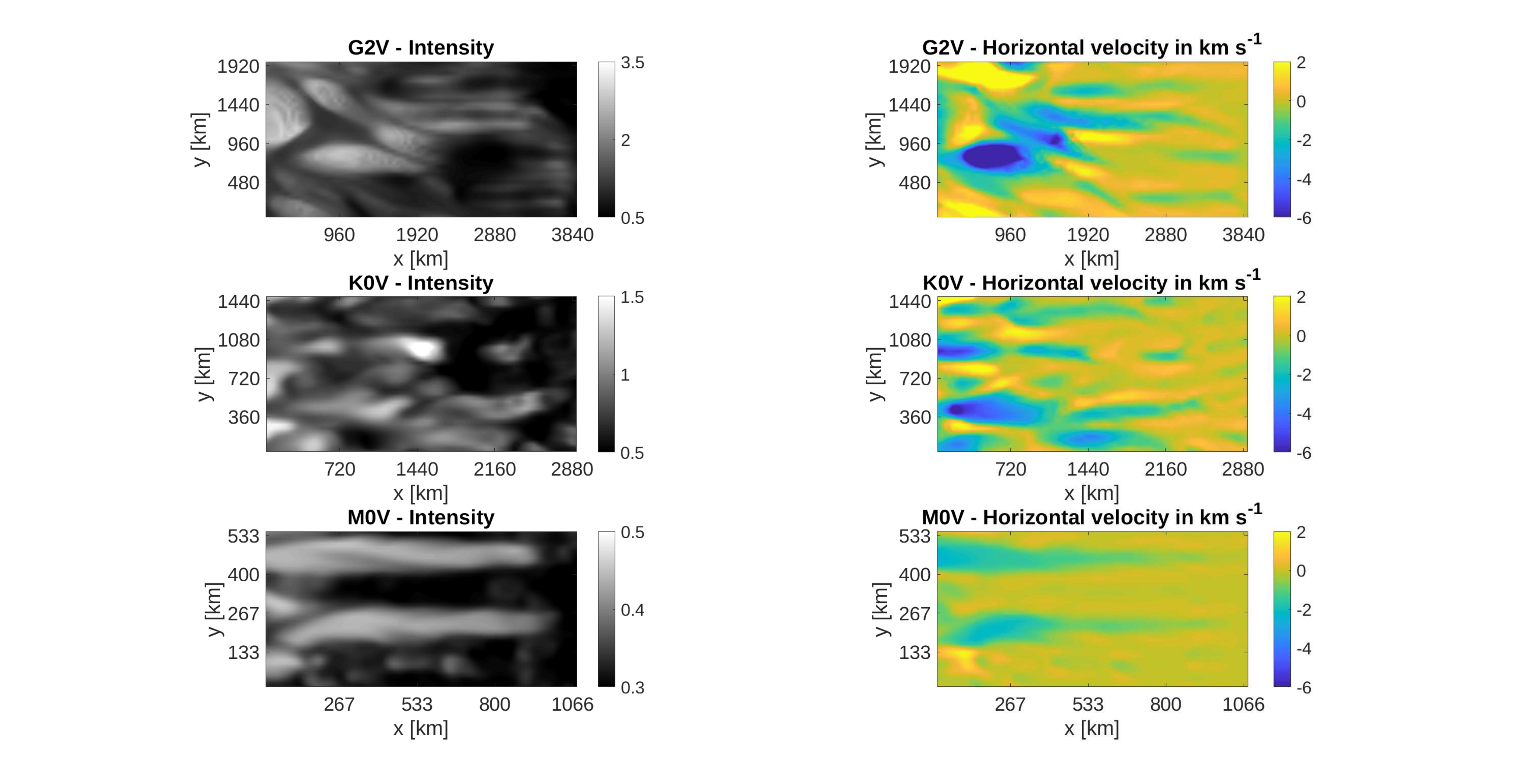}
           \caption{The panels on the left show intensity images of sections of penumbral regions, with the umbrae being to the right of the selected regions. The intensities are in units of 10$^{10}$ erg cm$^{-2}$ ster$^{-1}$ s$^{-1}$  . The panels on the right show corresponding horizontal velocities in km s$^{-1}$ for the same regions. The color blue indicates flows going towards the left, and yellow indicates flows going towards the right. 
               }
        \label{fig:fig12}
    \end{figure*}

\section{2D Results} \label{sec:2d_res}
Figure \ref{fig:fig4} summarizes the results obtained from our 2D simulations, where we have plotted the obtained umbral relative intensities(I$_\mathrm{umbra}$/I$_\mathrm{quiet}$) and the average umbral magnetic field strengths at the stellar surface for the three different spectral types. 

We see a steep decrease in umbral relative intensity (or increase in umbral contrast relative to the quiet star) with increasing T$_\mathrm{eff}$ in our 2D simulations. 
Interestingly, the relative intensities do not show much variation ($<$ 10\%) within a spectral type when we change the initial magnetic field strengths of the flux tube. This is important as it demonstrates that the relative brightness of spots of a certain size is largely determined by the stellar surface properties and does not depend significantly on our choice of initial conditions. This allows us to compare our results on spot brightness to observations with some degree of certainty.

All of the simulated spots have field strengths in the 4-6 kG range. The K0V spots, on average, have slightly lower umbral field strengths compared to the G2V and M0V spots, and the M0V spots reach the highest field strengths.The M0V umbral field strengths also have the maximum dependence on $B_\mathrm{bot}$ and therefore the widest distribution in field strength. Nevertheless, we find that the obtained umbral field strengths do not vary by more than 25$\%$ even when we change $B_\mathrm{bot}$ by a factor of 3 and $B_\mathrm{opt}$ by a factor of 4. Although the umbral field strengths obtained in our 2D simulations are higher than 3D simulations \citep{rempel09a}, their only weak dependence on initial conditions is reassuring. 
A detailed analysis of the 2D runs has been presented in the Appendix.
\section{3D Results}  \label{sec:3d_res}
\subsection{Surface Properties}

   Figure \ref{fig:fig5} shows the bolometric intensity images of the three simulated spots.  There are significant differences between the quiet star regions of the three spectral types, the most conspicuous of them being - 1) the contrast in the intensity between the granules and the intergranular regions is the most pronounced in the G star and is progressively lower in the cooler K and M stars, 2) the average granule size decreases from the G to the M stars and 3) there are almost no bright magnetic    
      features on the M star. These findings are in line with the results of earlier studies focusing on quiet-star magnetoconvection simulations. Detailed analyses have been done by \cite{Beeck2,Beeck3} and \cite{cobold}, to which we direct the reader. It is important to note that even though we have performed grey simulations, our results for the quiet star regions are similar to the results obtained by both these sets of papers. Here we focus solely on spot properties.
   
 
\subsubsection{\textit{Identifying the umbra}}
For all of the three simulated starspots, we first smoothed the intensity images and then applied intensity thresholds to identify the penumbrae. The lower limits (upper limits) of the intensity thresholds, normalized by their average quiet star intensities were - 0.4 (0.8), 0.6 (0.85) and 0.8 (0.94) for the G2V, K0V and the M0V spot respectively. All points within the inner boundaries of the penumbrae were considered as part of the umbrae. We obtained these thresholds by trial and error, using visual inspection to determine what thresholds work the best. The results of the intensity thresholds are shown in Figure \ref{fig:fig5}, where the penumbral boundaries are marked in red.

%
\begin{table}
\caption{ Spatial average of umbral properties.}             
\label{table:3}      
 \centering                       
\begin{tabular}{c c c c c }        
\hline\hline                 
Star & $I_{umbra}$/I$_{quiet}$ & $|B|_{tau=1}$(G) & $T_{\tau=1}$(K) &  $T_{eff}$(K)\\    
\hline                        
   G2V & 0.33 & 3406 & 4462 & 4610.8\\
       &(0.1)&(505)&(419.53)& (102.6)\\
   K0V & 0.52  & 3254 & 4150 & 4262.5 \\
       &(0.09)& (561.3)&(233.6)& (83.6)\\
   M0V & 0.71 & 4187  & 3627 & 3622.8 \\     
       &(0.05) & (426)&(75) &(58.8)\\
\hline                                   
\end{tabular}
\tablecomments{The averages were computed at the time of our snapshots, with the standard deviations displayed inside brackets.}
\end{table}

\subsubsection{\textit{Relative intensity and temperature of starspots}}
Table \ref{table:3} lists the spatially averaged umbral properties at the time of our selected snapshots. The most striking difference between the three simulated starspots is in the intensity contrast between the umbrae of the spots and the surrounding quiet star regions. The spot contrast decreases progressively from the G starspot to the M starspot. The $I_{\rm Umbral}/I_{\rm Quiet}$ ratios for the three stars are 0.3, 0.5 and 0.7 for the G, K, and M stars respectively. Not surprisingly, the temperature maps (Figure \ref{fig:fig6}) correlate well with the intensity maps. Therefore, following the trend in intensity contrasts, the temperature differences between the spots and the quiet star regions decrease from spectral type G to M. 

The umbral dots are noticeably numerous in the maps of G-star and K-star spot temperature and intensity maps than in those of the M-star spot. This points towards the existence of small-scale convective processes underneath the visible surface in the G and K spots. Such processes appear to be comparatively suppressed below the M star umbra.

In Figure \ref{fig:fig7}, we have plotted the average umbral, penumbral and quiet stars temperatures at different $\tau$ levels. As expected, there is a monotonic increase in temperature with increasing $\tau$. For $\tau>$1, the M0V atmospheres (spot and quiet star) show a more gentle increase with optical depth than the G2V and K0V atmospheres.

In the top panel of Figure \ref{fig:fig8}, we have plotted the temperature differences between the quiet star photosphere at $\tau$=1 and the spot at the $\tau$ levels 1 (blue) and 0.01(red). Naturally, at $\tau$=0.01 the spots are colder and the temperature contrast between the quiet star photosphere at $\tau$=1 and the spot at $\tau$=0.01 is higher. In our simulations, the periodicity of our boundary conditions inhibits the growth of an expansive penumbra, as the field is forced to point vertically in the upper part of the box near its boundary in the x-direction (due to the virtual presence of another spot with the same polarity outside the domain). In addition, in slab geometry, the penumbra to umbra area ratio is always underestimated, for purely geometrical reasons. Thus, we estimate penumbra-to-umbra area ratios of approximately 1, 0.5, 1 for the G2V, K0V, and M0V spots respectively, while for the Sun observations typically give a ratio of 4-5 \citep{solanki_review}. This means that those spot temperatures which include contributions from the penumbra (boxes), although warmer, are not significantly different from umbral temperatures(circles).  The green line shows a fit to observed starspot temperature contrasts and has been reproduced from Figure 7 of \cite{Berdyugina2005}. Simulated spot temperature contrasts at  $\tau$ height 0.01, compare well with the fit to the observed data. However, the spot temperature differences at $\tau$=1 show a systematic offset to the line fitted to the observed data points. Note, however, that starspot temperature differences obtained from transit mapping are considerably smaller (e.g. \cite{Espinoza19, Mancini2013}), even below the simulated temperature differences at $\tau=1$. In any case, the observations and simulations display a remarkably similar dependence of the temperature difference on the host star's effective temperature.

Assuming the solar umbra:penumbra area ratio of 1:4 to hold true for all the three simulated spectral types and using temperatures at $\tau$=1 obtained from our simulations, we predict spot temperatures of 4900 K, 4360 K and 3790 K and therefore quiet star to spot temperature contrasts of 980 K, 560 K, and 190 K for the G2V, K0V, and M0V spots respectively. Further, we can calculate spot to umbra temperature contrasts of 440 K (G2V), 210 K (K0V), and 160 K (M0V).

This increase in spot temperature contrast with stellar surface temperature is an effect of the strong dependence of opacity on temperature in the near-surface layers of our simulated spectral types. The opacity dependence of temperature is also responsible for the increase in contrast seen between the granular and intergranular regions with increasing T$_\mathrm{eff}$. We have expanded on this in the Discussion (Section 5).

\subsubsection{\textit{Spot Magnetic Field Strengths}}

Figure \ref{fig:fig9} depicts magnetic field strength maps at the optical surface of the three stars. At the time when we took the snapshots, the M-star spot had the highest umbral average field strength, around 4200 Gauss, and the G and K starspots have average umbral field strengths of around 3400 Gauss and 3200 Gauss respectively.
 The umbral magnetic field distribution is highly non-uniform in both the G and K spots and shows a lot of fine structure, which is related to the fine structure also seen in the surface temperature maps (Figure \ref{fig:fig6}) and is caused by the presence of umbral dots.   

Figure \ref{fig:fig10} shows the umbral magnetic field strengths averaged over different iso-$\tau$ surfaces. The M0V spot shows a slight but steady increase in field strength with increasing optical depth. The magnetic field strengths for the G and K spots do not show much variation with increasing $\tau$ and becomes almost constant below $\tau$=1. 

In the bottom panel of Figure \ref{fig:fig8} we have plotted our obtained average spot field strengths at the heights where $\tau$ = 1 and 0.01. The red line is a fit to field strengths measured on different stars and has been reproduced from Figure 8 of \cite{Berdyugina2005}. It is important to note that the observations are of field strengths averaged over large portions of stellar surfaces and therefore have, probably large, contributions from magnetic fields outside starspots as well. Also, the linear fit in  Figure 8 of \cite{Berdyugina2005} does not include the solar umbral field strength. Therefore, it is not surprising that our simulation results do not agree very well, although the general trend does show some similarity. 

\begin{figure}
  
   \centering
   \hspace*{-0.2cm}\includegraphics[scale=0.5]{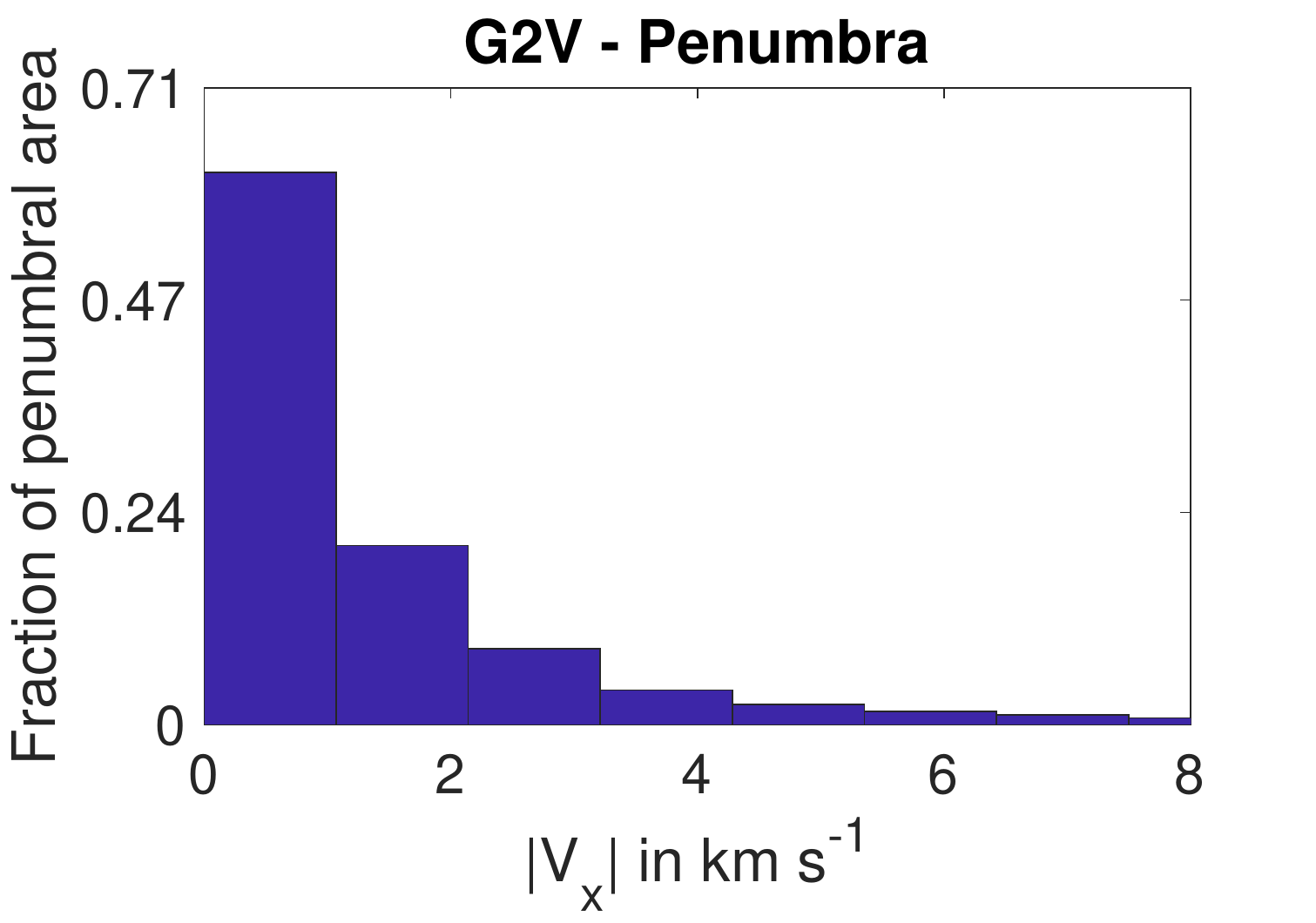}
\hspace*{-0.2cm}\includegraphics[scale=0.5]{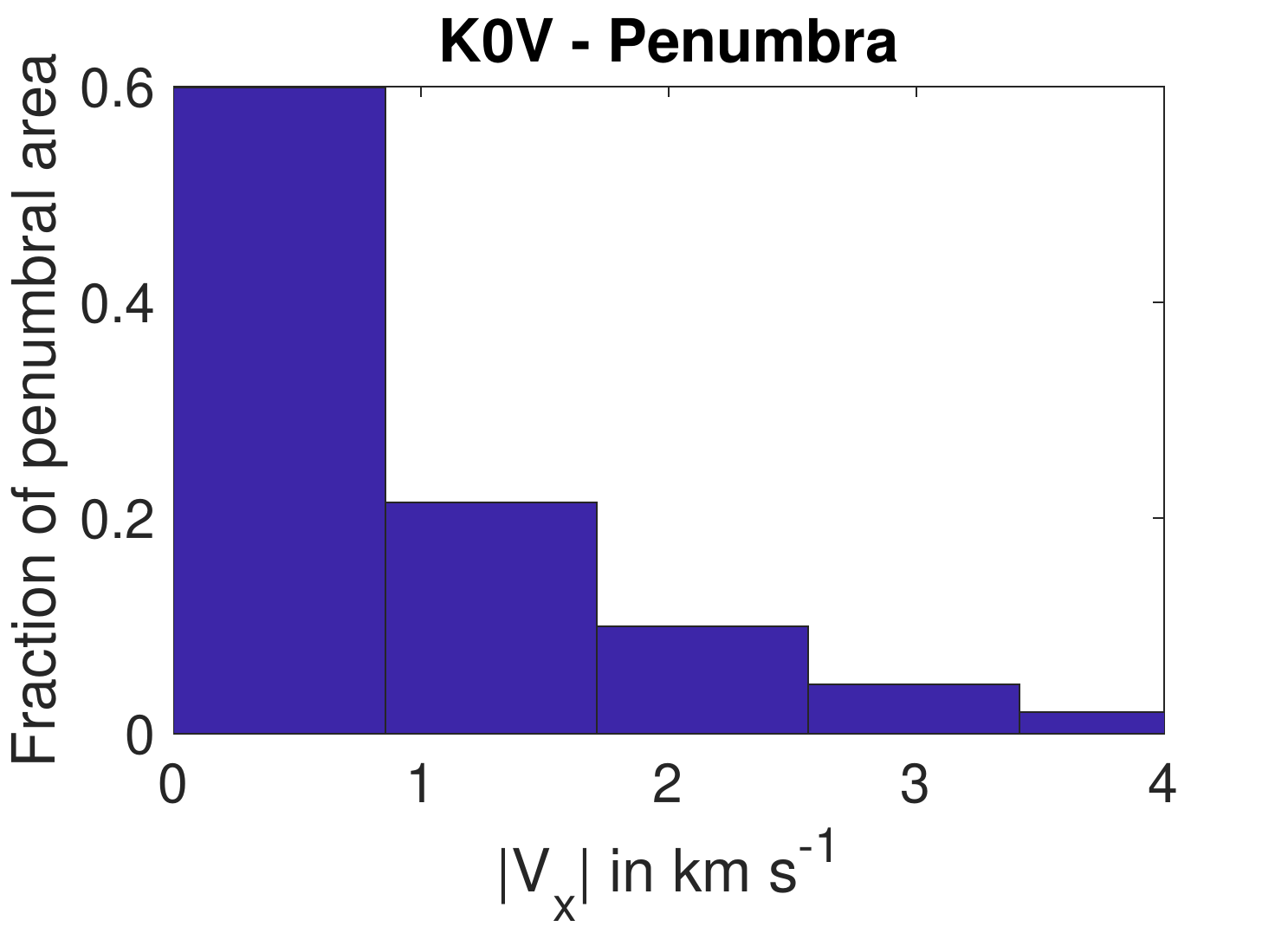}
 \hspace*{-0.2cm}\includegraphics[scale=0.5]{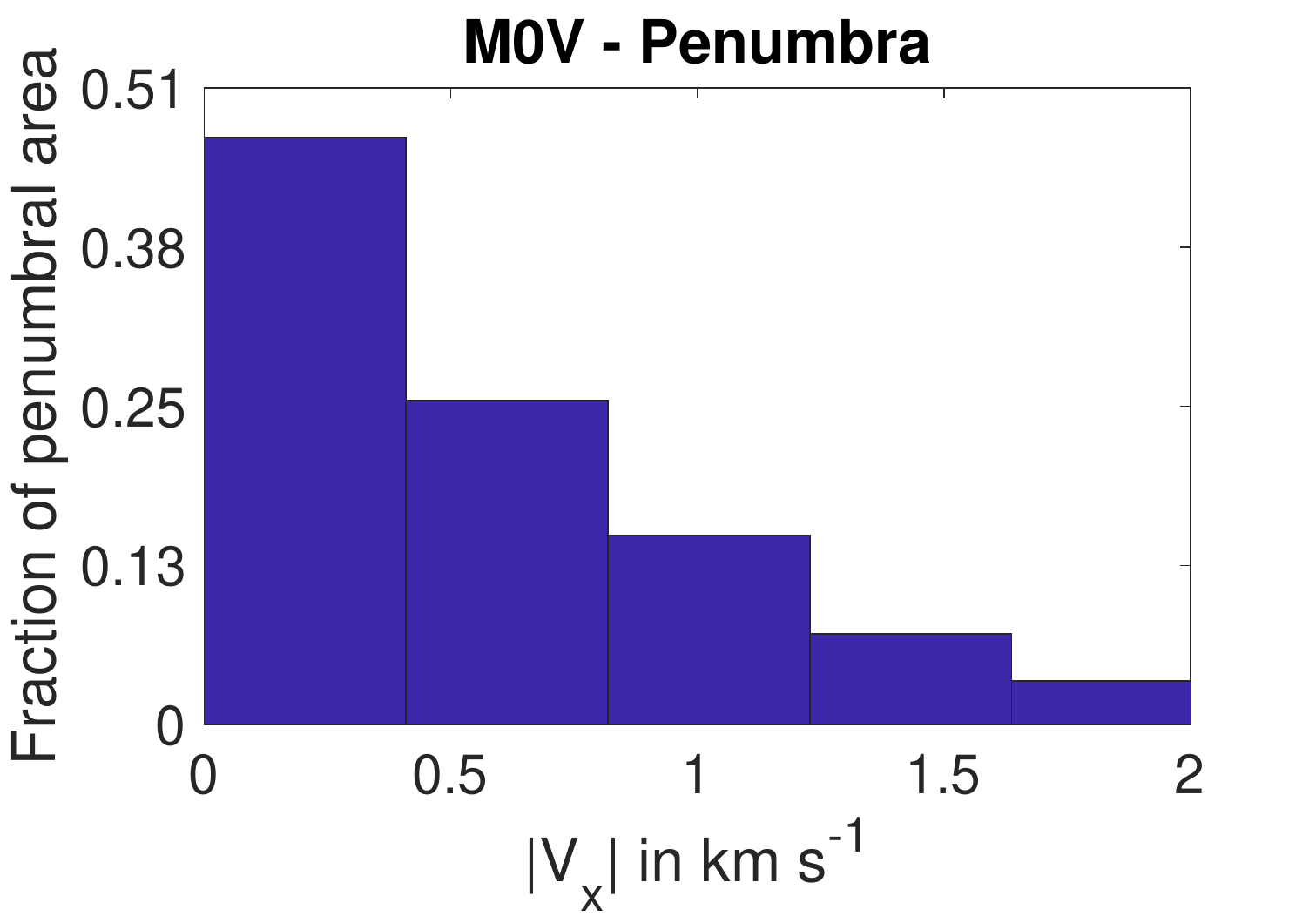}
   
\caption{Histograms of Evershed flow speeds, in, from top to bottom, the G2V, the K0V and the M0V starspots, calculated using the areas marked as penumbrae in Figure \ref{fig:fig5}. }
         \label{fig:fig13}     
    \end{figure}
    
\subsubsection{\textit{Wilson Depression}}
The presence of strong magnetic fields causes a reduction in the local gas pressure and consequently a reduction in the gas density. The reduction in temperature, because of the inhibition of convection, also causes a drop in the opacity. The absorption coefficient($\kappa\rho$), which is the product of the gas density and the opacity, naturally drops. This allows us to see deeper into the star within the starspot and this depression of the optical surface is called the Wilson depression. Figure \ref{fig:fig11} shows the depth of the Wilson depressions of the three spots. The optical surface of the G2V spot is the most depressed and the value of the Wilson depression in the umbra drops by roughly an order of magnitude from the G2V to the M0V spot. The average Wilson depressions of the G, K and M umbral regions are 515, 161 and 34 km respectively. This difference in Wilson depression is a combination of several factors - the difference in pressure scale heights among the stars, the plasma-$\beta$ ratio and the temperature dependence of opacity. We have explored this further in the Discussion. 

\subsubsection{\textit{Penumbral Filaments}}
 The periodicity of our boundary conditions in the horizontal directions implies that our positive spots effectively sit between two other spots of the same magnetic polarity. This hinders the full development of penumbrae in all three cases. Nevertheless, we present here some results from our obtained penumbrae.
 
 The left-hand panel of Figure \ref{fig:fig12} shows the intensity images of segments from the penumbral regions, while the right-hand panel shows the associated horizontal flow velocities in the same regions. All of the three spots show penumbra like features. The K0V penumbral filaments are similar to filaments observed on the Sun - elongated with thin dark cores in the middle. The M0V filaments do not develop dark cores and are more homogeneous. The average intensities of the penumbral regions, normalized by their quiet star intensities, and their average  Evershed flow speeds have been listed in Table \ref{table:4}. Like the umbra, the relative intensity of the penumbra increases from the G2V to the M0V star. 
 Plotted in Figure \ref{fig:fig13} are the histograms of the Evershed flow speeds. For all the three spectral types, the distributions are skewed with only a very small fraction of the penumbra having high horizontal velocities. The G2V penumbral Evershed flows are the fastest with the maximum value reaching almost 8 km s$^{-1}$. In the K0V penumbra, they reach 4 km s$^{-1}$ while the M0V penumbra has the slowest Evershed flows with the maximum value reaching only $\sim$ 3 km s$^{-1}$. The average sound speed at the photospheres is roughly 8 km s$^{-1}$, 7 km s$^{-1}$ and 5 km s$^{-1}$ for the G, K and M stars, respectively. So, while penumbral flows on the G2V star reach supersonic speeds, penumbral flows on the K0V and M0V stars are always subsonic.
 

\begin{table}
\caption{Average penumbral properties with spectral type.}             
\label{table:4}      
 \centering                       
\begin{tabular}{c c c c c c}        
\hline\hline                 
Star & $I_{penumbra}$/I$_{quiet}$ & Evershed velocity(km s$^{-1}$) & $T_{tau=1}$(K)\\    
\hline                        
   G2V & 0.57 & 1.32 & 5007 \\
       &(0.2)&(1.58)&(564)\\
   K0V & 0.71  & 1.00 & 4412 \\
       &(0.15)& (1.06)&(298))\\
   M0V & 0.88 & 0.59  & 3833 \\     
       &(0.07) & (0.52)&(115) \\
\hline                                   
\end{tabular}
\tablecomments{The standard deviations are in brackets. The areas marked as penumbral regions in Figure \ref{fig:fig5} were used to compute the averages.}
\end{table}
     

\subsection{Subsurface Properties}
 Figures 14 to 19 show the variation of starspot properties with depth, averaged horizontally along the width of the simulation boxes. 
 
 We have shown only the first few relevant pressure scale heights in the following figures.

\begin{figure}
   \centering
	\hspace*{-1.42cm}\includegraphics[width=10.5cm]{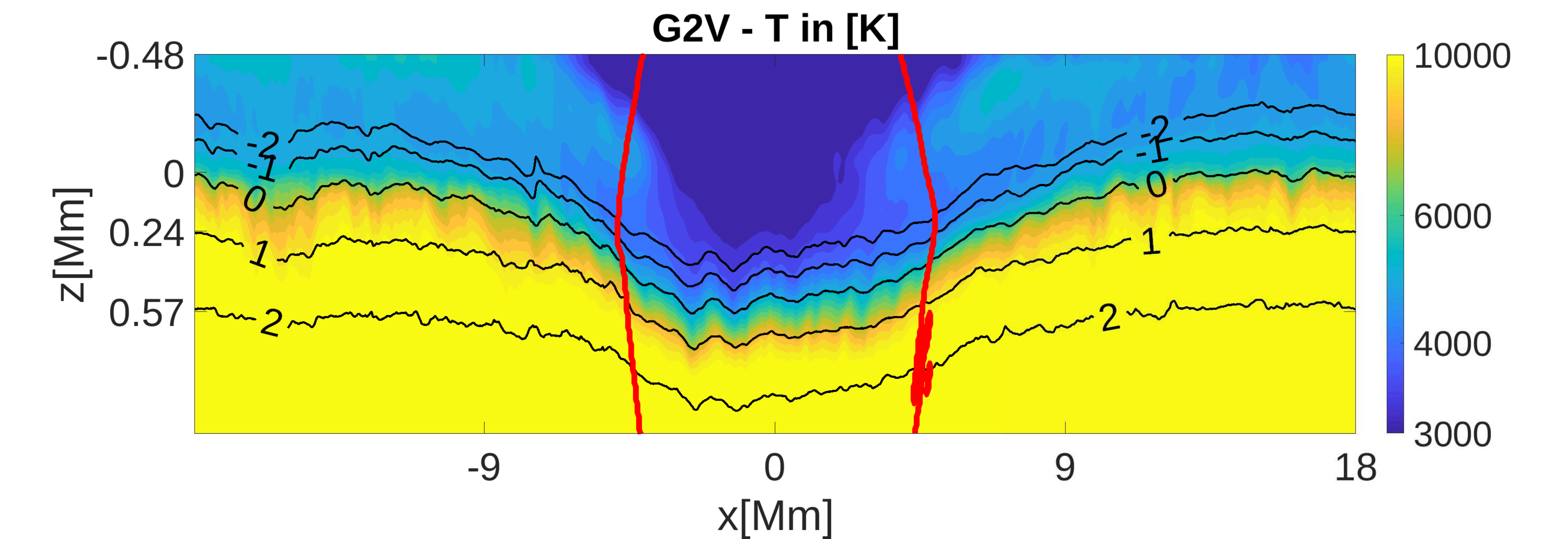}
    \hspace*{-1.42cm}\includegraphics[width=10.5cm]{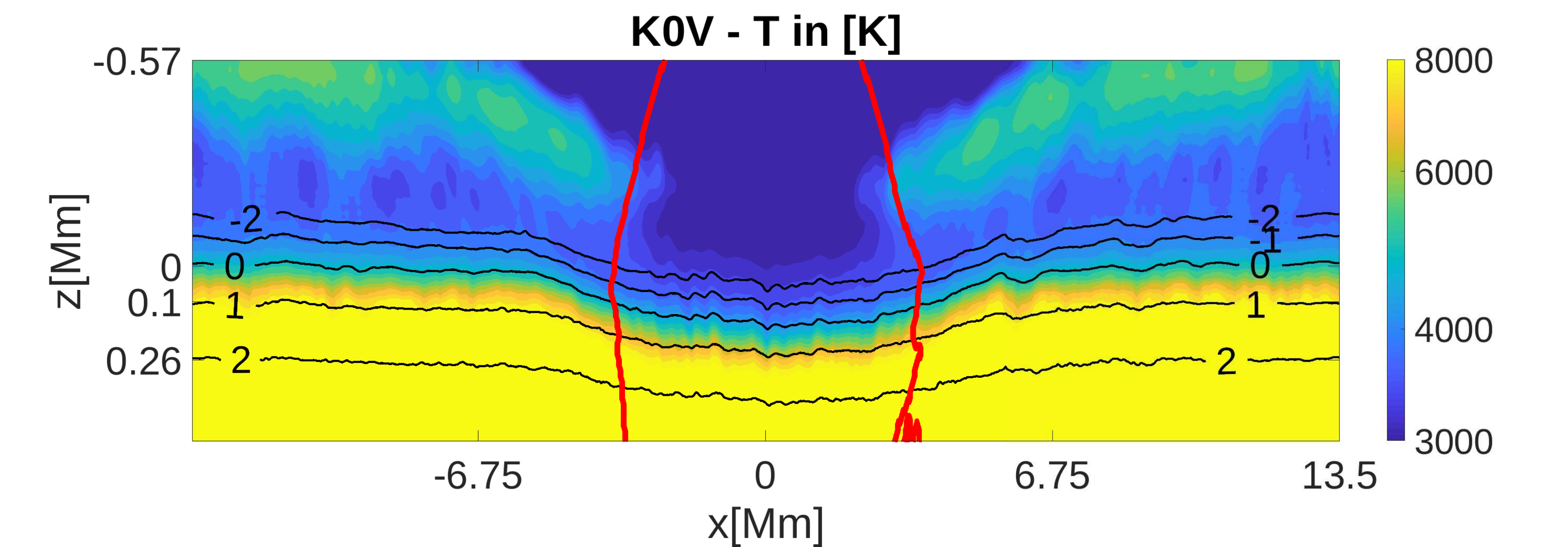}
    \hspace*{-1.42cm}\includegraphics[width=10.5cm]{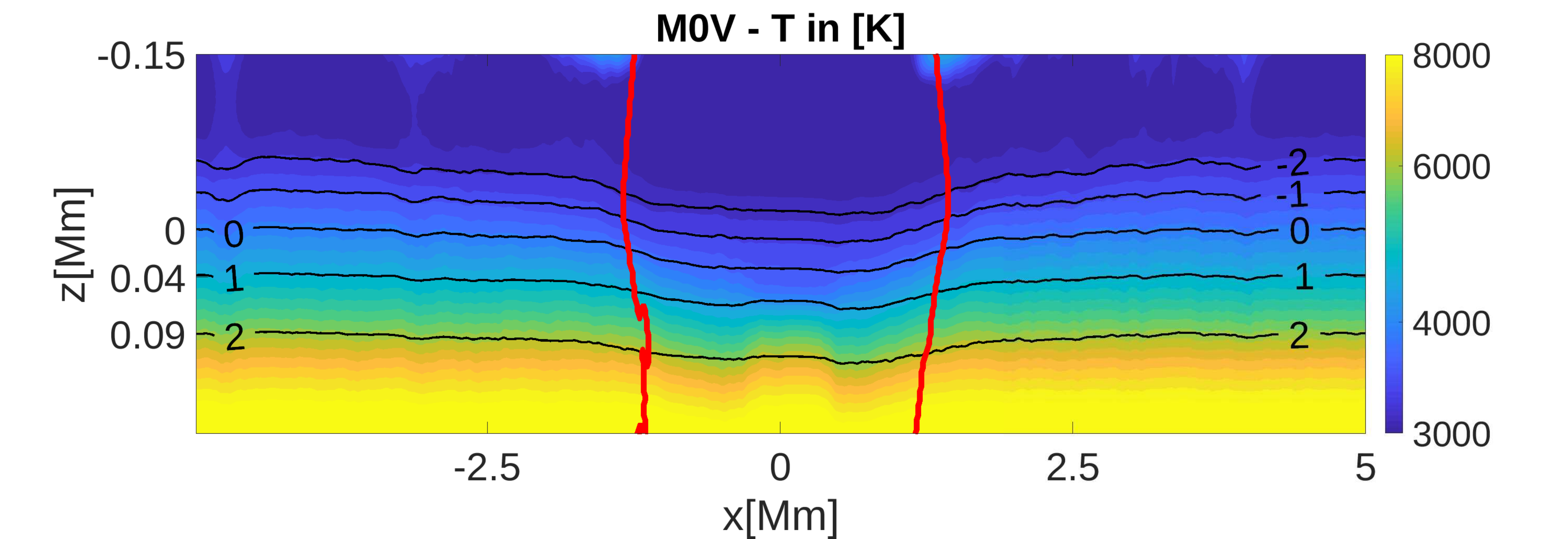}
    \hspace*{-1.42cm}\includegraphics[width=9.5cm]{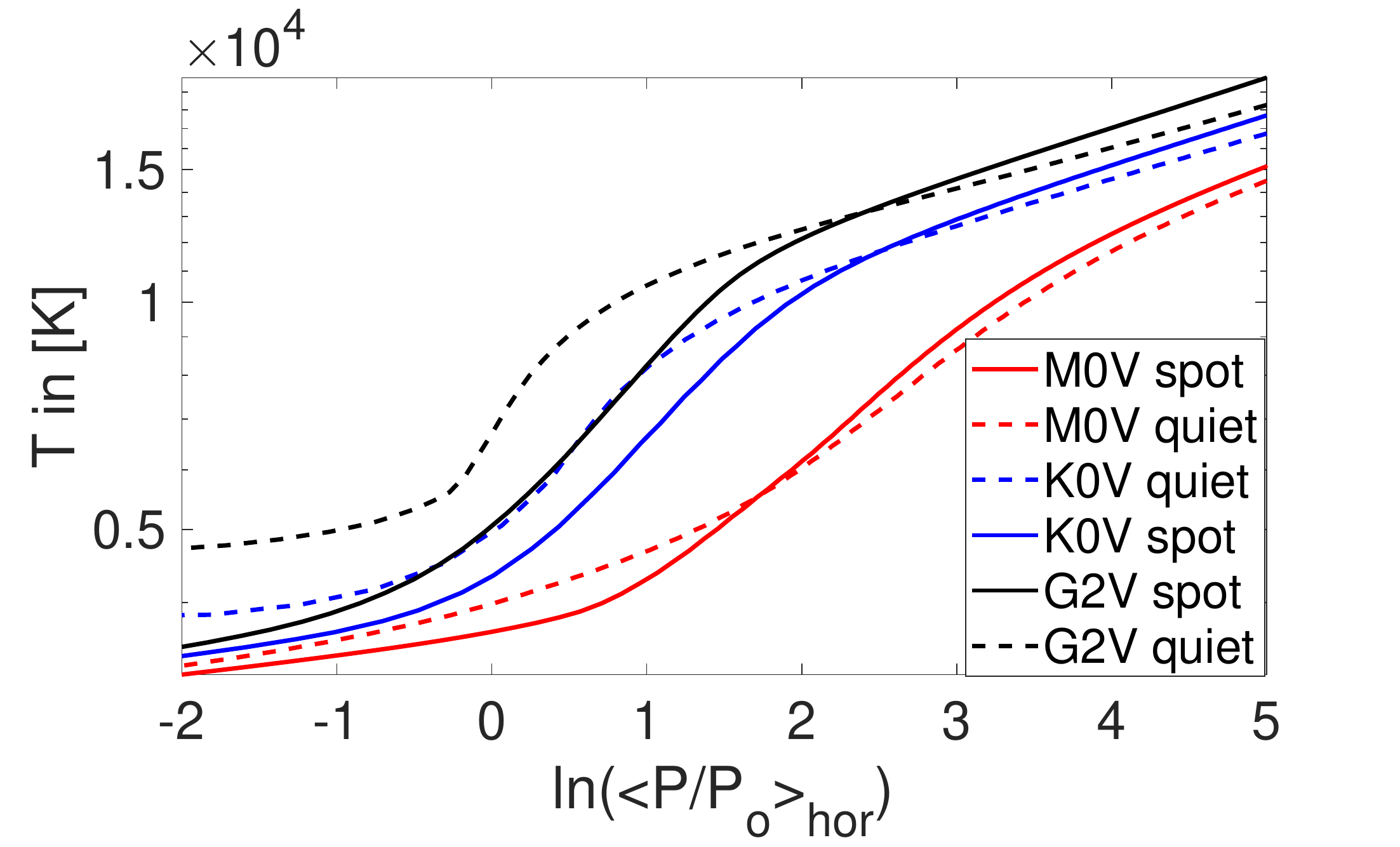}
    \caption{First 3 panels from the top: Temperatures, in Kelvin, plotted with depth, averaged horizontally over iso-z surfaces along the y-axis of the box. Overplotted on all of the figures are contours of    $\log_{e}(<(\frac{p}{p_{0}})>_{y})$, where p is the pressure and p$_{0}$ is the pressure at $\tau$ = 1 at every column of the atmosphere. The red contours mark the boundary of $\vert$B$\vert$ = 2000 Gauss.  The x-axes show the distances in Mm along the length of the box, while the z-axes mark the geometrical heights with respect to the optical surface. We have shown only the first few relevant pressure scale heights.
    Bottom panel: 1D horizontal averages (along x and y) of umbral and quiet star temperatures plotted against pressure scale height.  
               }
       \label{fig:fig14}
    \end{figure}
    
    \begin{figure}
   \centering
	\hspace*{-0.80cm}\includegraphics[width=10.5cm]{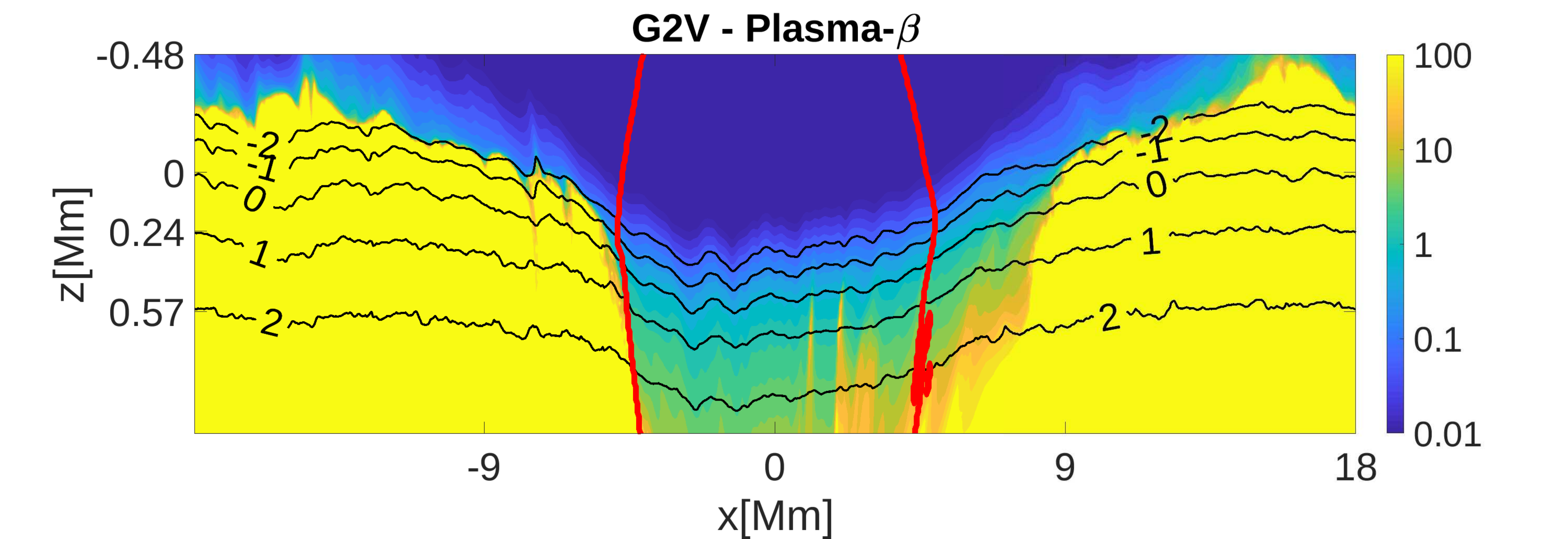}
    \hspace*{-0.80cm}\includegraphics[width=10.5cm]{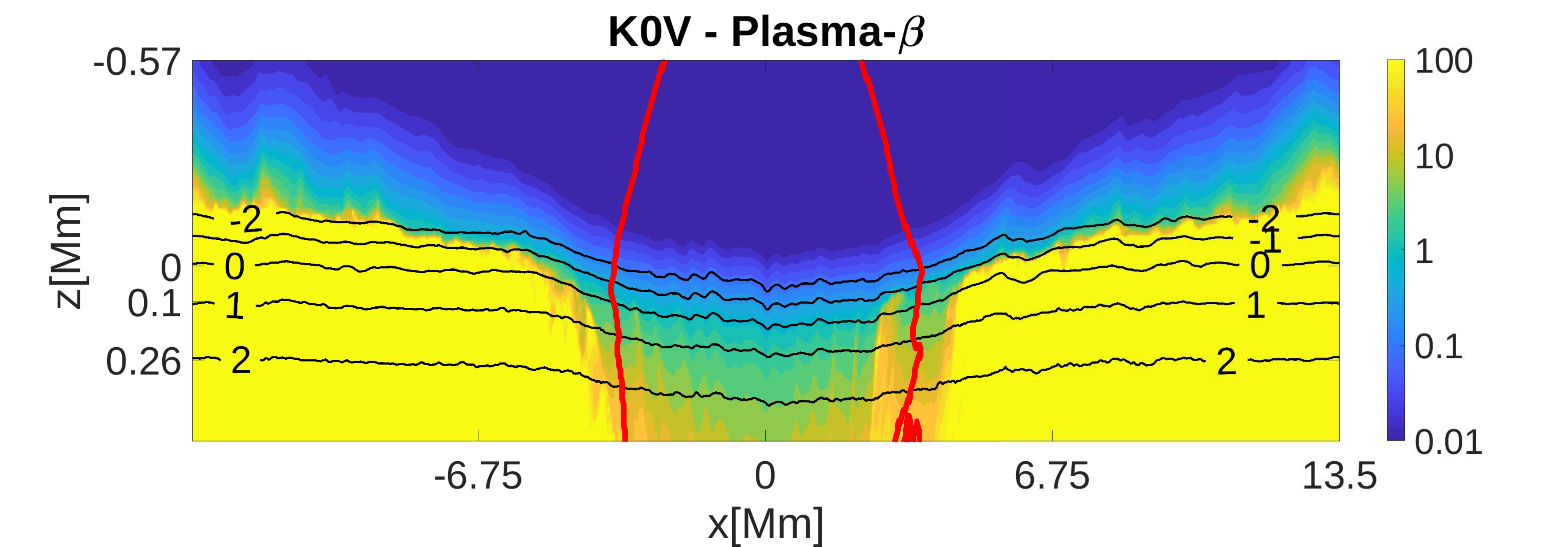}
    \hspace*{-0.80cm}\includegraphics[width=10.5cm]{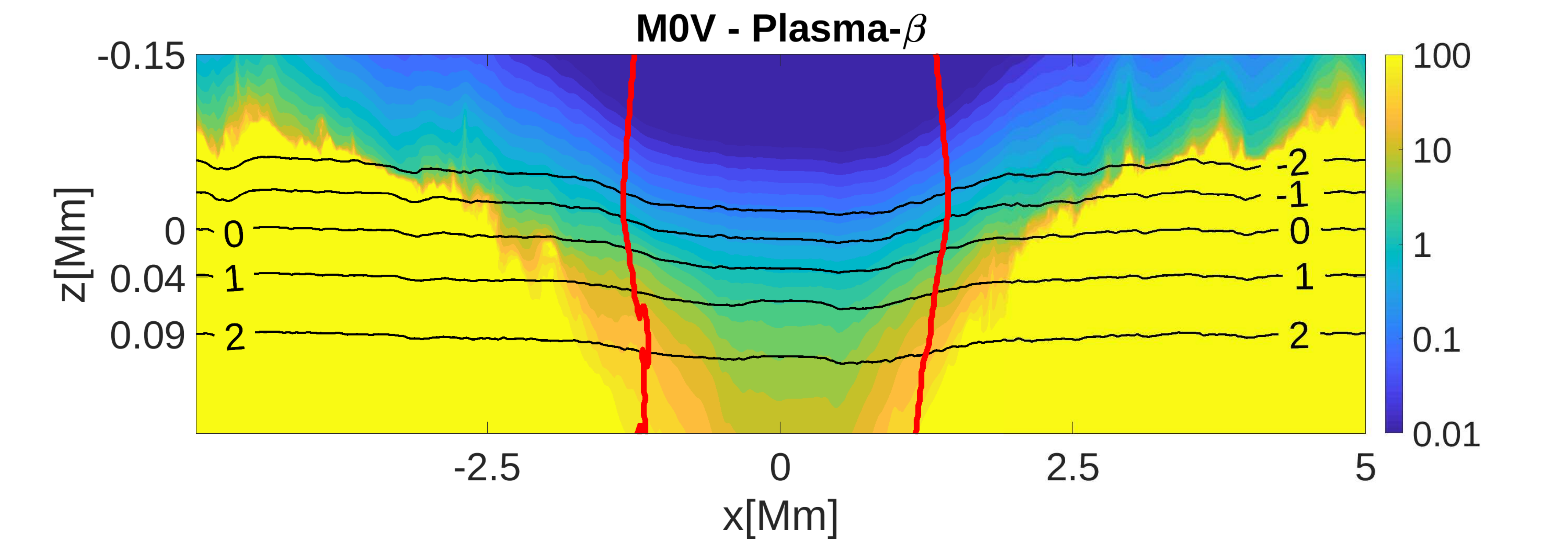}
      \caption{The, horizontally averaged, plasma-$\beta$ ratio - $\frac{8\pi P}{B^{2}}$. The black and red contours are the same as in Fig: \ref{fig:fig14}. 
              }
        \label{fig:fig15}
    \end{figure}

     \begin{figure}
   \centering
	\hspace*{-1.00cm}\includegraphics[width=10.5cm]{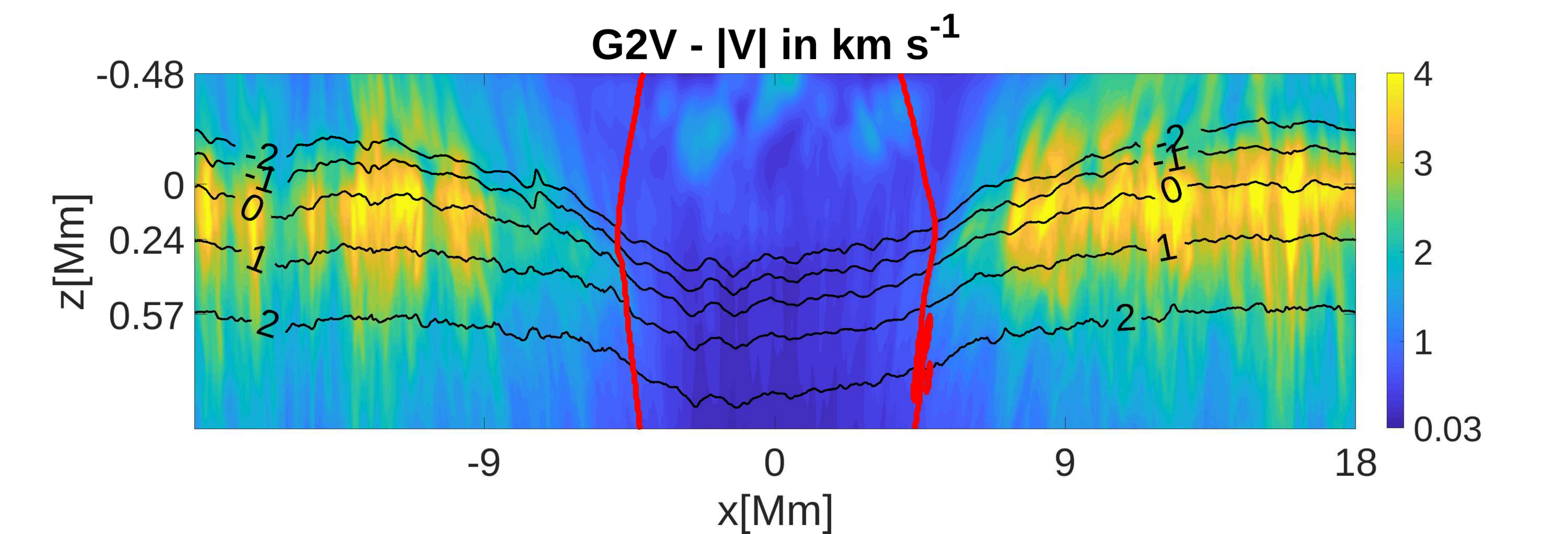}
    \hspace*{-1.00cm}\includegraphics[width=10.5cm]{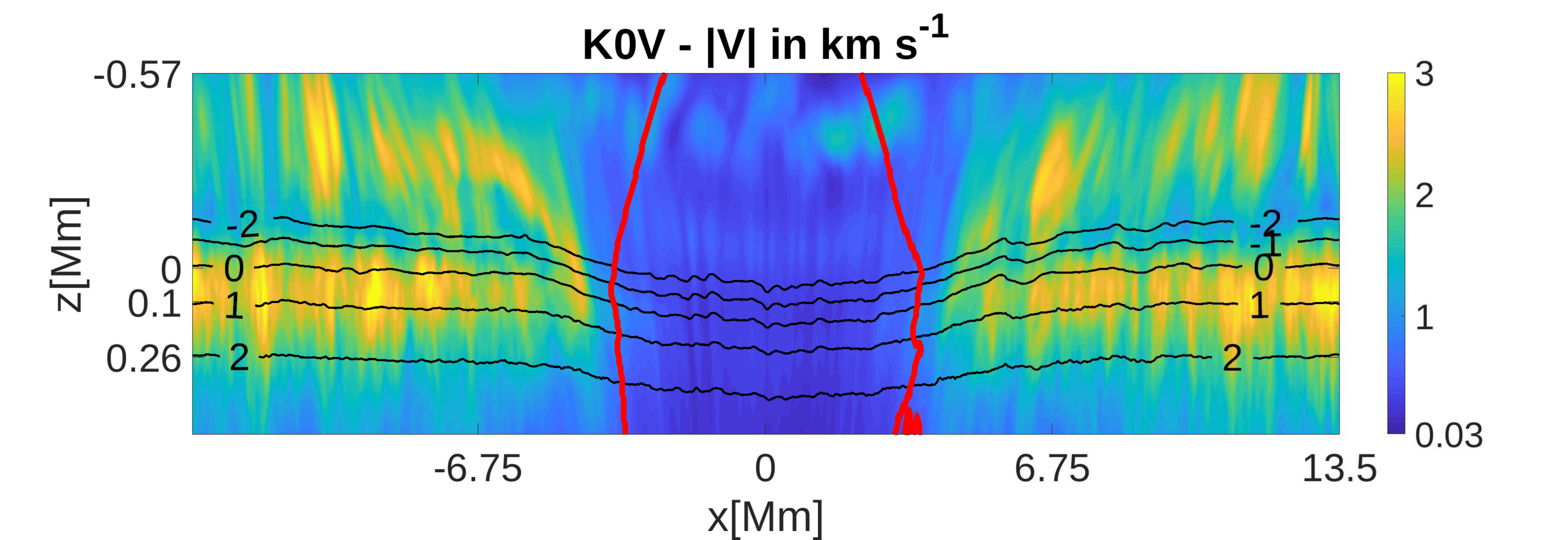}
    \hspace*{-1.00cm}\includegraphics[width=10.5cm]{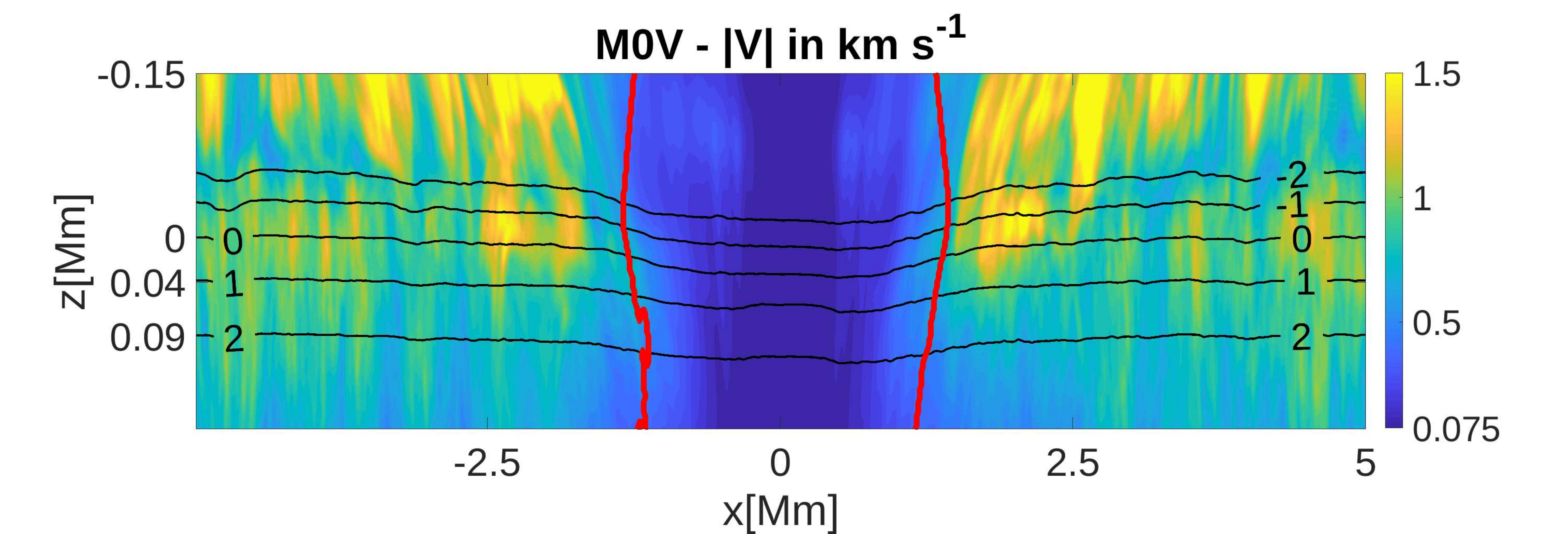}
      \caption{The, horizontally averaged, magnitude of the velocity field ($|v| = (v_{x}^2+v_{y}^2+v_{z}^2)^\frac{1}{2}$) in units of km s$^{-1}$. The black and red contours are the same as in Fig: \ref{fig:fig14}. 
              }
        \label{fig:fig16}
    \end{figure}
    

\subsubsection{\textit{Temperature Structure}}
 Figure \ref{fig:fig14} shows the variation of temperature with depth. In the quiet star regions, there is a sharp vertical gradient in temperature with depth, at the photosphere of the G star, where the temperature rises from $\sim$ 6000 K at the surface to $\sim$ 9000 K within a span of half a pressure scale height. This gradient is weaker for the K star and smoothest for the M star. In the M star, the temperature rises by a mere 2000 K over 2 pressure scale heights, from $\sim$ 4000 K at the photosphere to $\sim$ 6000 K at a depth of 2 pressure scale heights. It is important to note that the opacity due to the ionization of H$^{-}$ is the main source of continuum opacity in the surface layers of cool main sequence stars. The H$^{-}$ opacity shows a steep temperature dependence in the range 3000 - 8000 K and therefore, the vertical temperature gradient plays an important role in determining the observed intensity contrasts. 
 
 Inside the spots, the reduced temperatures also diminish the pressure scale heights and this is evident in Figure \ref{fig:fig14}.
 Below a certain depth, the temperatures inside the spots of all three spectral types become indistinguishable from their surroundings.
 
\subsubsection{\textit{Plasma-Beta and Convection}}
The ratio of the gas pressure (P) to the magnetic pressure ($B^{2}/8\pi$), called the plasma-$\beta$ ratio, has been plotted in Figure \ref{fig:fig15}. Inside all of the spots, the plasma-$\beta$ ratio is close to unity around the surface. The atmosphere above the M0V spot is less evacuated (the ratio is around 0.1 just above the surface) than the G2V spot, where the ratio drops to below 0.01 within a span of 2 pressure scale heights. This plays a role in the G2V spot having a much more depressed optical surface. 

The depth dependence of the magnitude of the velocity field, $|v| = (v_{x}^2+v_{y}^2+v_{z}^2)^\frac{1}{2}$, averaged along the y-axis of the box is shown in the Figure \ref{fig:fig16}. Convection is suppressed because of the presence of strong magnetic fields and the average velocities inside the umbra drop roughly by a factor of 10, for all the three spots. 
The G spot has average photospheric velocities of around 4 km s$^{-1}$ and inside the umbra the convection is reduced to around 0.3 km s$^{-1}$. Similarly for the M star, the photospheric convective velocity of around 1 km s$^{-1}$ is reduced to less than 0.075 km s$^{-1}$. The K star has velocity fields of around 2.5 km s$^{-1}$ at the surface which are reduced to around 0.3 km s$^{-1}$ inside the spot umbra.


\subsubsection{\textit{Radiation Field}}

\begin{figure}
   \centering
	\hspace*{-1.0cm}\includegraphics[width=10.5cm]{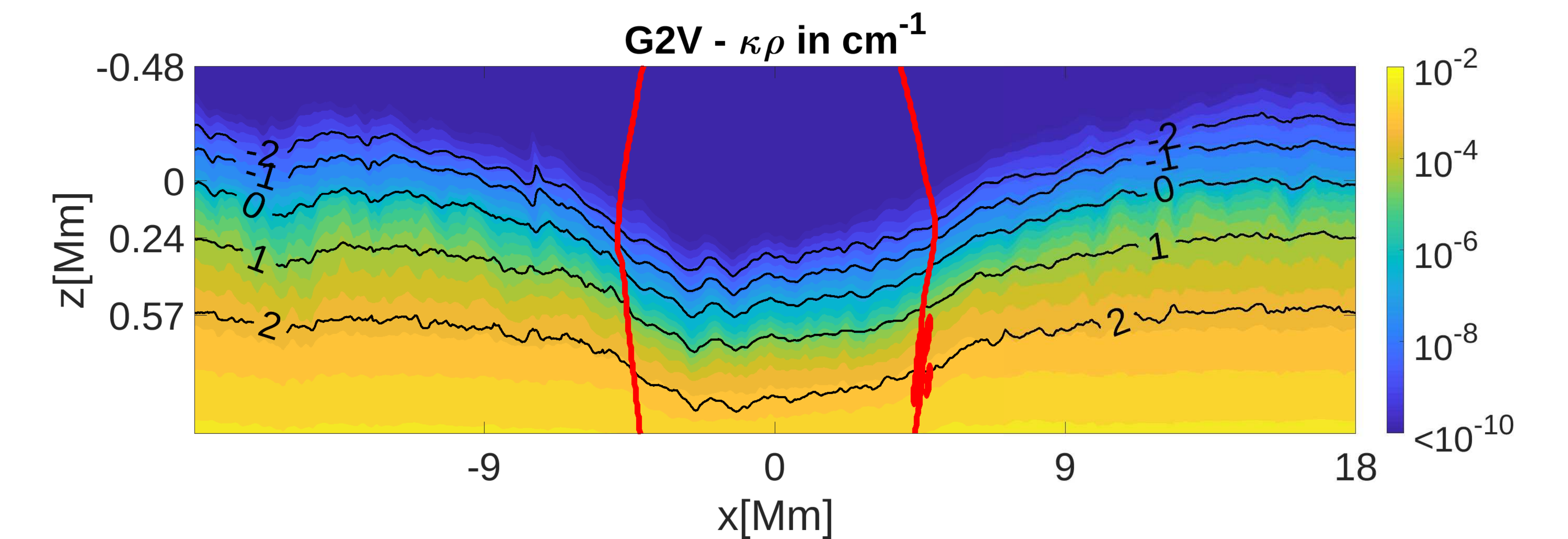}
    \hspace*{-1.0cm}\includegraphics[width=10.5cm]{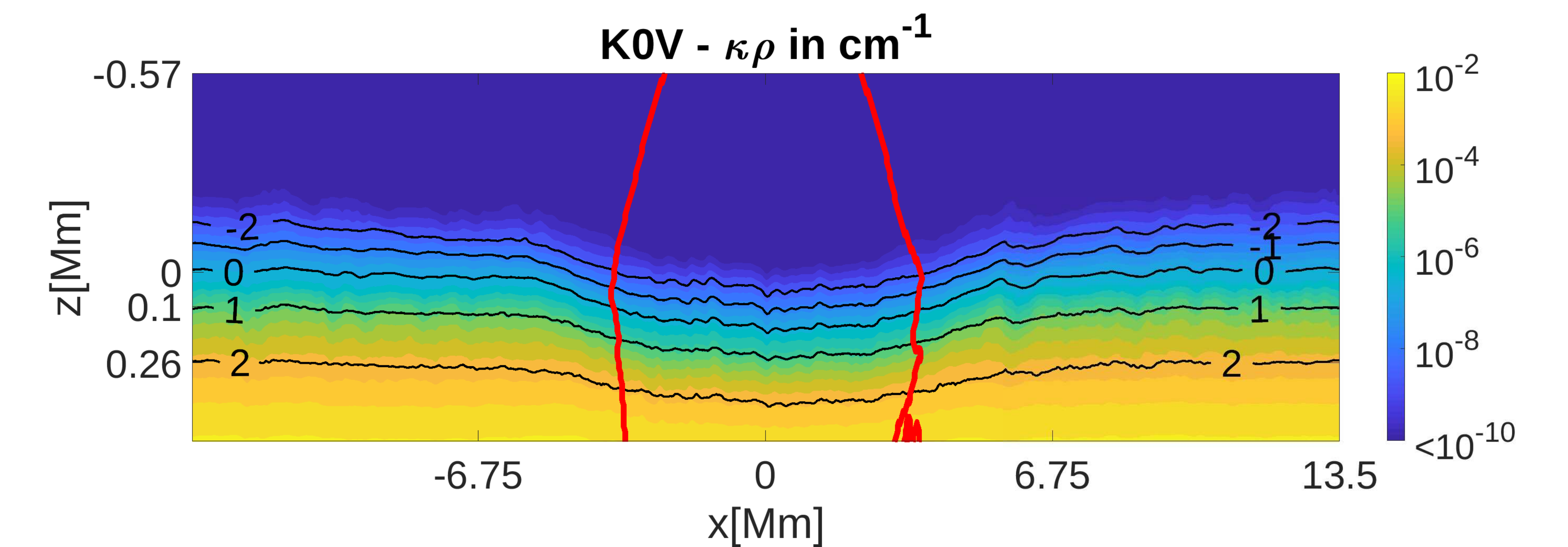}
    \hspace*{-1.0cm}\includegraphics[width=10.5cm]{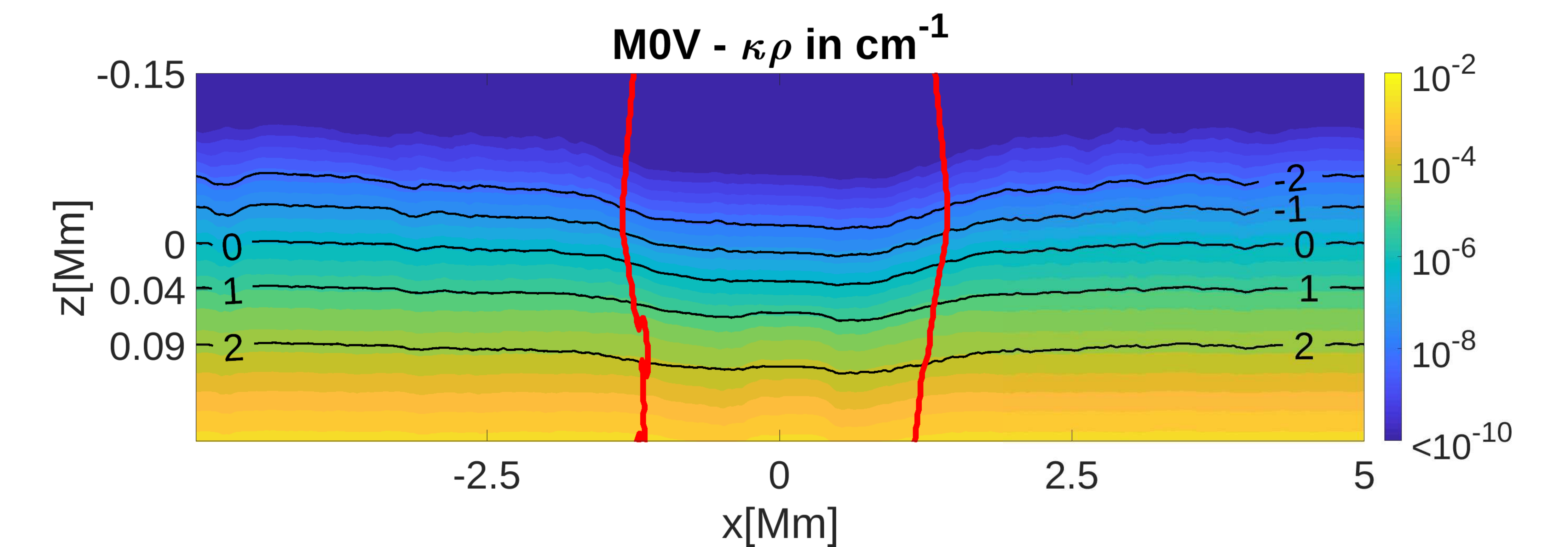}
      \caption{The absorption coefficient ($\kappa\rho$), averaged horizontally, computed in units of cm$^{-1}$. The black and red contours are the same as in Fig: \ref{fig:fig14}. 
              }
        \label{fig:fig17}
    \end{figure}

\begin{figure}
   \centering
	\hspace*{-0.5cm}\includegraphics[width=10.5cm]{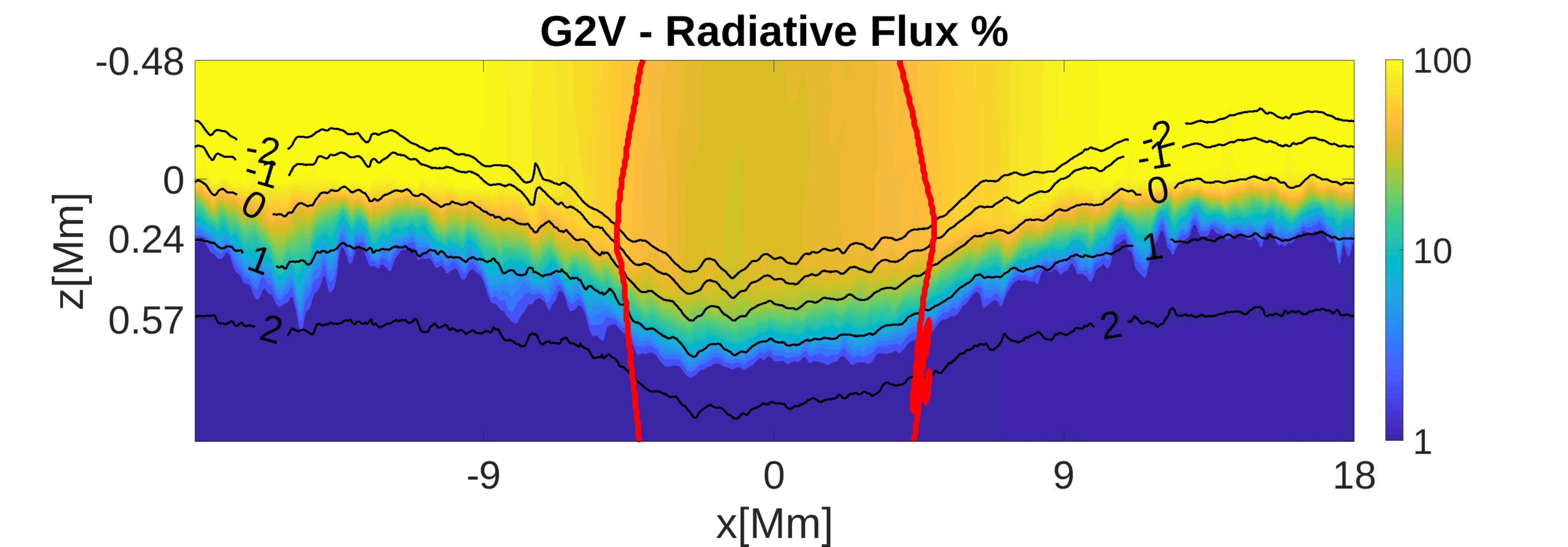}
    \hspace*{-0.5cm}\includegraphics[width=10.5cm]{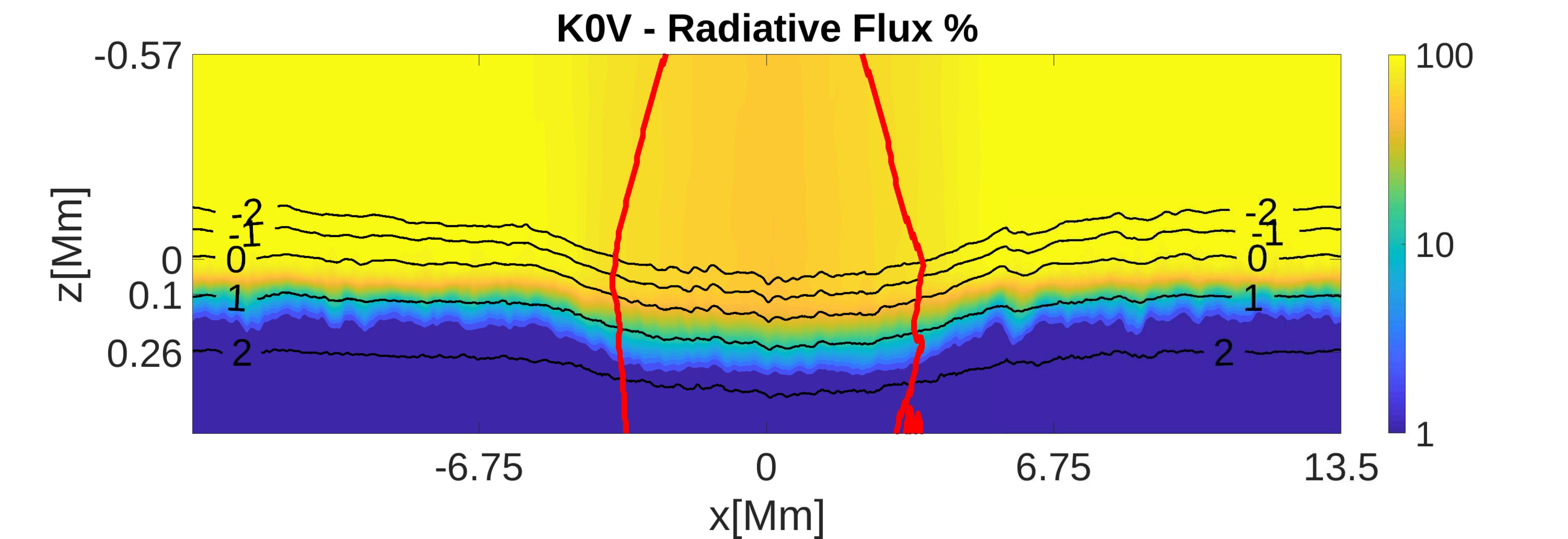}
    \hspace*{-0.5cm}\includegraphics[width=10.5cm]{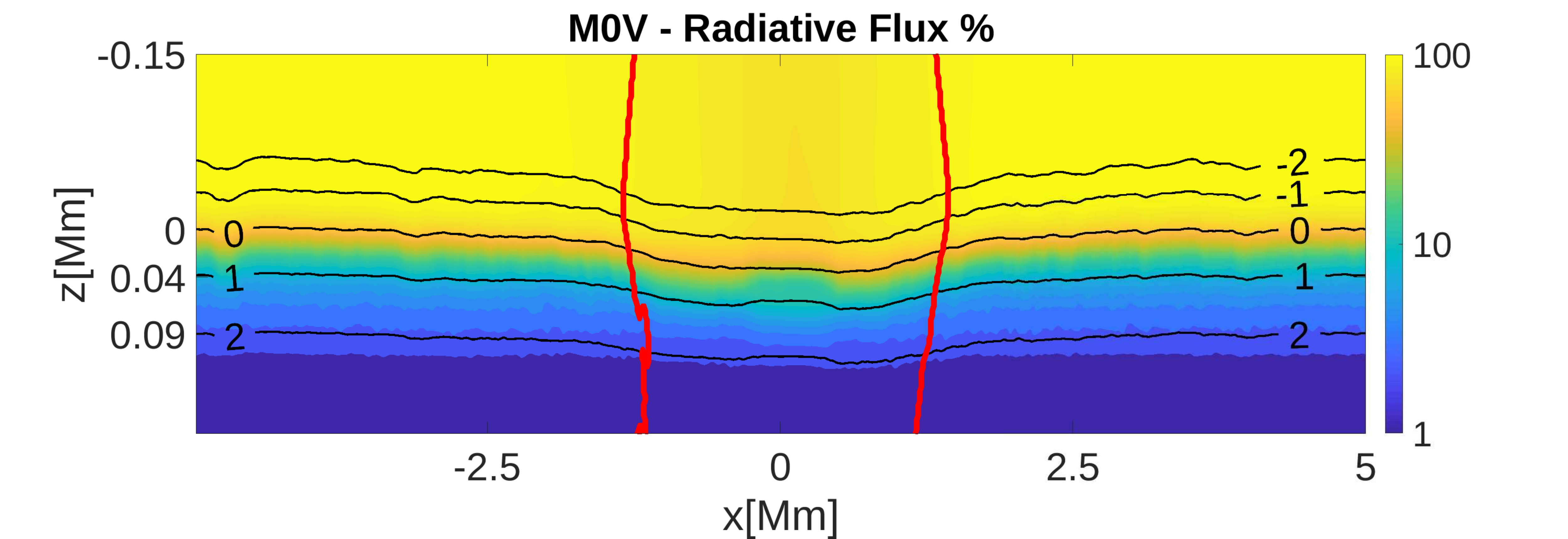}
    \caption{The vertical component of the radiative flux, expressed a percentage of the average radiative flux leaving the box above the quiet star regions. The black and red contours are the same as in Fig: \ref{fig:fig14}. 
          }
        \label{fig:fig18}
    \end{figure}

 Figure \ref{fig:fig17} shows the horizontally averaged absorption coefficient($\kappa\rho$) of the stellar atmospheres. In the quiet star regions, we find that the transition from optically thin to thick takes place over a larger number of pressure scale heights for the cool M star, while for the G star this change is much more rapid. This is also reflected in the vertical component of the radiative flux which is plotted in Figure \ref{fig:fig18}. The radiative flux has been normalized at every point by the final radiative flux leaving the box above the quiet star regions. This value rises from less than 1 \% to nearly 100 \% over a single pressure scale height for the quiet G2V atmosphere. The transition from radiative to convective energy transport is much more gradual for the cooler M star and takes place over several pressure scale heights. For the K star the transition rate lies between the M and G stars. 

In the quiet G2V atmosphere, the energy transport by radiation is negligible ($<$ 1\% of the flux leaving the box) at a depth of 1 pressure scale height, whereas in the spot atmosphere the radiative flux is already at 10\% at a similar depth. Clearly, the radiative properties inside the G starspot are very different from the radiative properties of the G  quiet star atmosphere. However, the differences between the M spot atmosphere and its surroundings are not that pronounced. The radiative properties of the K spot lie between the M and G starspots. 

\begin{figure}[ht]
   \centering
	\hspace*{-1cm}\includegraphics[width=10.5cm]{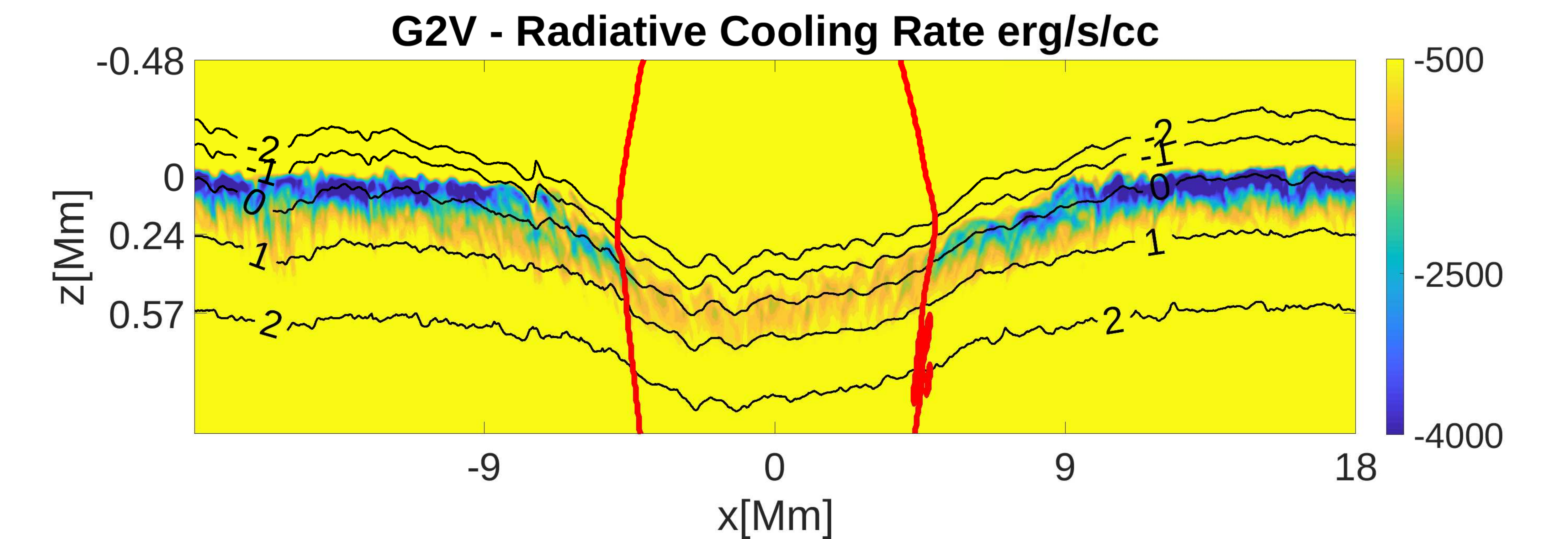}
    \hspace*{-1cm}\includegraphics[width=10.5cm]{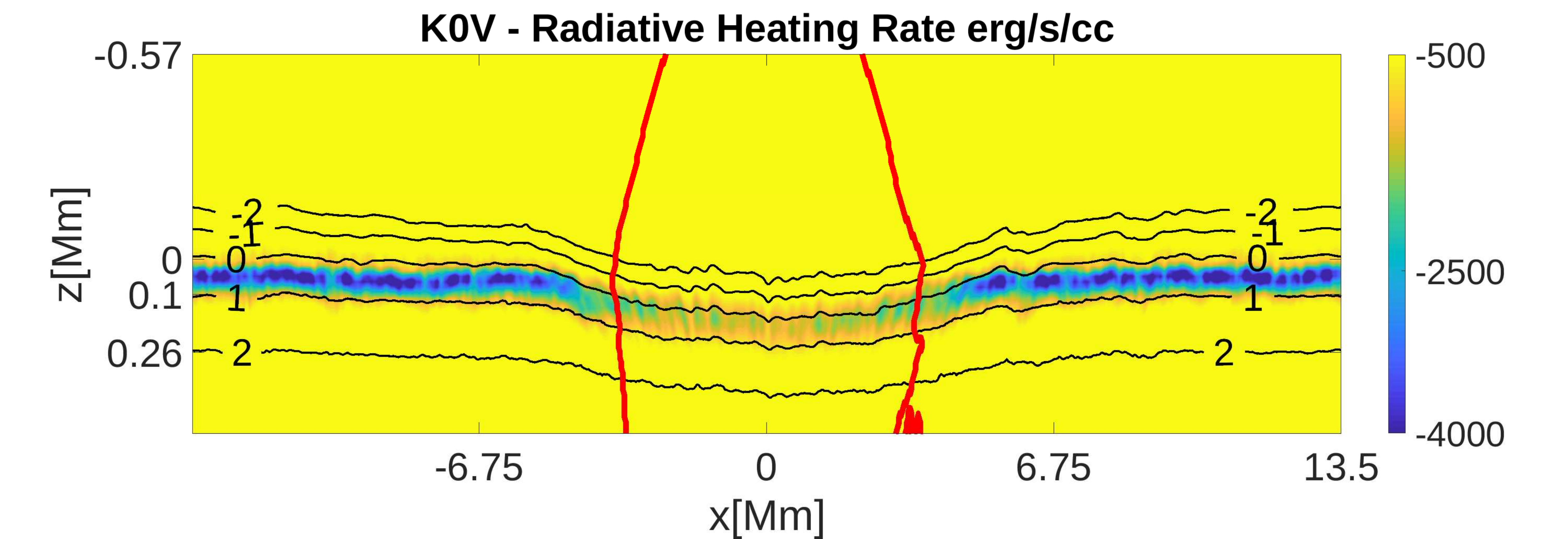}
    \hspace*{-1cm}\includegraphics[width=10.5cm]{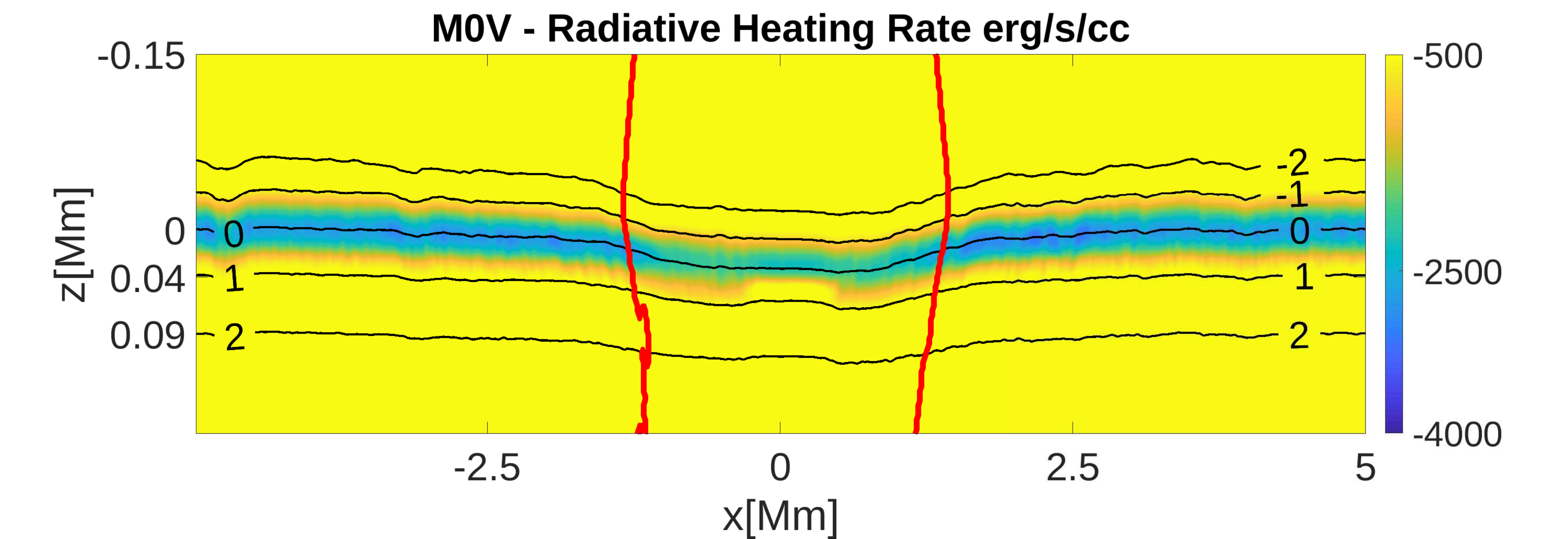}
       \caption{Radiative heating rates, averaged horizontally, in erg cm$^{-3}$ s$^{-1}$. The black and red contours are the same as in Fig: \ref{fig:fig14}. 
               }
        \label{fig:fig19}
    \end{figure}

Figure \ref{fig:fig19} plots the radiative heating rates for the different stars with depth. In the G star the cooling is much more concentrated and intense near the photosphere, while the radiative cooling for the M star is spread out over a larger vertical extent. However inside the spots, the radiative cooling is spread out over almost 2 pressure scale heights for all of the spots. The spots are at much lower temperatures and therefore their cooling rates are lower as well.

\begin{figure*}[!htbp]
  
   \centering
   \hspace*{-0.2cm}\includegraphics[scale=0.25]{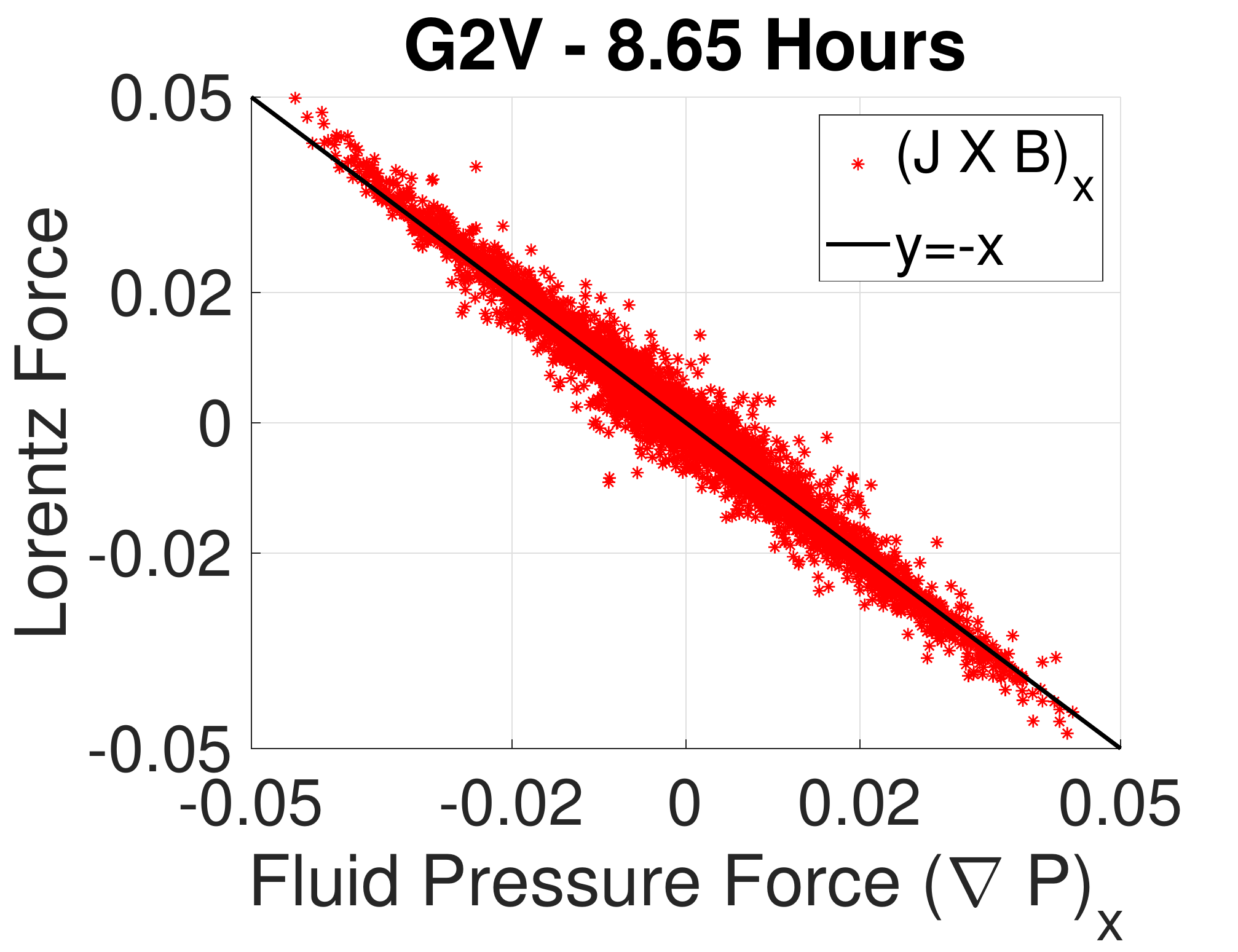}
\hspace*{-0.2cm}\includegraphics[scale=0.25]{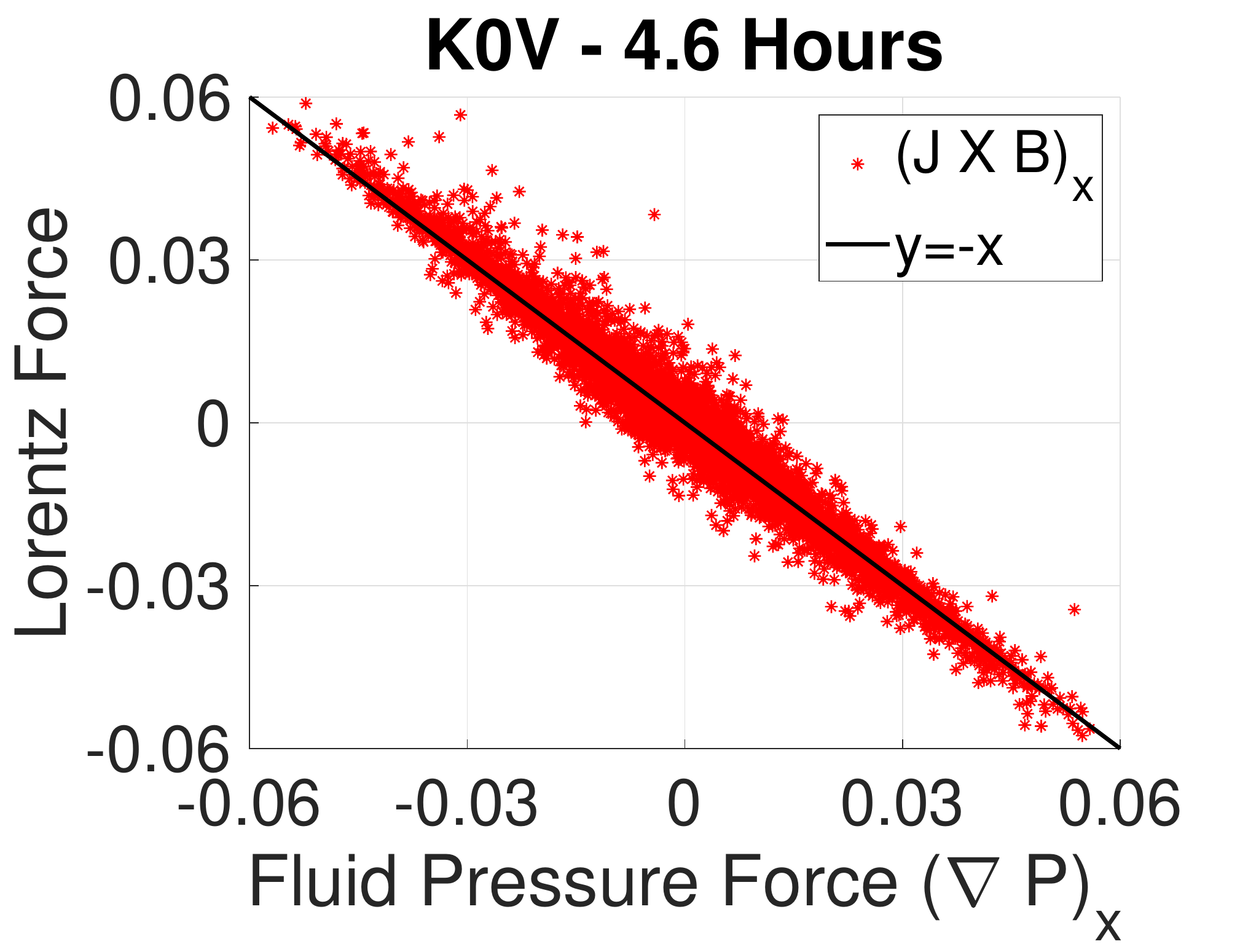}
 \hspace*{-0.2cm}\includegraphics[scale=0.25]{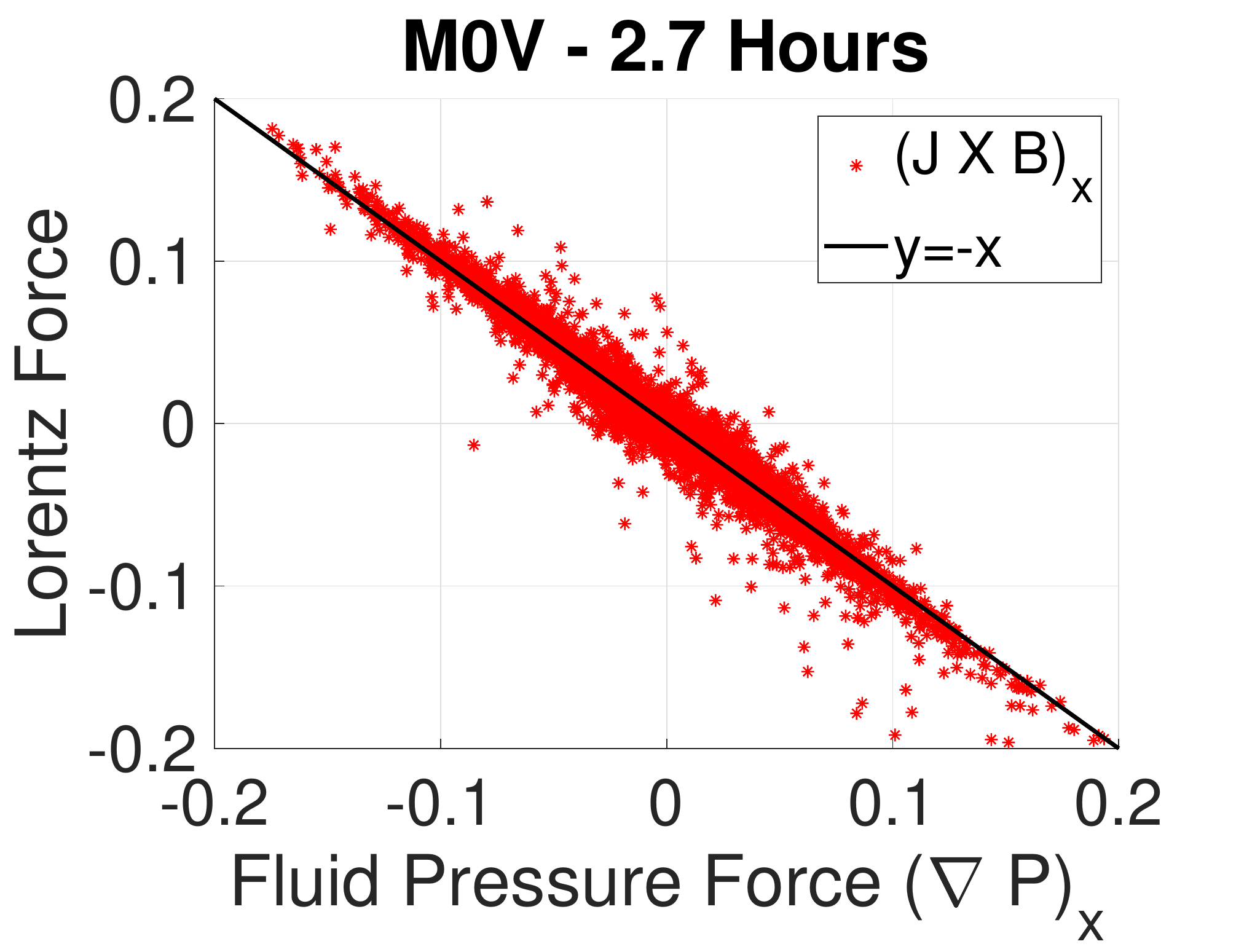}
   
\caption{Scatter plots of the magnetic forces against the fluid pressure forces at constant geometrical depths. The horizontal cuts were taken at depths which corresponded to the average Wilson depression of the umbrae}.
              \label{fig:fig20}
    \end{figure*}
    
    \begin{figure*}[!htbp]
   \centering
	\hspace*{-0.80cm}\includegraphics[width=10.5cm]{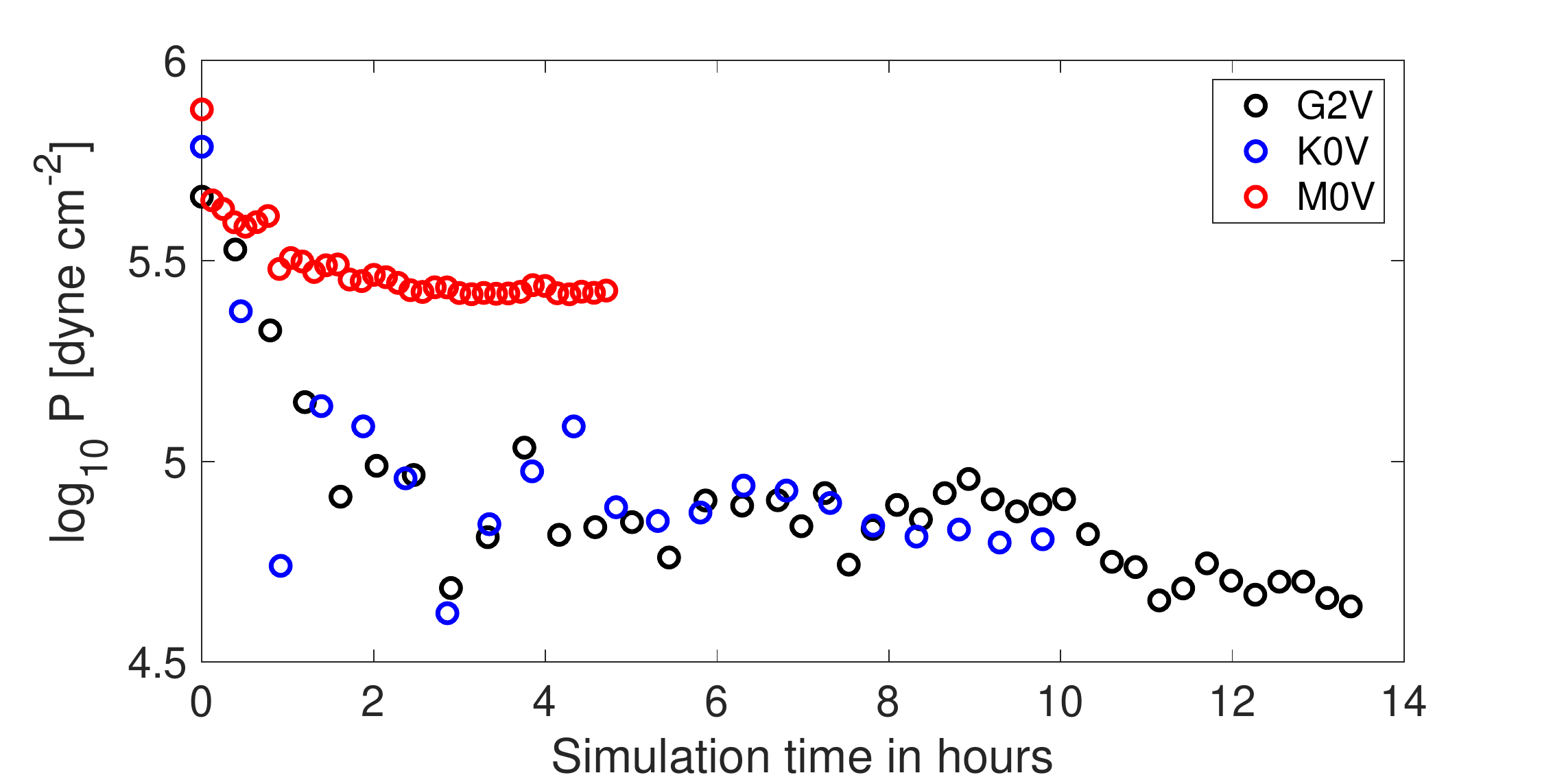}
    \hspace*{-0.80cm}\includegraphics[width=10.5cm]{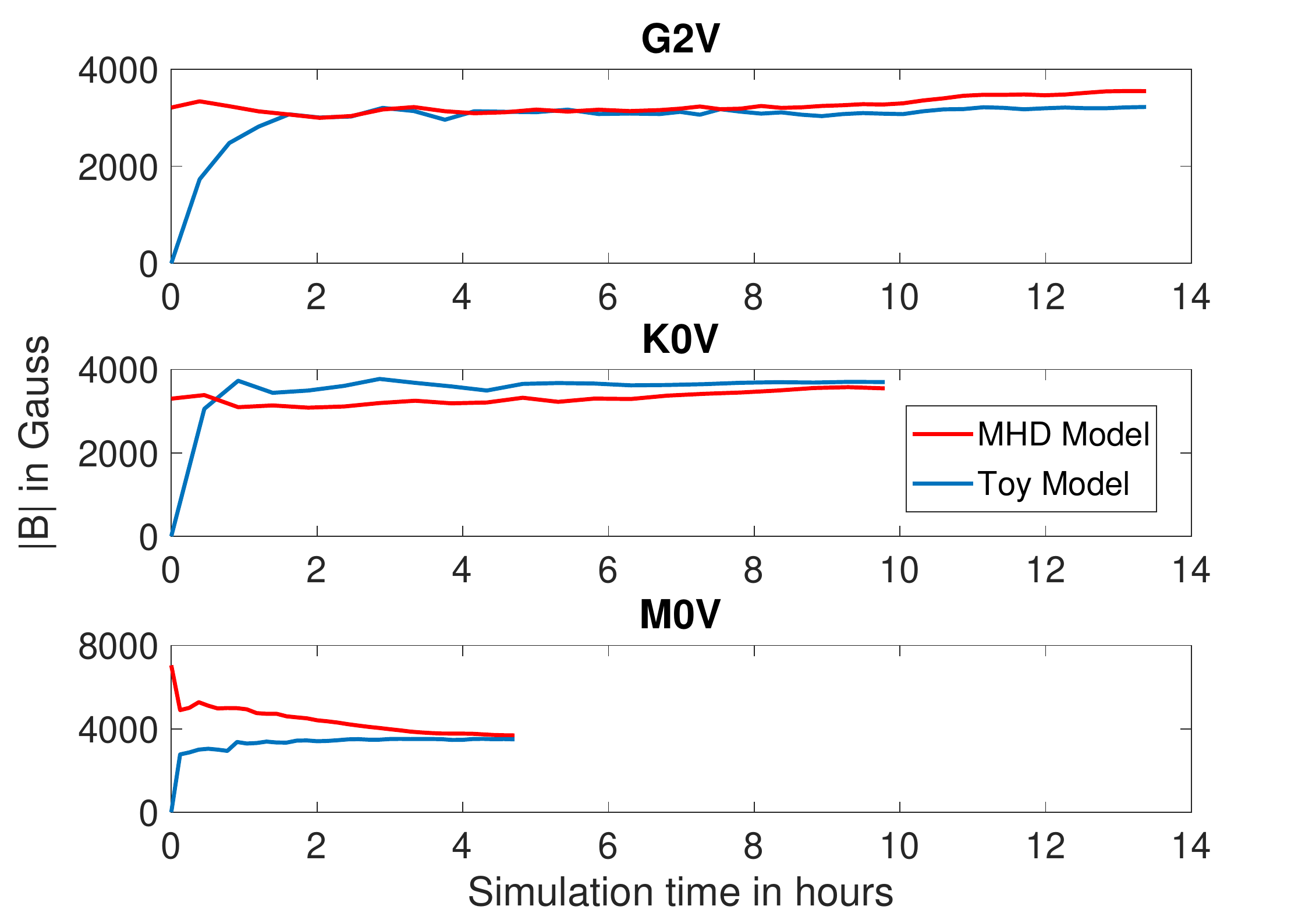}
   
      \caption{The first panel shows the evolution of average fluid pressure with time in a box close to the optical surface inside the starspots. The three bottom panels plot the average magnetic field strength in the box (red curves) and the field strength one would expect (blue curves) from the drop in fluid pressure ($\sqrt{8\pi(P_{t} -P_{initial})} $).  
              }
        \label{fig:fig21}
    \end{figure*}

\section{Discussion}  \label{sec:disc}

\subsection{Spot Temperature Contrast}

The results of our simulations reveal a clear pattern in starspot intensity contrasts. There is a monotonic decrease in contrast from the hottest to the coolest star. This is also seen in the list of starspot temperature measurements compiled by \cite{Berdyugina2005}.
This trend can be explained by the dependence of opacity on temperature. In the context of stellar near-surface simulations, this has been described by \cite{NordandDravis90,Beeck1,STAGGER_I,cobold}. The opacity at the surface of cooler stars is largely governed by H$^{-}$ ions and in the temperature range 3000 - 6000 K, the H$^{-}$ opacity is approximately proportional to $T^{9}$ \citep{opacity_book}. Thus, the opacity of H$^{-}$ increases much more sharply with temperature in the hotter G2V surface (6000 K) in contrast to the cooler photospheres of K (5000 K) and M (4000 K) stars. Therefore, in the hotter G2V quiet star atmosphere, a slight increase in temperature with depth causes the opacity to rise sharply. This results in a sudden change from convective energy transport to radiative energy transport in the hotter G2V star within a span of half a pressure scale height, whereas this change, from convective to radiative energy transport, is the most gradual for the M0V case and is spread over nearly 2 pressure scale heights

When we introduce strong magnetic fields in the G2V star, the transport of energy upwards by convection is hindered. Since convection is the primary mode of heat transport below the photosphere in the G2V star, the temperature of the umbra is lowered substantially. As a consequence, the opacity falls sharply and the increased effectiveness of radiation also contributes to lowering the temperature. On the other hand, in the M0V star, where radiation already plays a substantial role in energy transport below the photosphere, the magnetic fields have a smaller impact on the  energy transport. Also at 4000 K, the surface temperature of the M0V star, the opacity is less dependent on temperature and on introducing magnetic fields, there is only a small change in radiative properties of the medium as seen in Figures \ref{fig:fig17}, \ref{fig:fig18} and \ref{fig:fig19}. 

This explains why the G starspot has the highest temperature contrast, and the M spot is not as cool as one would naively expect it to be. The radiative properties of the K star atmosphere lie in between the M and the G star and this is reflected in the K spot temperature contrast as well.

\subsection{Wilson Depression}

The Wilson Depression of the starspots varies significantly with spectral type - 500 km for the G2V spot to around 30 km for the M0V spot. This can be explained largely by the difference in pressure scale height between the stars. The pressure scale heights, near the surface, of the quiet star atmospheres are  230 (G2V), 100 (K0V) and 40 km(M0V).
However, the Wilson depression when expressed in terms of pressure scale height also exhibits significant differences -  the G2V umbra is more than 2 pressure scale heights deep, whereas the M0V umbra is less than a pressure scale height deep. Similar variations of the Wilson depression with spectral type, were also observed in small flux tubes embedded in the intergranular regions in the simulations of \cite{Beeck2} and \cite{cobold}.

The substantial depression of the optical surface in the G2V spot compared to the M0V spot can be explained by the difference in the absorption coefficient inside and outside the spot.  At the same geometrical height, the absorption coefficient (plotted in Figure \ref{fig:fig17}) drops by several orders of magnitude inside the G2V spot. This is not the case for the M0V spot. 
The M0V star has a higher surface pressure ($\sim$ 5 times that of the G2V star) and therefore the gas is less evacuated in the M spot atmosphere, in spite of its larger field strength. This is reflected in the plasma-$\beta$ ratio - $8\pi P/B^{2}$ (see Figure \ref{fig:fig15}) above the M0V spot. In addition, the weaker dependence of opacity on temperature in the 3000 - 4000 K range means that the opacity inside and outside the M0V spot are comparable. This results in the much smaller depression of the M-spot optical surface. 

\subsection{Umbral Magnetic Field Strength}

The umbral field strengths of the starspots lie between 3 and 4.5 kG. The difference between the average umbral field strengths of the M0V spot and the G2V spot is only around 700 Gauss despite the surface pressure of the M star being 5 times higher. This is related to the change in the magnitude of the Wilson Depression with spectral type. 

In Figure \ref{fig:fig20} we have plotted the $x$-component of the Lorentz force against the $x$-component of the fluid pressure force at a constant geometrical depth close to the optical surface of the spots. If we assume magnetostatic equilibrium, $\nabla(P)$ and $J \times B$ should have equal values and opposite signs. The $y=-x$ line shows a good fit to the points and this shows the simulated spots are close to magnetostatic equilibrium at their respective optical surfaces. 

Further, we have constructed a simple model to predict the umbral field strength assuming pressure balance. For each of the three spots, we take a small region inside the starspots near the optical surface and plot the fluid pressure with time (top panel in Figure \ref{fig:fig21}).

If $P_{0}$ is the initial pressure, then $P_{0}$ - $P_{t}$ (pressure at time t) would give us the pressure of the displaced gas. Equating $P_{0}$ - $P_{t}$ with $\frac{B^2}{8\pi}$ would give us an expected magnetic field strength. In the lower three panels of Figure \ref{fig:fig21}, we have plotted the expected magnetic field strength and the actual field strength obtained in this box, with time. After an initial transient phase, the field strengths we obtain from our simulations are comparable to what is predicted by a simple pressure equilibrium model.


The measured umbral field strength is thus a result of two competing effects – the gas pressure of the star and how deep we are seeing into the star. As we move from the G2V to the M0V star the surface pressure increases, which would mean higher umbral field strengths, while the atmosphere becomes less vacuous and more opaque, which would lower the measured umbral field strengths. In the case of the K0V star, the effect of the absorption coefficient becoming higher wins over the competing effect of  increased surface pressure (~1.8 times the G2V case) and therefore the umbral field strength of the K0V starspots (2D and 3D) is lower than the G2V starspots. Whereas for the M0V star, the gas pressure is high enough (~5 times the G2V case) that we still record higher umbral field strengths. We have further demonstrated this with a simple calculation.

Let the pressure at the surface of the G2V star be Po. So the surface pressures of the K0V and the M0V stars would roughly be 1.8Po and 5Po respectively. The Wilson depressions of the G2V, K0V, and M0V starspots, when expressed in terms of pressure scale height (Hp) are – 2.25, 1.61, and 0.85 respectively. This means, at the heights where the optical surfaces of the three spots form, the ambient gas pressures would be approximately Po exp(2.25), 1.8Po exp(1.61) and 5Po exp(0.85). This yields values of 9.5Po, 9.0Po, and 11.7Po  for the G2V, K0V, and M0V stars respectively. Although this calculation is very simplistic, it explains the trend in umbral field strengths very well.


\section{Summary and Conclusion} \label{sec:summary}
We have performed the first-ever, ab-initio radiative MHD simulations of spots on cool main sequence stars of the spectral types G2V, K0V and M0V.  We investigated the variation of fundamental spot properties - temperature, intensity and magnetic field strength with spectral type. Our main findings can be summarized as follows - \\

1) Our simulations show that the temperature contrast between a starspot and its surrounding photospheric environment is a function of the stellar surface temperature. The hotter the stellar surface, the higher is the spot temperature contrast. Obviously, this trend is reproduced in the intensity contrast as well. Our analysis has revealed that the variation in spot contrast with spectral type can be attributed to radiative processes playing an increasingly dominant role in the atmospheres of cooler stars. This is because as we progress to cooler stars the ionisation of H$^{-}$ takes place at greater pressure scale depths, making the atmosphere near the surface more transparent, thereby smoothing out both horizontal and vertical fluctuations in temperature. Our starspot simulation results are consistent with the larger picture that cooler the stellar surface, lower are the variations in temperature and intensity. 

2) Our simulated umbral field strengths for all the spectral types lie in the 3-4.5 kG range. The umbral field strength is largely determined the fluid pressure at the depth where the $\tau = 1$ surface of the spots form. The optical surface of the G2V starspot is more than 2 pressure scale heights deep whereas the optical surface of the M0V starspot is less than 1 pressure scale height deep. This explains why M stars and G stars have umbral field strengths that are not too different.

3) Prior to conducting 3D simulations, we carried out 2D studies, where we varied the initial conditions of our magnetic flux tubes extensively. Although our 2D starspots display slightly different values of temperature and field strength, they reproduce the trends seen in the intensity contrasts in our 3D simulations very well. In addition, they also show that our obtained trends in spot temperatures and magnetic field strengths do not depend crucially on our choice of initial conditions. 

4) All the simulated starspots develop penumbral filament-like structures. The filaments of the K0V spot look similar to solar penumbral fiaments, with thin dark cores running along the centre of the filaments. The M0V penumbral filaments are more homogeneous and do not develop such dark cores. The Evershed speeds decrease progressively from the G2V spot to the M0V spot.

We expect our calculations to help understand the variability of G-M main sequence stars, which is largely determined by starspots on rotational timescales. Conversely, our starspot models provide useful priors for reconstructing stellar spottedness based on light-curve modeling and (Zeeman) Doppler Imaging. Our results will also aid the hunting of exo-planets. Starspot intensity contrasts are important in the exoplanet detection process as spots and planets both reduce the amount of light we receive from a star. Our constraints on spot temperatures should also be useful in interpreting maps of stellar magnetic fields derived using Zeeman Doppler Imaging. In addition, we expect our starspot properties to help improve estimates of the total magnetic flux on lower main-sequence stars, thus setting improved constraints on the efficiencies of stellar dynamos.

Future work may involve using non-grey radiative transfer to enable the synthesis of spectral lines. Additionally, simulating spot pairs of opposite polarities would facilitate the formation of more expansive penumbrae and therefore better constraints on starspot properties.

\acknowledgments

We thank Pradeep Chitta, Ivan Milic and  Yvonne Unruh for useful discussions. MP acknowledges support  by  the  International  Max-Planck  Research  School  (IMPRS)  for  Solar System Science at the University of Göttingen. This project has received funding from the European Research Council (ERC) under the European Union’s Horizon 2020 research and innovation program (grant agreement No 695075) and  has been supported by the BK21 plus program through the National Research Foundation (NRF) funded by the Ministry of Education of Korea. The simulations have been carried out on supercomputers at GWDG and on the Max Planck supercomputer at RZG in Garching.

\appendix

\section{2D Simulations}
\begin{figure*}
   \centering
	\hspace*{-1.0cm}\includegraphics[width=20.0cm]{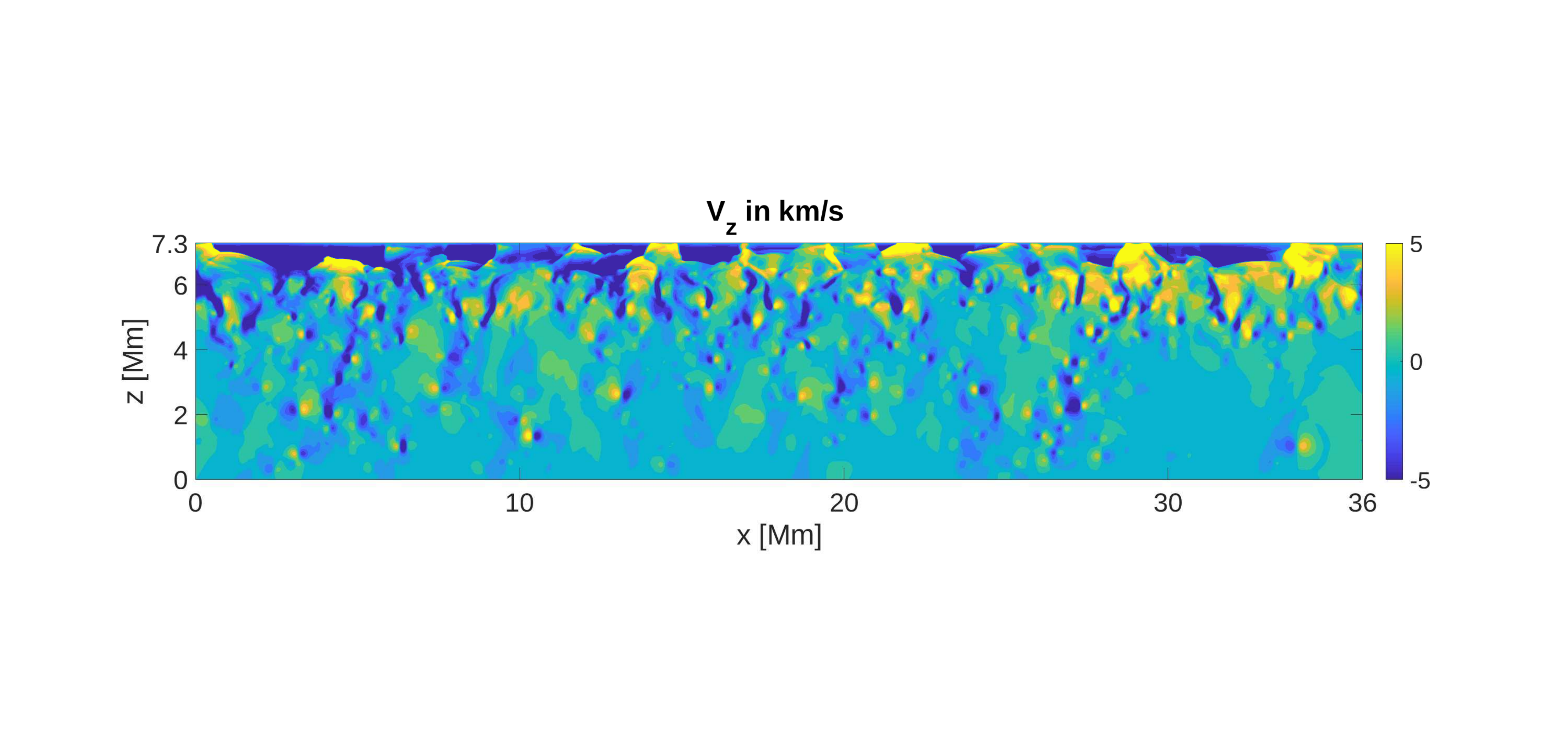}
           \caption{Vertical velocity in the 2D simulation setup before the magentic field was introduced. The colors show V$_{z}$ in km s$^{-1}$.
             }
       \label{fig:fig22}
    \end{figure*}

\begin{figure*}
   \centering
	\hspace*{-1.0cm}\includegraphics[width=20.0cm]{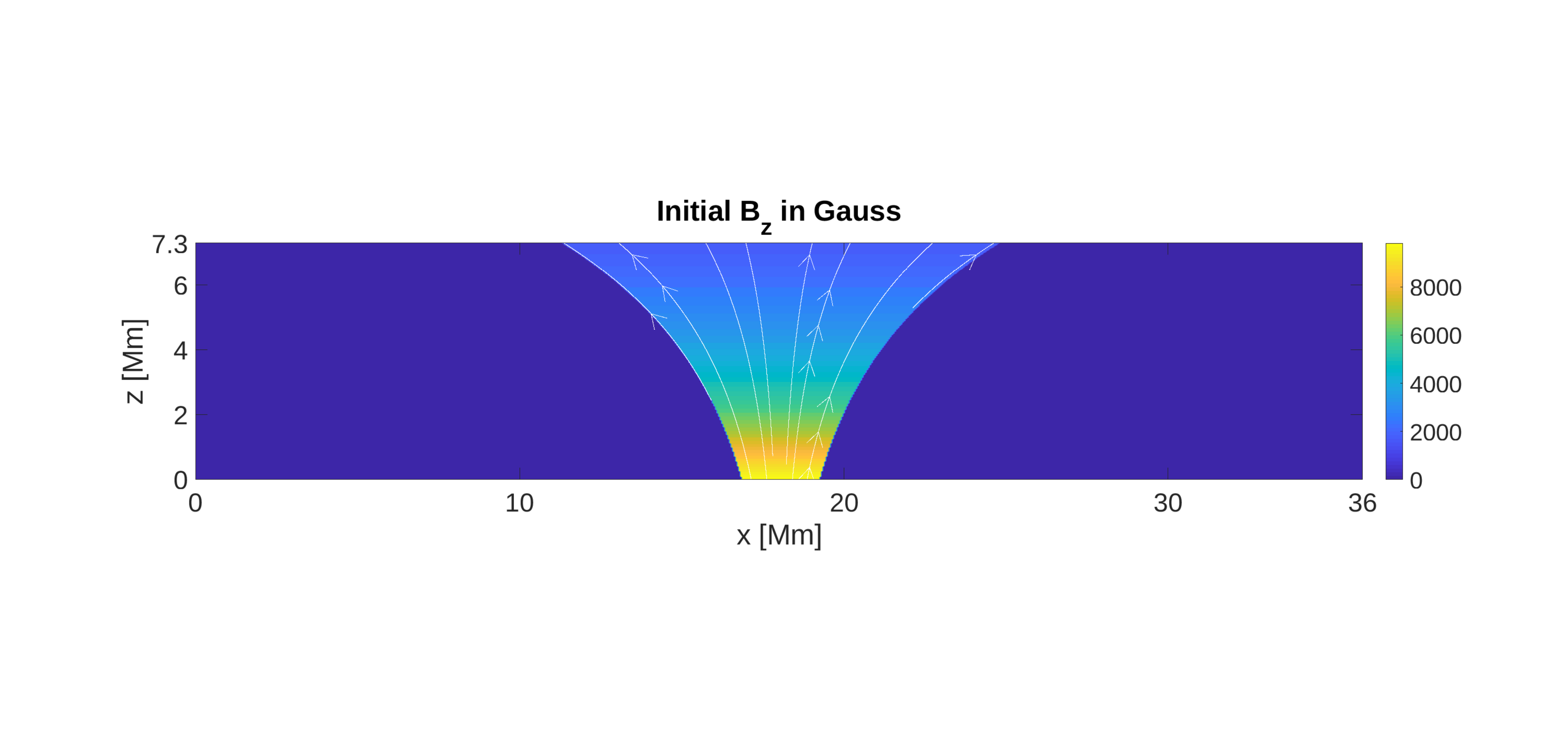}
           \caption{Initial magnetic field configuration for a 2D G2V spot simulation. The colors show B$_{z}$ in Gauss and the white lines with arrows mark sample magnetic field lines.
               }
       \label{fig:fig23}
    \end{figure*}

This section details the results of our 2D simulations which we used to explore the parameter space more extensively. As discussed before, the two parameters that determine the shape of our flux tubes are the vertical components of the magnetic field at the lower boundary and at the optical surface - $B_\mathrm{bot}$ and $B_\mathrm{opt}$. The initial conditions used for the 2D simulations were the 2D analogs of the conditions used for the 3D runs. Table \ref{table:A1} summarizes our 2D runs. Figure \ref{fig:fig22} shows the vertical velocity field before the magnetic field is put in. The upflow areas are in yellow and the downflow areas are in blue. Figure \ref{fig:fig23} shows an example of a magnetic field configuration inserted into the hydrodynamic simulation.

\begin{table}
\caption{Summary of the results of the 2D runs.}             
\label{table:A1}      
\centering                          
\begin{tabular}{c |c c c c  }        
\hline\hline                 
Sp. Type & $B_\mathrm{opt}$ (kG)& $B_\mathrm{bot}$ (kG) & $I_{umbra}$/I$_{quiet}$ &  $|B|_\mathrm{umbra}$(kG) \\ 
\hline\hline  
     & & & & \\
   G2V & 2.2 & 4 & 0.22, \textcolor{red}{0.30}  & 4.15, \textcolor{red}{4.00} \\
       & 2.2   &  6  & 0.22, \textcolor{red}{0.27} & 4.70, \textcolor{red}{4.58} \\
       & 2.2    &  8  &  0.18, \textcolor{red}{0.20}  & 5.07, \textcolor{red}{5.03} \\
       & 2.2    &  10  &  0.20, \textcolor{red}{0.22} & 4.89, \textcolor{red}{4.84} \\
       & 2.2    &  12  &  0.19, \textcolor{red}{0.23} & 4.95, \textcolor{red}{4.87} \\
       & 2.2    &  14  &  0.19, \textcolor{red}{0.22}  & 5.16, \textcolor{red}{5.06}            \\
    & & & & \\
     & 4.4    &  12  &  0.18, \textcolor{red}{0.19} & 4.82, \textcolor{red}{4.78} \\
     & 8.8    &  12  &  0.19, \textcolor{red}{0.19} & 4.98, \textcolor{red}{4.96} \\
    & & & & \\
\hline\hline 
 & & & & \\
 K0V   & 3.2    & 6    & 0.34, \textcolor{red}{0.40}  & 4.32, \textcolor{red}{4.16} \\
       & 3.2    &  8   & 0.35, \textcolor{red}{0.39} & 4.45, \textcolor{red}{4.23} \\
       & 3.2    &  10  &  0.34, \textcolor{red}{0.38}  & 4.22, \textcolor{red}{4.07} \\
       & 3.2    &  12  &  0.33, \textcolor{red}{0.47} & 4.25, \textcolor{red}{3.74} \\
       & 3.2    &  14  &  0.32, \textcolor{red}{0.35} & 4.43, \textcolor{red}{4.33} \\
       & 3.2    &  16  &  0.33, \textcolor{red}{0.37}  & 4.33, \textcolor{red}{4.24}            \\
    & & & & \\
     & 6    &  16  &  0.32, \textcolor{red}{0.33} & 4.57, \textcolor{red}{4.52} \\
     & 12    &  16  &  0.32, \textcolor{red}{0.35} & 4.53, \textcolor{red}{4.43} \\
     & & & & \\
 \hline\hline 
 & & & & \\    
   M0V & 3.0    & 5    & 0.68, \textcolor{red}{0.74}  & 4.50, \textcolor{red}{4.03} \\
       & 3.0    & 8    & 0.67, \textcolor{red}{0.69} & 4.66, \textcolor{red}{4.54} \\
       & 3.0    & 10   &  0.67, \textcolor{red}{0.69}  & 4.91, \textcolor{red}{4.76} \\
       & 3.0    & 12   &  0.68, \textcolor{red}{0.71} & 5.06, \textcolor{red}{4.87} \\
       & 3.0    & 15   &  0.66, \textcolor{red}{0.69} & 5.30, \textcolor{red}{5.05} \\
       & 3.0    & 18   &  0.66, \textcolor{red}{0.71}  & 5.40, \textcolor{red}{4.82}            \\
    & & & & \\
     & 6    &  15  &  0.65, \textcolor{red}{0.69} & 5.36, \textcolor{red}{5.01} \\
     & 12    &  15  &  0.65, \textcolor{red}{0.67} & 5.57, \textcolor{red}{5.42} \\
     & & & & \\
  
\hline                                   
\end{tabular}
\tablecomments{The numbers in black indicate averages computed using simple intensity and magnetic field thresholds to define umbral regions.The numbers in red indicate averages computed without ignoring the sharp intensity peaks seen in the umbral regions of our 2D spot simulations. See A.1 for more details.}
\end{table}

\subsection{Selecting the umbra}
\begin{figure}
   \centering
	\hspace*{-1.0cm}\includegraphics[width=10.0cm]{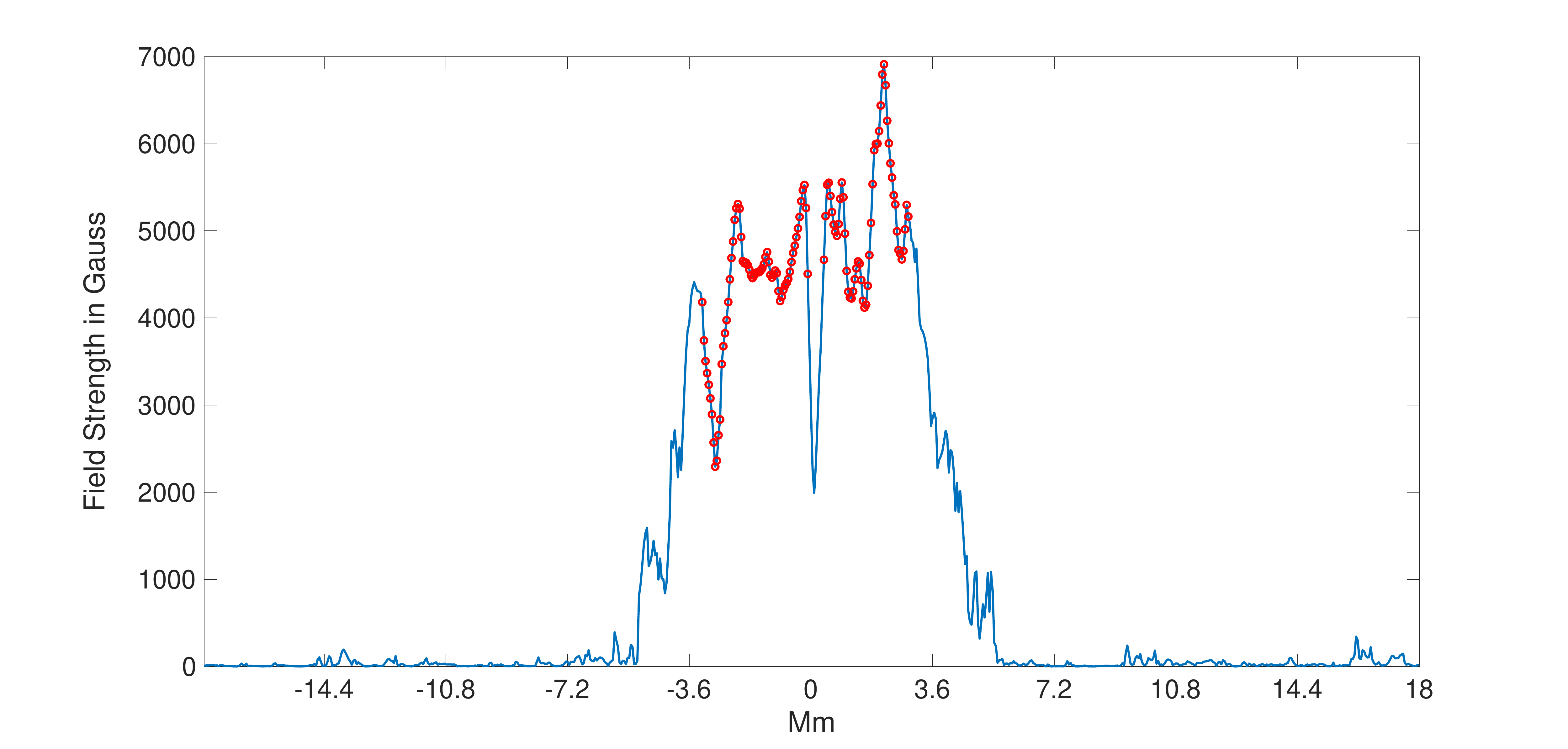}
    \hspace*{-1.0cm}\includegraphics[width=10.0cm]{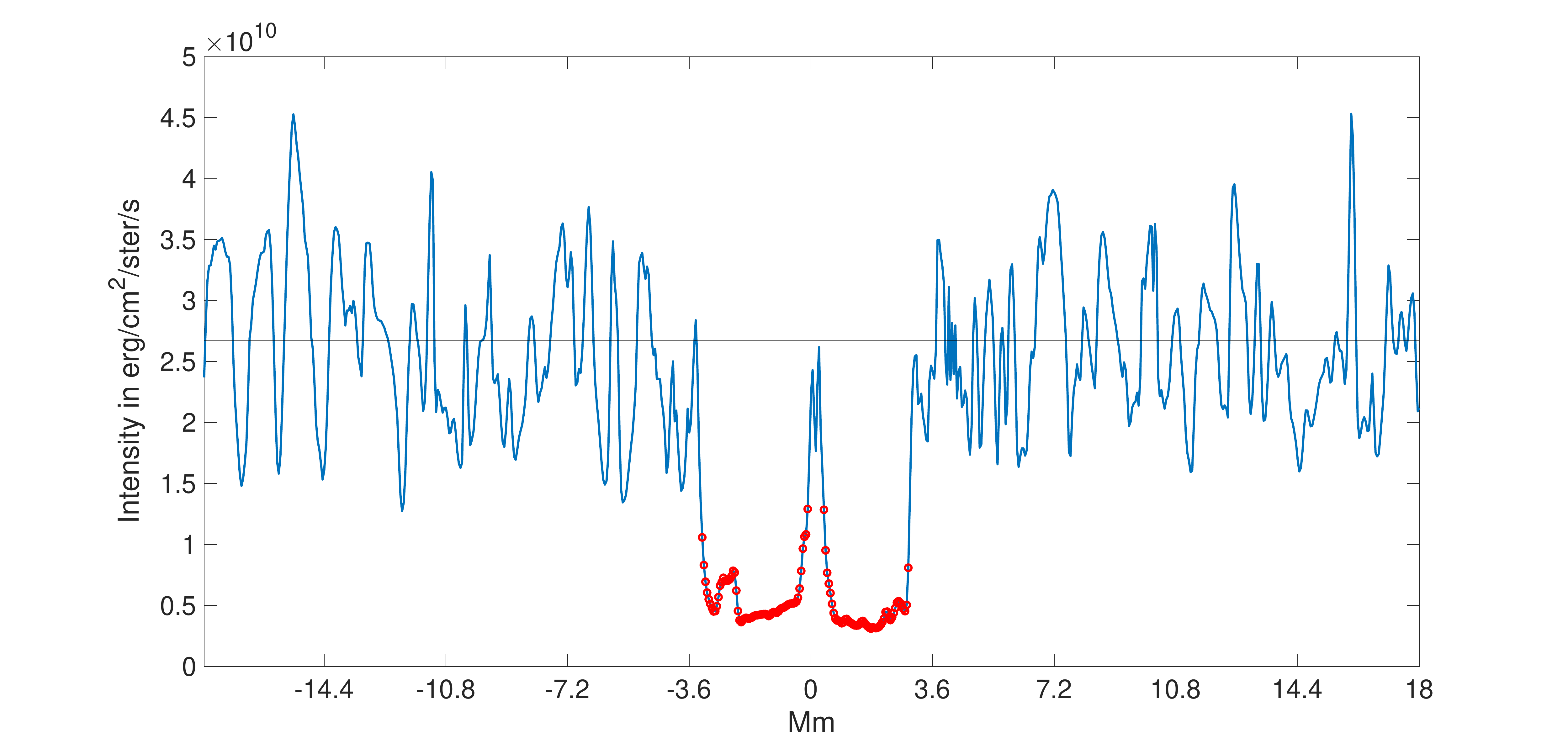}
           \caption{Umbral field strength at the optical surface and bolometric intensity for a sample G2V spot. The parts in red mark the region defined as the umbra. This definition of the umbra excludes the spike observed in both the intensity and field strength. The black horizontal line represents the average quiet star intensity.
               }
        \label{fig:fig24}
    \end{figure}

\begin{figure}
   \centering
	\hspace*{-1.0cm}\includegraphics[width=10.0cm]{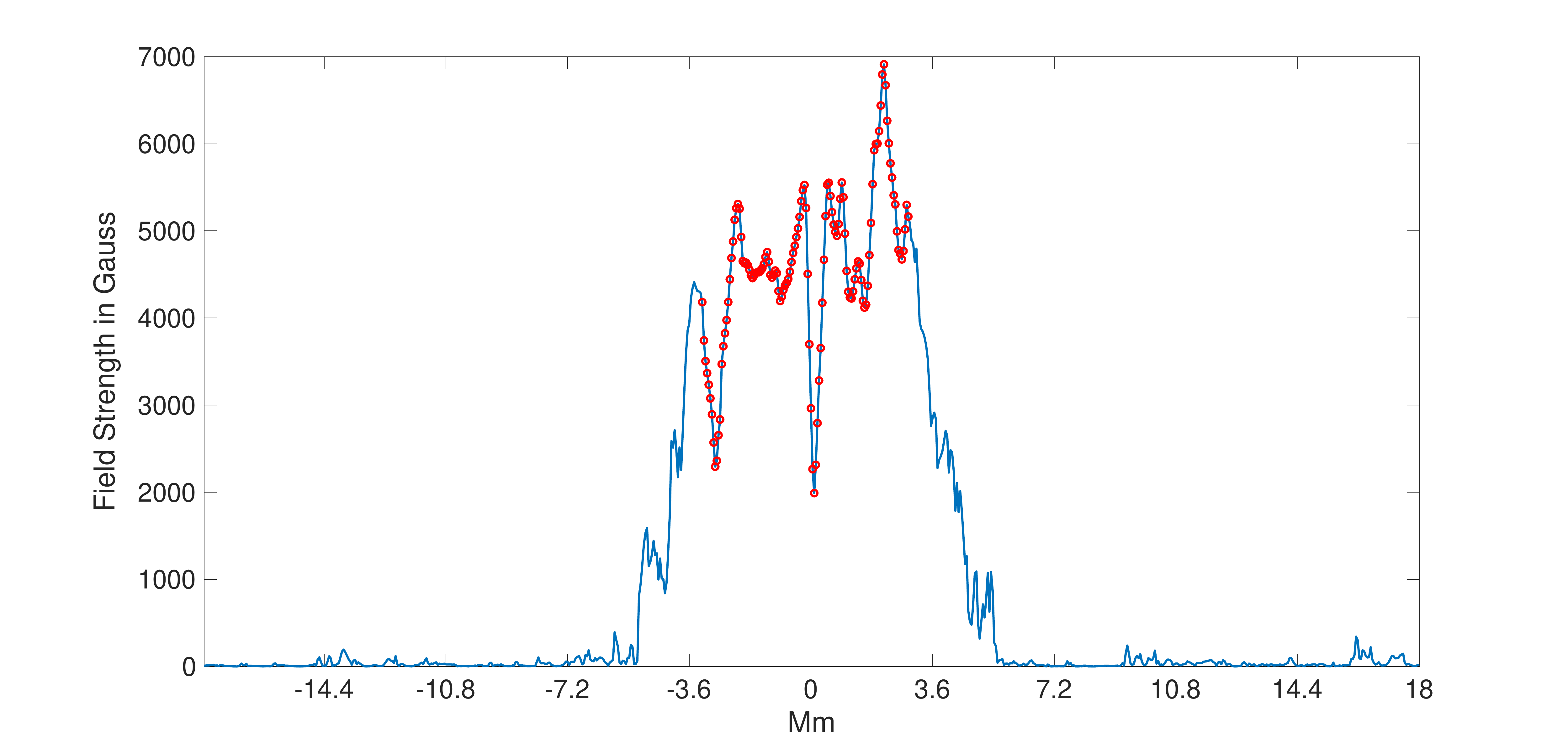}
    \hspace*{-1.0cm}\includegraphics[width=10.0cm]{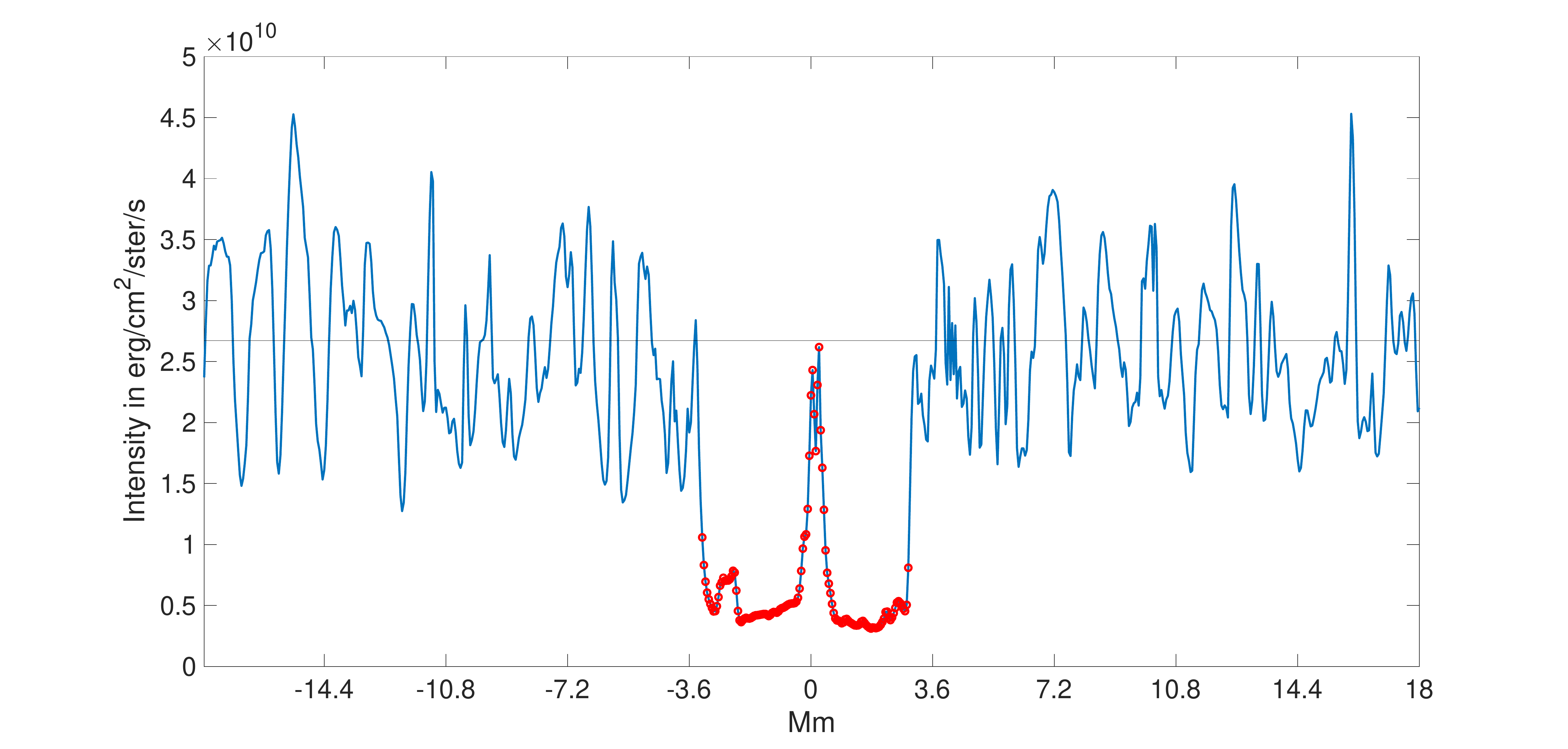}
           \caption{Umbral field strength at the optical surface and bolometric intensity for the G2V spot shown in Fig. \ref{fig:fig24} . The parts in red mark the region defined as the umbra. This definition of the umbra includes the spike observed in both the intensity and field strength. The black horizontal line represents the average quiet star intensity.
               }
       \label{fig:fig25}
    \end{figure}

\begin{figure}
   \centering
	\hspace*{-1.0cm}\includegraphics[width=10.5cm]{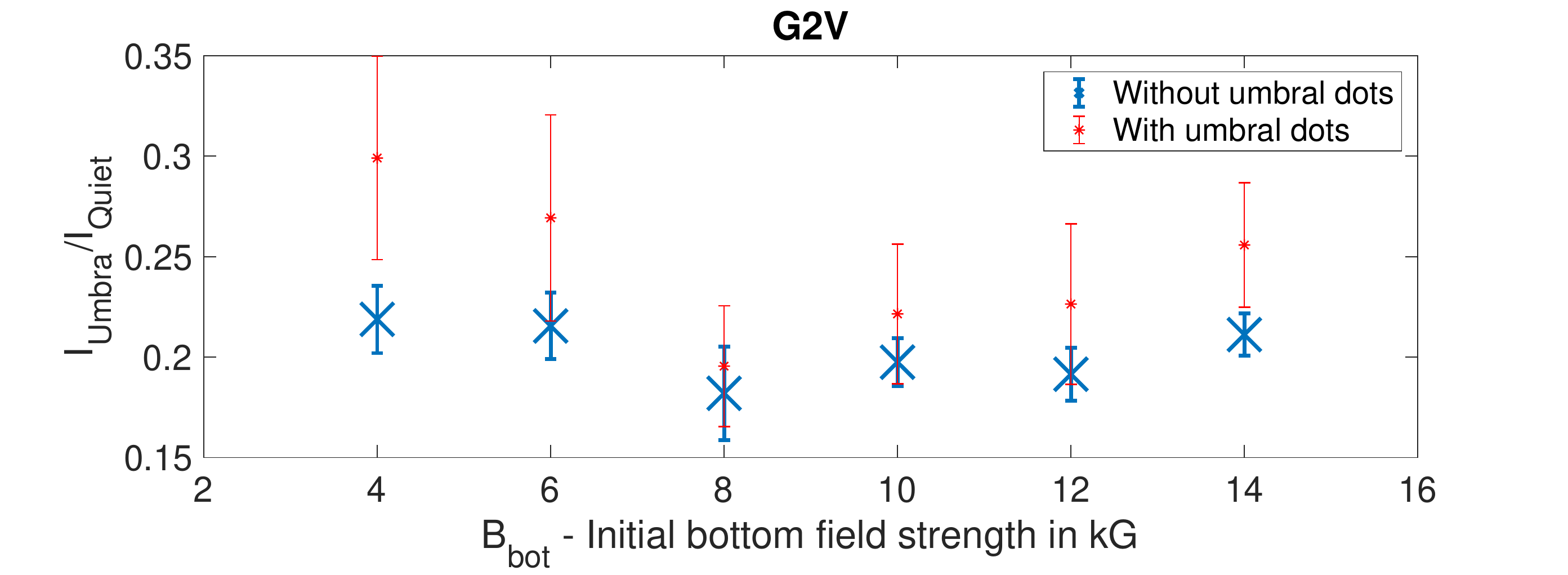}      
    \hspace*{-1.0cm}\includegraphics[width=10.5cm]{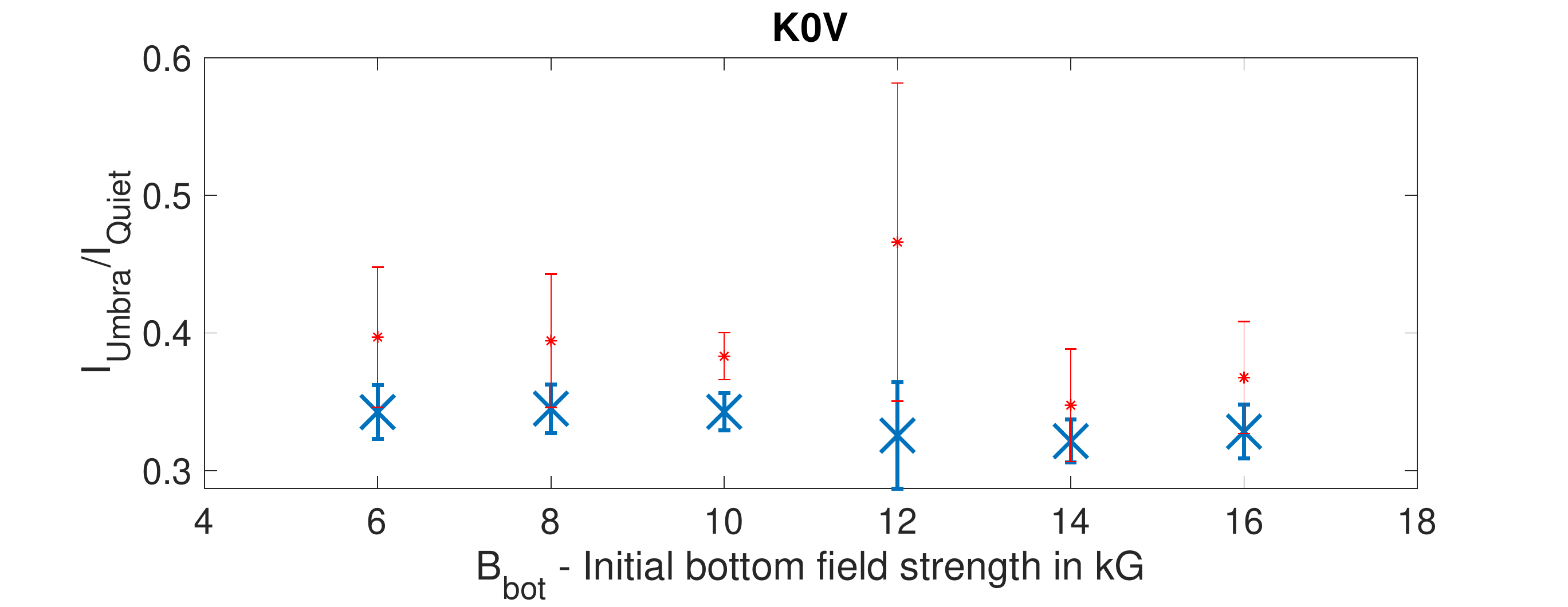}
    \hspace*{-1.0cm}\includegraphics[width=10.5cm]{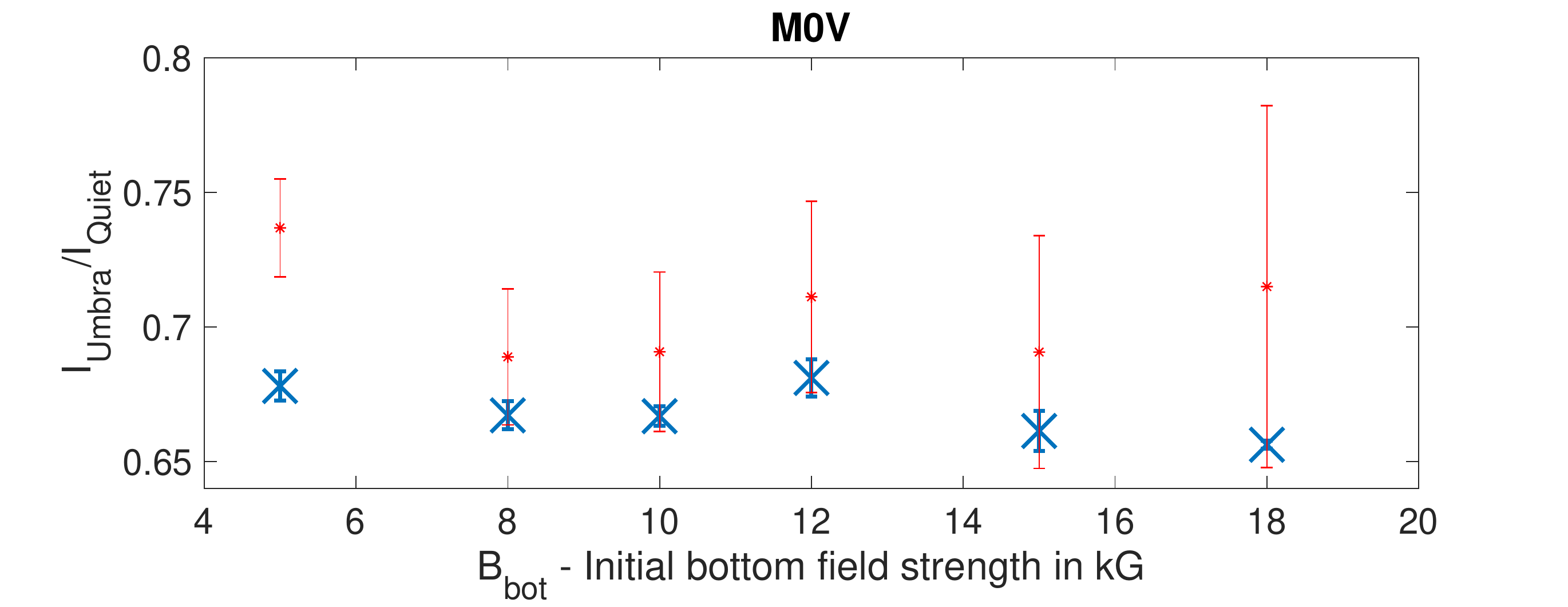}
      \caption{Relative intensity of spots plotted against initial field strengths at the lower boundary. Top to bottom: G2V, K0V and M0V. Red: with umbral dots. Blue: without umbral dots.
              }
        \label{fig:fig26}
    \end{figure}

\begin{figure}[!htb]
   \centering
	\hspace*{-0.5cm}\includegraphics[width=10.5cm]{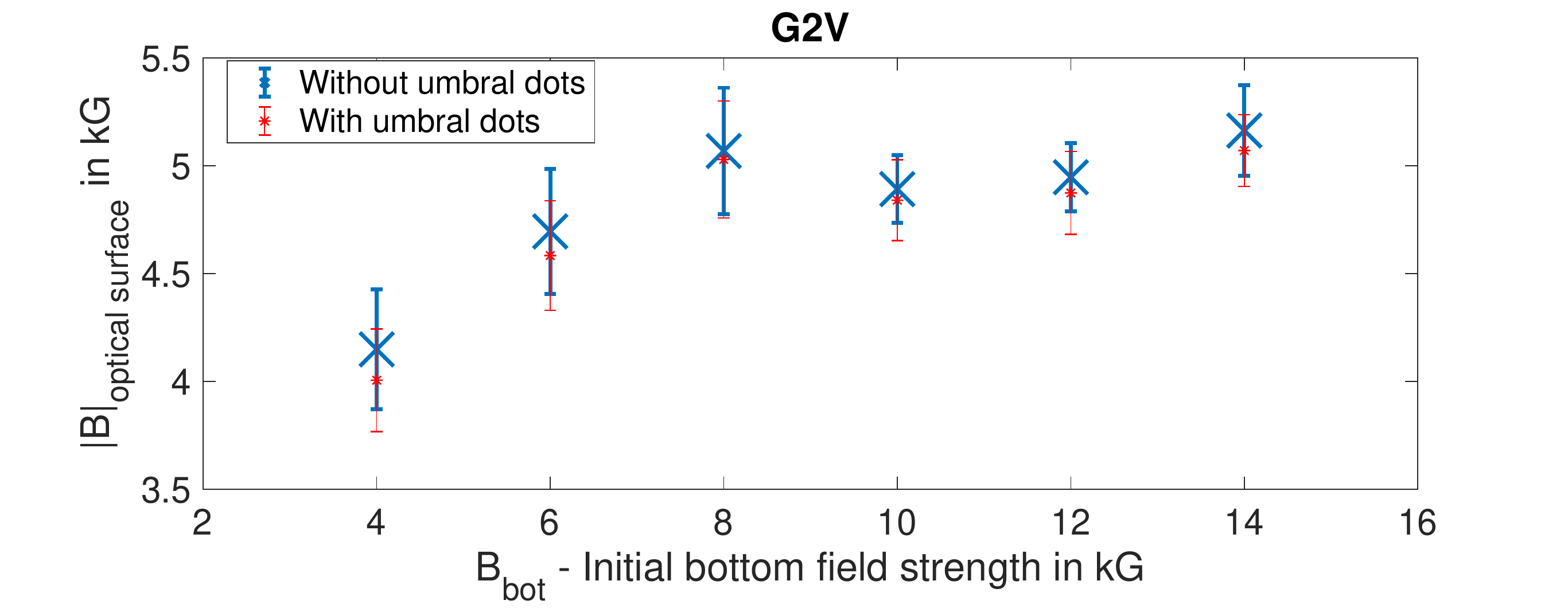}
    \hspace*{-0.5cm}\includegraphics[width=10.5cm]{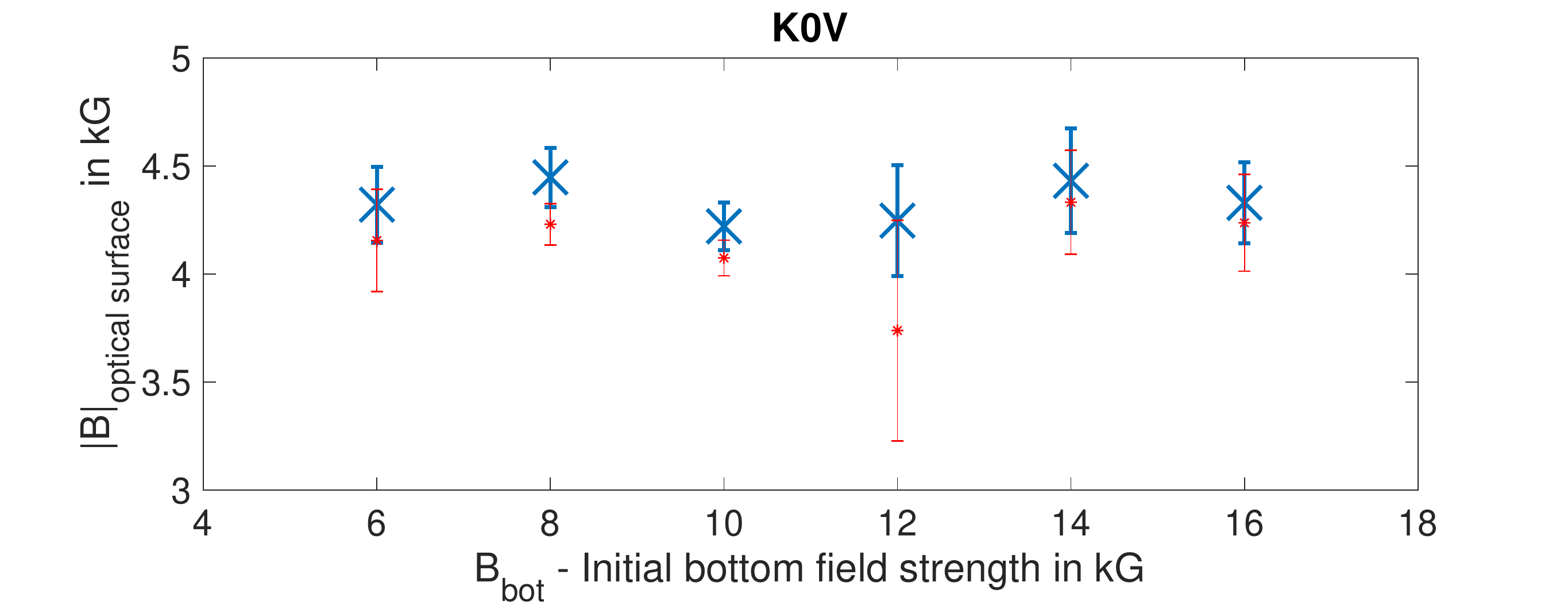}
    \hspace*{-0.5cm}\includegraphics[width=10.5cm]{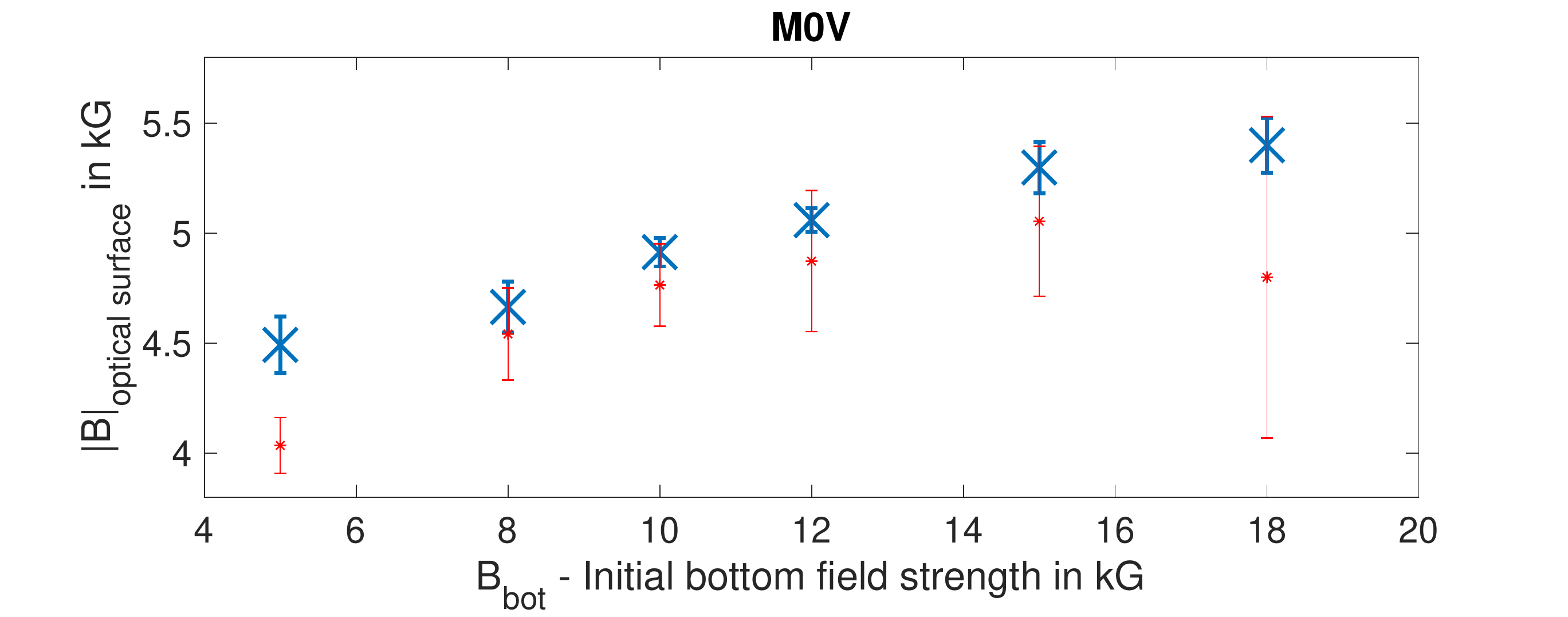}
      \caption{Final umbral field strength at the optical surface plotted against initial field strengths at the lower boundary. Top to bottom: G2V, K0V and M0V. Red: with umbral dots. Blue: without umbral dots.
              }
        \label{fig:fig27}
    \end{figure}

 We have selected the umbral region using two different methods. In the first method we simply set thresholds for the intensity and magnetic field strengths, and all points that satisfy the criteria are considered to be part of the umbra. We set a threshold of 1500 Gauss for the magnetic field strength in all three stars. 
For the G2V and K0V spots we used a relative intensity threshold of I$_{Umbra}$/I$_{Quiet}$ $<$ 0.5. Since in the M0V spot there were no regions with such low intensities we chose a threshold of I$_{Umbra}$/I$_{Quiet}$ $<$ 0.75. This method which chooses just those points that satisfy the above  mentioned criteria excludes the peaks in the intensity inside the spot region as seen in  Figure \ref{fig:fig24}. These intensity peaks seem to be the 2D equivalent of umbral dots. However they are significantly larger and brighter than umbral dots typically seen in 3D simulations. The second method of selecting the umbral region does not ignore these intensity peaks. We choose the first point and last point that satisfy the thresholds and take all points in between as shown in Figure \ref{fig:fig25}. \\
 Once we have chosen the umbra we average the properties over space and time, such that several granule lifetimes are covered. Each spot was averaged over a few hours.\\
 In all of the plots from \ref{fig:fig26} to \ref{fig:fig31}, the data points in blue exclude the umbral dots, and the data points in red include the umbral dots. The error bars show the standard deviation of the computed spatio-temporal averages.

\subsection{Varying $B_\mathrm{bot}$} 
 For this numerical experiment, we chose a $B_\mathrm{opt}$ of 2.2 kiloGauss for the G2V spots and slightly higher field strengths of 3.2 kiloGauss  and 3 kiloGauss for the K0V and M0V starspots respectively, and varied $B_\mathrm{bot}$, the field strength at the lower boundary. The choices for $B_\mathrm{opt}$ were motivated by the fact that average sunspots have field strengths in the 2-3 KiloGauss range, and we began with the assumption that starspot field strengths would not be drastically different. 
 
 Plotted in Figure \ref{fig:fig26} is the variation in I$_{Umbra}$/I$_{Quiet}$ with $B_\mathrm{bot}$ for the all three stars. The initial $B_\mathrm{opt}$ was the same for spots of the same spectral type. It is clear, in all three spectral types, that the relative intensity of the spots does not have a clear dependence on the field strength at the lower boundary especially when we do not consider the umbral dots. Naturally, the cases where we do not consider the umbral dots (marked in blue) have lower relative intensities. In none of the cases, the I$_{Umbra}$/I$_{Quiet}$ ratio changes by more than 0.1 even when $B_\mathrm{bot}$ is changed by a factor of 3. 
 
Figure \ref{fig:fig27} shows the variation of the magnetic field strength, finally obtained at the optical surface, with initial $B_\mathrm{bot}$. Like the relative intensity, this shows no significant dependence on the initially chosen $B_\mathrm{bot}$. In the G2V case,  after an increase in the first 3 cases (i.e. from $B_\mathrm{bot}$ = 4 kG to 8 kG), the final umbral field strength ceases to be sensitive to an increase in $B_\mathrm{bot}$. When the field strength at the lower boundary is too weak, we get a lot more spikes in the intensity which weakens the umbral field strength. When we increase $B_\mathrm{bot}$, the spikes in the intensity become rarer and the magnetic field is largely determined by the surface pressure. The M0V spots show the maximum dependence on $B_\mathrm{bot}$ as seen in the lower panel of Figure 27. The magnetic pressure forces the field lines to fan out until they experience pushback from the ambient gas and the balance between magnetic pressure and fluid pressure determines the umbral field strength. In the case of the M0V star, owing to the pressure scale heights being very small, the vertical extent of the box is only 1.3 Mm compared to the 7.3 Mm of the 2D G2V box. The sharp drop in magnetic field strength with height means that the field lines of the M0V spots are already highly fanned out and the resulting magnetic tension limits how much they can fan out further. Nevertheless, the dependence is marginal, as the final field strength increases by only 25\% even when $B_\mathrm{bot}$ is increased by a factor of 3.6.

\begin{figure*}
   \centering
	\hspace*{-0cm}\includegraphics[width=10.5cm]{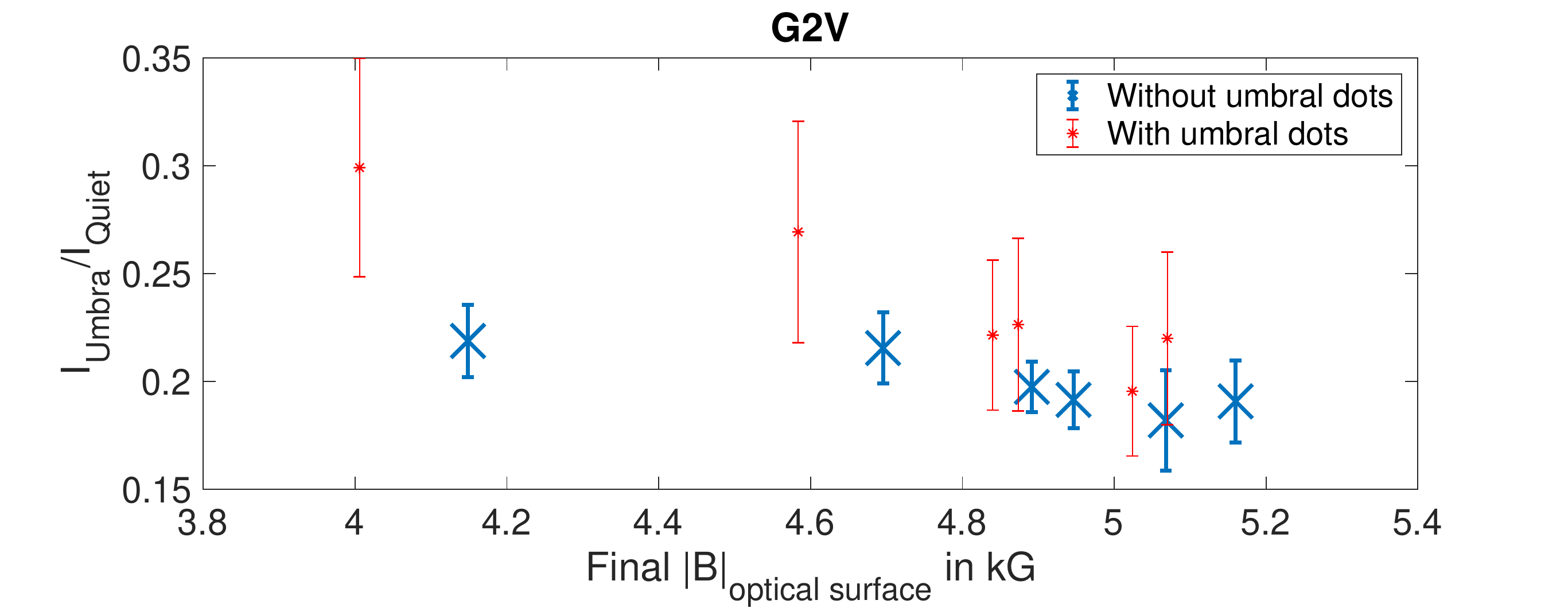}
    \hspace*{-0cm}\includegraphics[width=10.5cm]{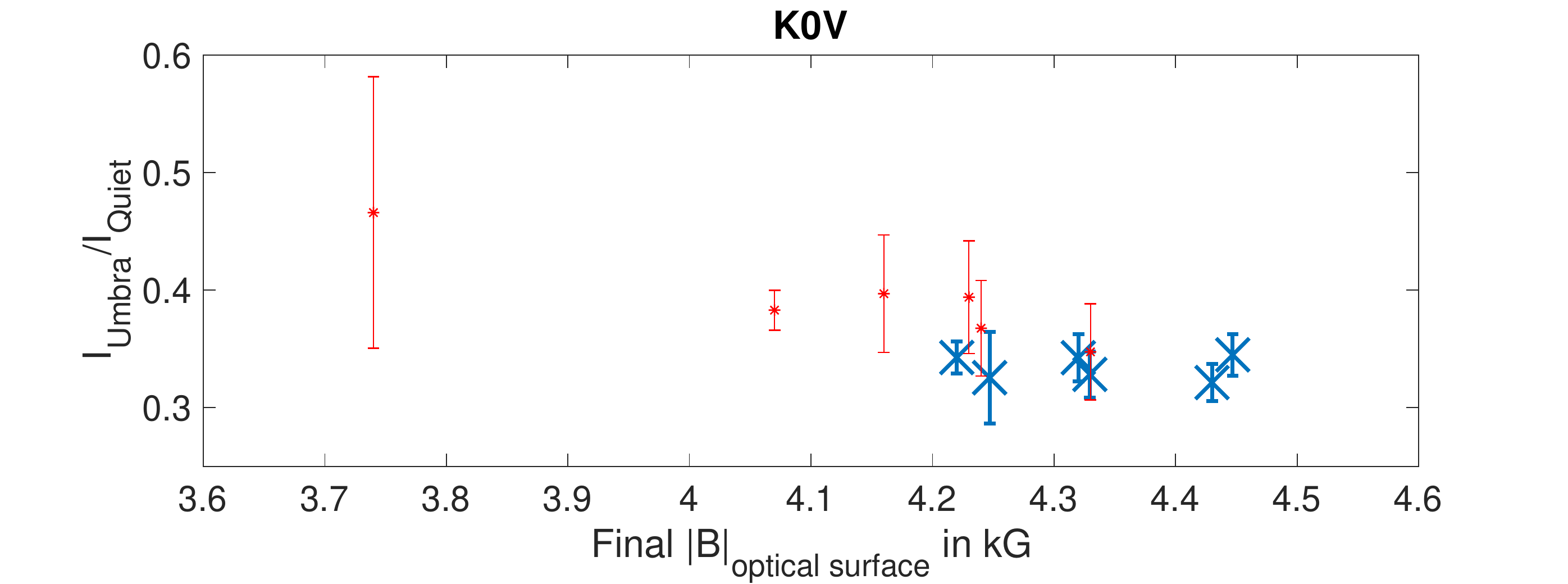}
    \hspace*{-0cm}\includegraphics[width=10.5cm]{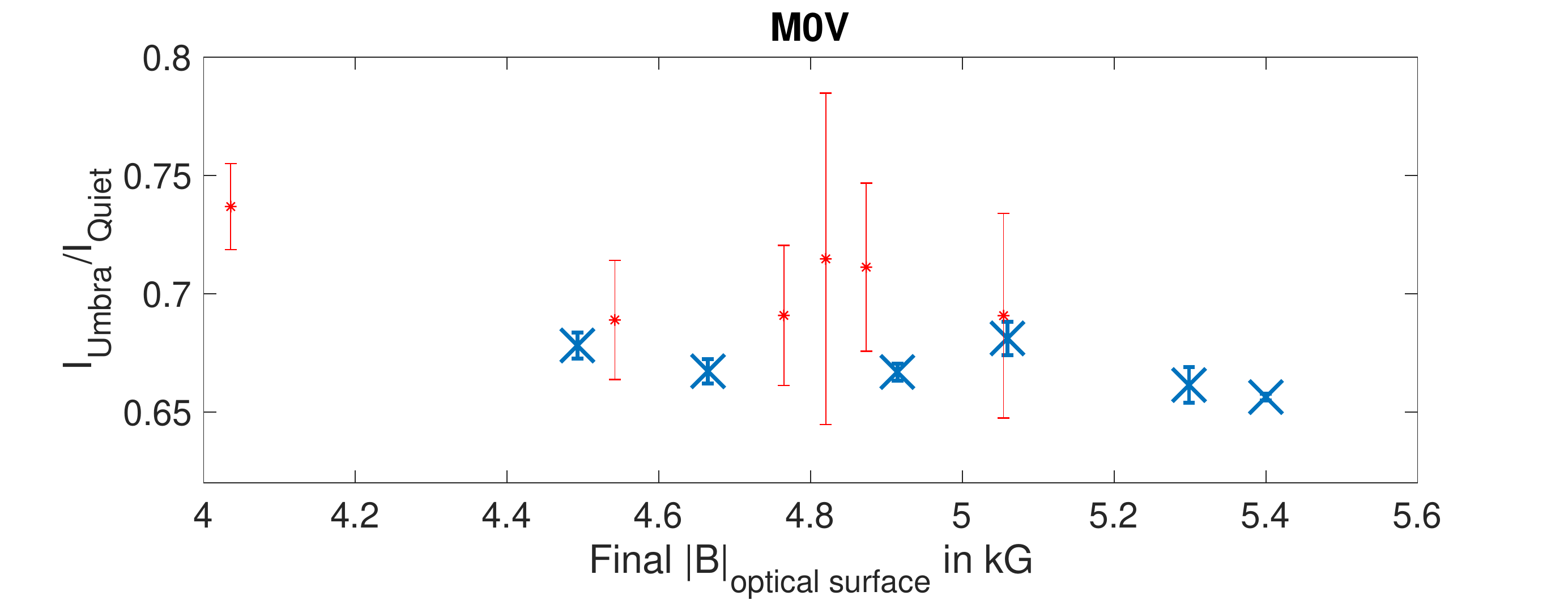}
      \caption{Relative intensity of spots plotted against final umbral field strengths at the optical surface. Top to bottom: G2V, K0V and M0V. Red: with umbral dots. Blue: without umbral dots.
              }
       \label{fig:fig28}
    \end{figure*}

Figure \ref{fig:fig28} shows the dependence of spot relative intensity on the final magnetic strength obtained at the optical surface. We see that for the G2V spot, the spots become darker with an increase in the photospheric field strength. However for the cooler K0 and M0 spots, the spot brightness shows no decrease with an increase in surface field strength. This is consistent with the fact the radiation plays a more important role in energy transport in the cooler K0 and M0 stars, thereby making the brightness of the spot less dependent on magnetic field strength. 

Although the relative intensities of the 2D spots are 5-10 \% lower than their 3D counterparts in all three stellar types, the decrease in spot relative intensity with stellar surface temperature is very well reproduced.  

\subsection{Varying $B_\mathrm{opt}$} 

Our very limited knowledge about surface field strengths on other stars, and the fact that M0V and K0V have higher surface pressures, prompted us to conduct further runs, where we kept the field strengths at the lower boundary ($B_\mathrm{bot}$) constant and increased the initial field strengths at the optical surface ($B_\mathrm{opt}$).  

\begin{figure}
   \centering
	\hspace*{-0.5cm}\includegraphics[width=10.5cm]{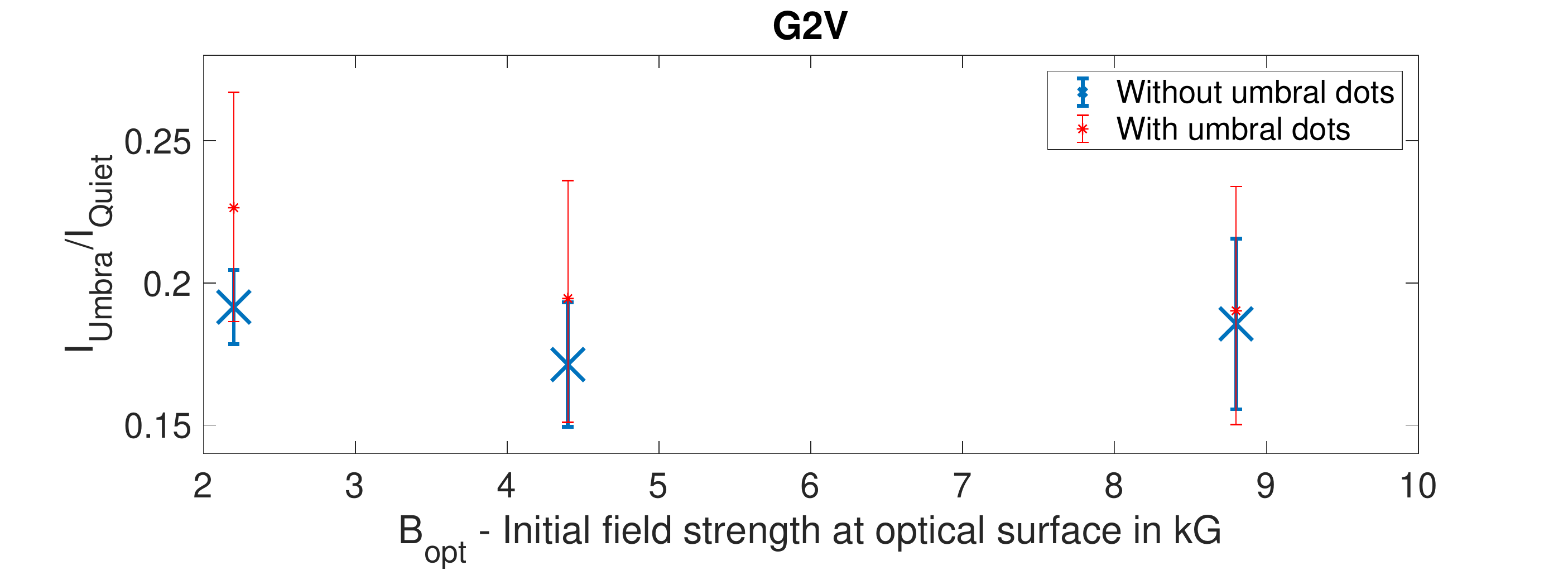}
    \hspace*{-0.50cm}\includegraphics[width=10.5cm]{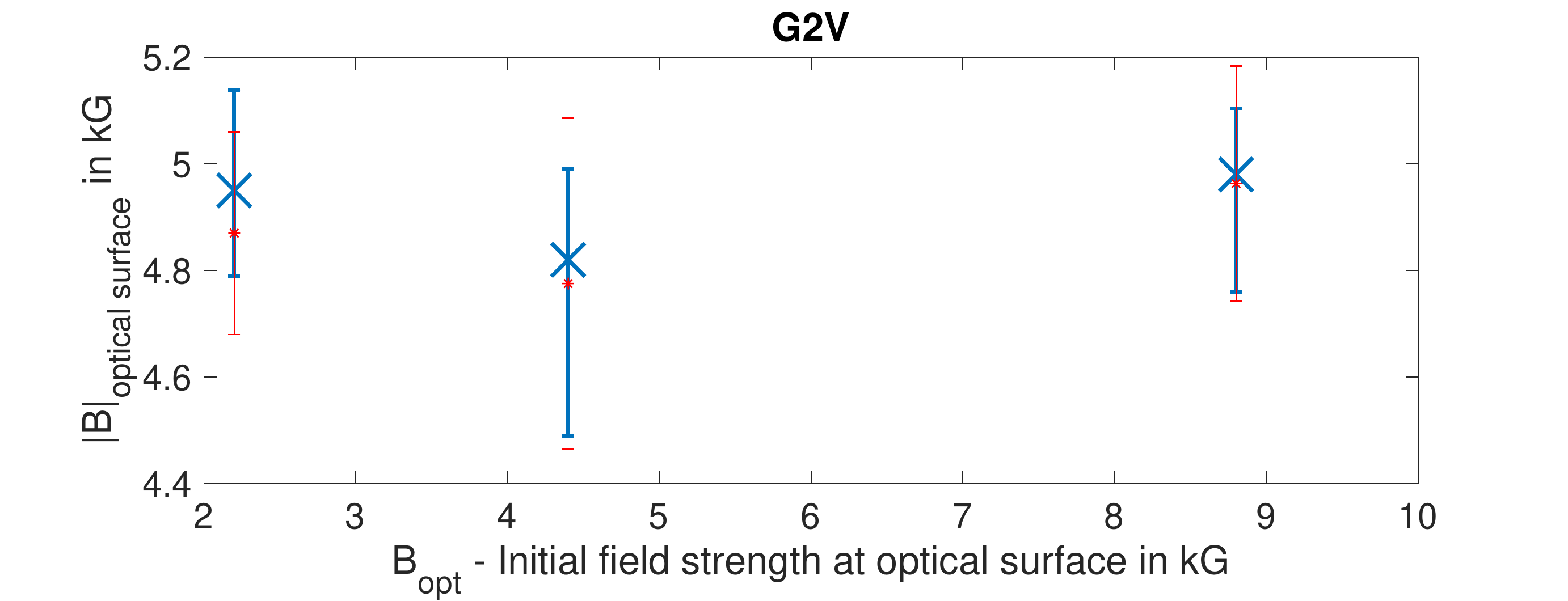}
    
      \caption{Both plots are for the G2V star. Relative intensity of spots (upper panel) and their final umbral field strengths (lower panel) plotted against initial field strengths at the optical surface.  Red: with umbral dots. Blue: without umbral dots.
              }
              \label{fig:fig29}
     \end{figure}

\begin{figure}
   \centering
	\hspace*{-0.0cm}\includegraphics[width=10.5cm]{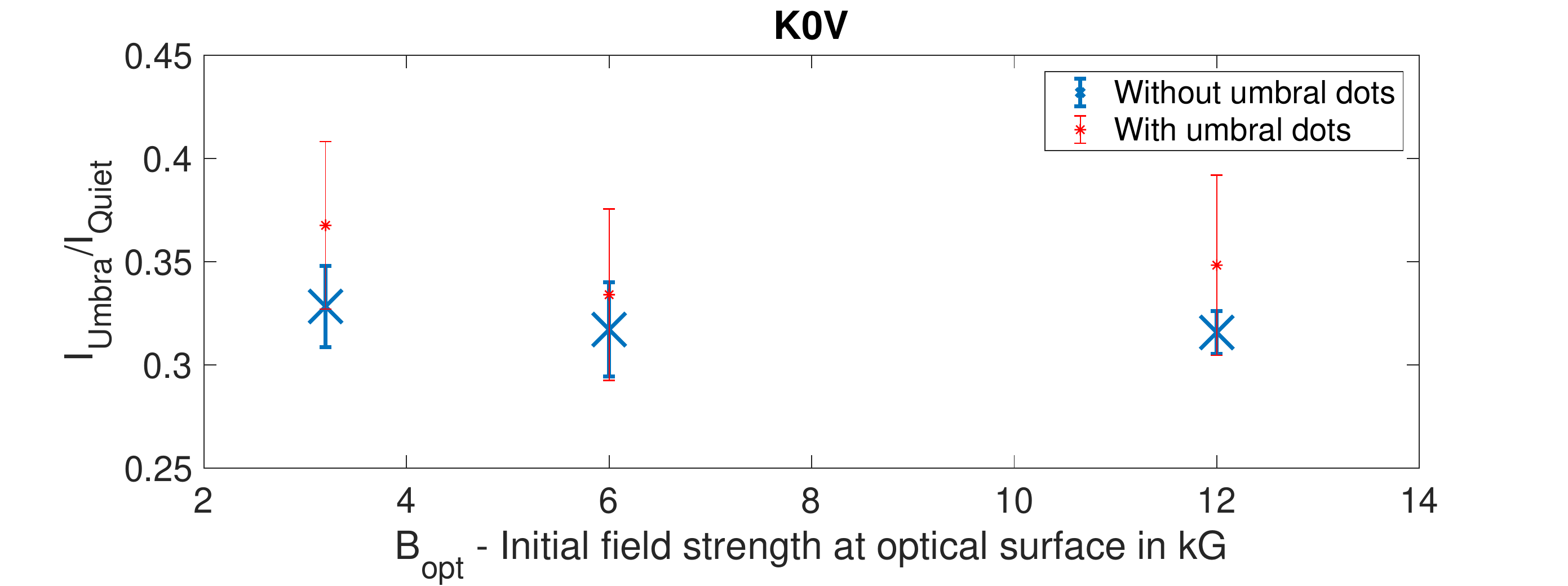}
    \hspace*{-0.0cm}\includegraphics[width=10.5cm]{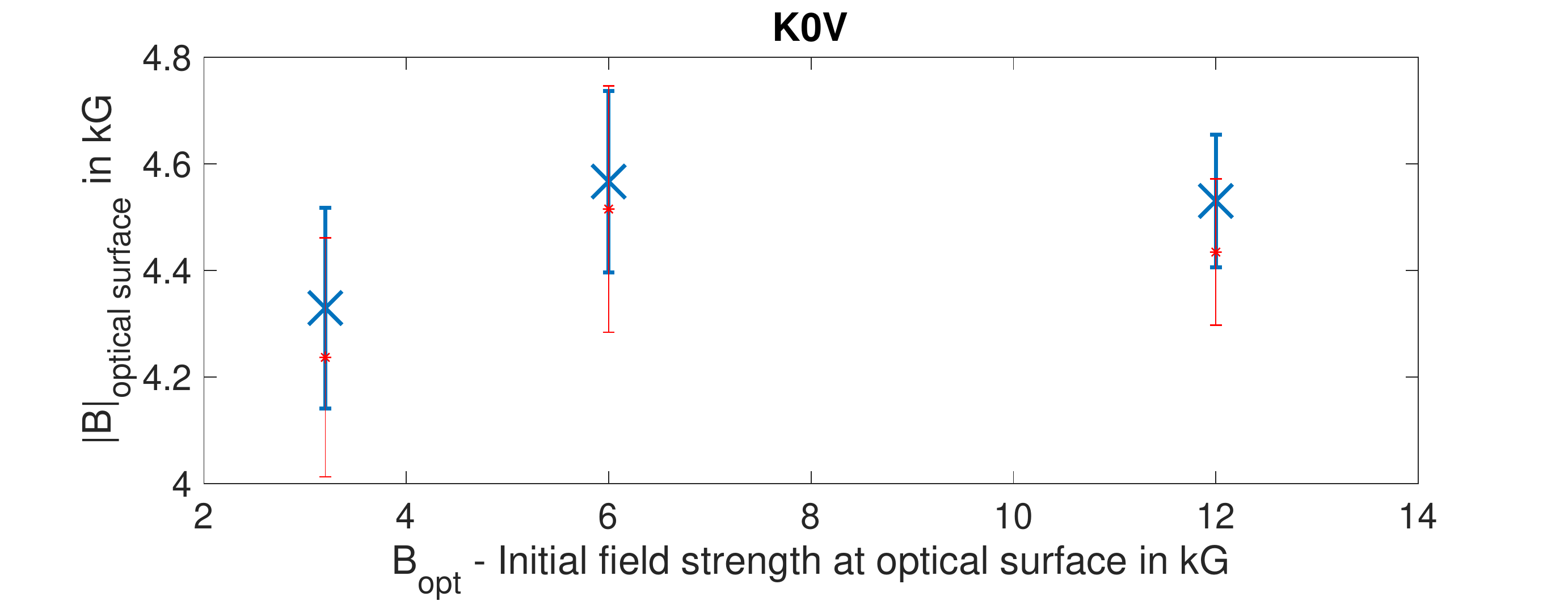}
    
      \caption{The same as Figure \ref{fig:fig29} but for the K0V star
              }
        \label{fig:fig30}
    \end{figure}

\begin{figure}
   \centering
	\hspace*{-0cm}\includegraphics[width=10.5cm]{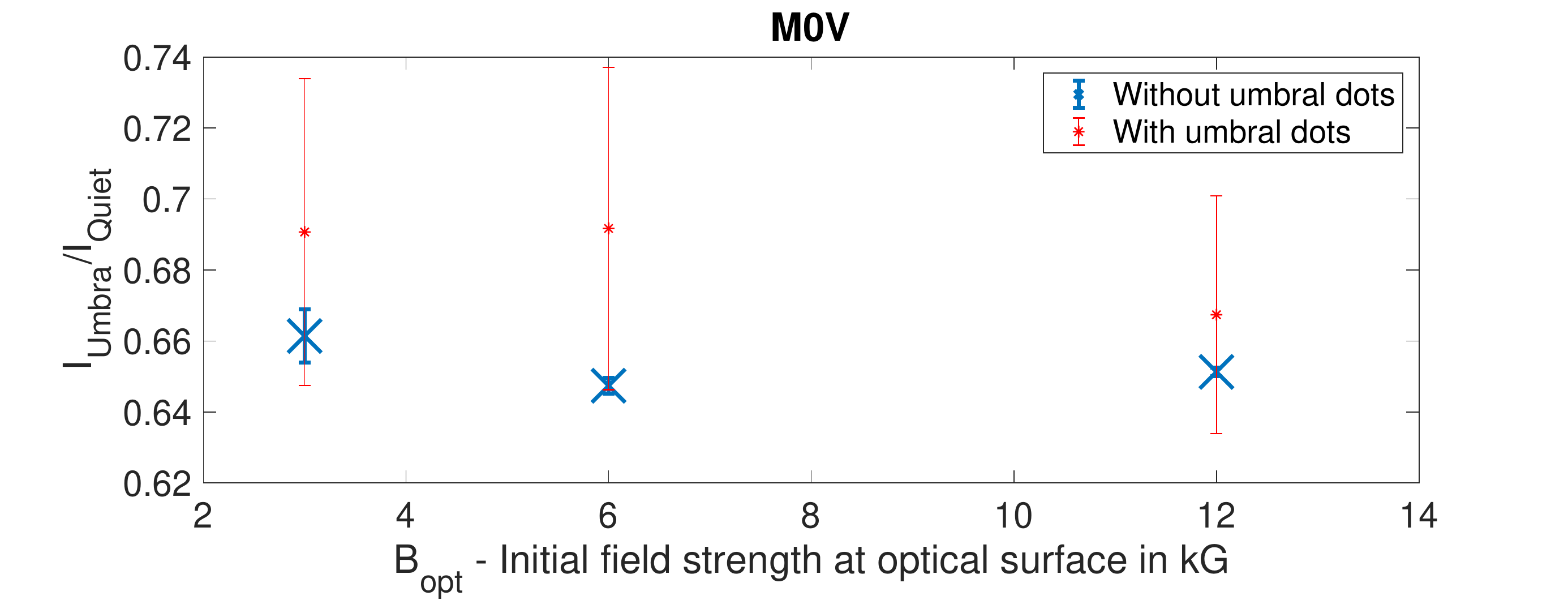}
    \hspace*{-0cm}\includegraphics[width=10.5cm]{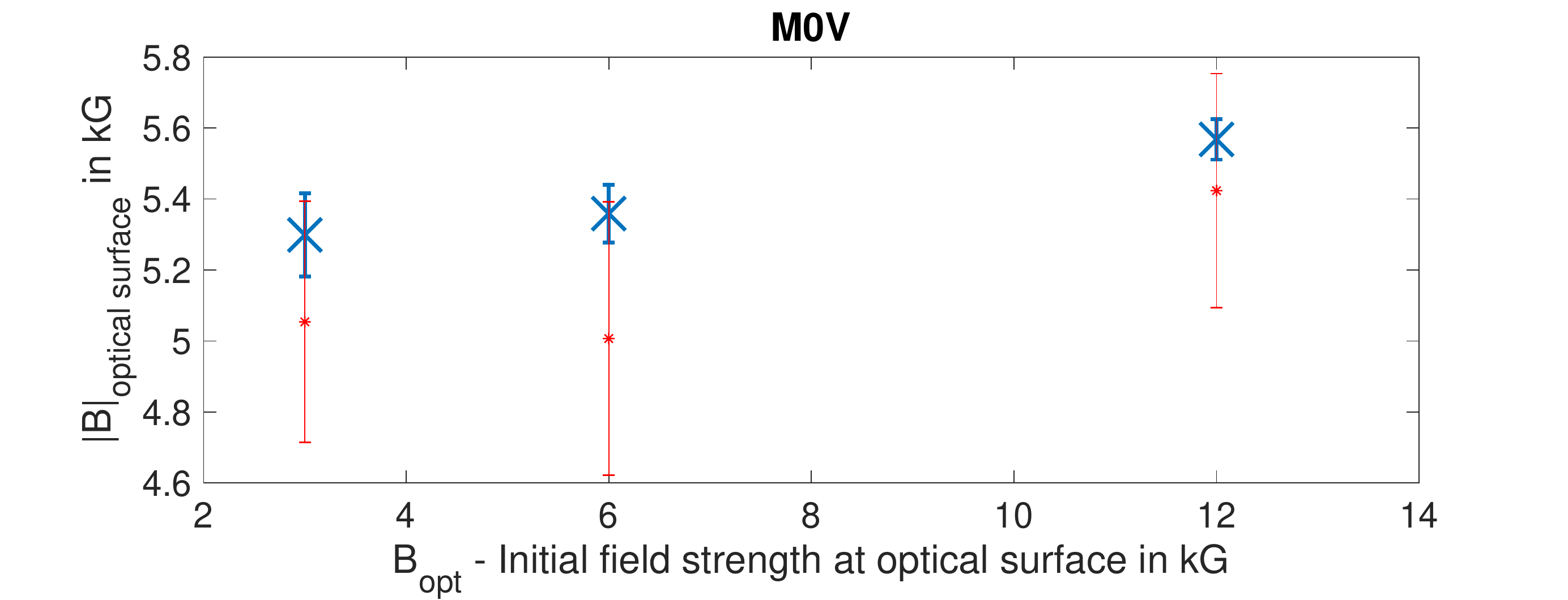}
    
      \caption{ The same as Figure \ref{fig:fig29} but for the M0V star
              }
       \label{fig:fig31}
    \end{figure}

For all of the spectral types we increased the initial $B_\mathrm{opt}$ to 2 and 4 times the magnitude used in our studies where we varied $B_\mathrm{bot}$. The $B_\mathrm{bot}$ used was 12 kiloGauss for the G2V runs, 16 kiloGauss for the K0V runs, and 15 kiloGauss for the M0V runs. We found that despite increasing the initial $B_\mathrm{opt}$ by a factor of 4, there is little change in the final relative intensity and magnetic field strength at the surface. This holds for all of the simulated stars as shown in Figures \ref{fig:fig29}, \ref{fig:fig30} and \ref{fig:fig31}.



\bibliography{starspots}{}
\bibliographystyle{aasjournal}



\end{document}